\documentclass[a4,11pt]{article}
\usepackage{graphicx} 
\usepackage{float}
\usepackage{amsmath,amsthm}
\usepackage{amsfonts}
\usepackage{latexsym}
\usepackage{amssymb}
\usepackage{setspace}
\usepackage{comment}
\usepackage{ascmac}
\usepackage{cases}
\usepackage{color}

\makeatletter
 
  \@addtoreset{equation}{section}
 \makeatother
\begin{document}
\title{{Equilibrium pricing of securities in the co-presence of \\cooperative and non-cooperative  populations}
\footnote{Forthcoming in {\it ESAIM: Control, Optimisation and Calculus of Variations}.}
}

\author{Masaaki Fujii\footnote{Quantitative Finance Course, Graduate School of Economics, The University of Tokyo. }
}
\date{ \footnotesize
The first version: 27 September, 2022\\
This version: 21 June, 2023.
}
\maketitle



\newtheorem{definition}{Definition}[section]
\newtheorem{assumption}{Assumption}[section]
\newtheorem{condition}{$[$ C}
\newtheorem{lemma}{Lemma}[section]
\newtheorem{proposition}{Proposition}[section]
\newtheorem{theorem}{Theorem}[section]
\newtheorem{remark}{Remark}[section]
\newtheorem{example}{Example}[section]
\newtheorem{corollary}{Corollary}[section]
%
\def\cala{{\cal A}}
\def\calb{{\cal B}}
\def\calc{{\cal C}}
\def\cald{{\cal D}}
\def\cale{{\cal E}}
\def\calf{{\cal F}}
\def\calg{{\cal G}}
\def\calh{{\cal H}}
\def\cali{{\cal I}}
\def\calj{{\cal J}}
\def\calk{{\cal K}}
\def\call{{\cal L}}
\def\calm{{\cal M}}
\def\caln{{\cal N}}
\def\calo{{\cal O}}
\def\calp{{\cal P}}
\def\calq{{\cal Q}}
\def\calr{{\cal R}}
\def\cals{{\cal S}}
\def\calt{{\cal T}}
\def\calu{{\cal U}}
\def\calv{{\cal V}}
\def\calw{{\cal W}}
\def\calx{{\cal X}}
\def\caly{{\cal Y}}
\def\calz{{\cal Z}}
%
\def\sskip{\hspace{0.5cm}}
\def\simleq{ \raisebox{-.7ex}{\em $\stackrel{{\textstyle <}}{\sim}$} }
\def\leqsim{ \raisebox{-.7ex}{\em $\stackrel{{\textstyle <}}{\sim}$} }
\def\nn{\nonumber}
\def\be{\begin{equation}}
\def\ee{\end{equation}}
\def\bea{\begin{eqnarray}}
\def\eea{\end{eqnarray}}
%

\def\calf{{\cal F}}
\def\wt{\widetilde}
\def\mbb{\mathbb}
\def\ol{\overline}
\def\ul{\underline}
\def\sign{{\rm{sign}}}
\def\wh{\widehat}
\def\mg{\mathfrak}
\def\display{\displaystyle}

\def\vr{\varrho}
\def\ep{\epsilon}

\def\Prb{\mbb{P}}
\def\del{\delta}
\def\Del{\Delta}

\def\deln{\delta_{\mathfrak{n}}}
\def\oldeln{\overline{\delta}_{\mathfrak{n}}}
\def\vep{\varepsilon}

\def\red{\textcolor{red}}
\def\Ito{It\^o}
\def\blangle{\bigl\langle}
\def\Blangle{\Bigl\langle}
\def\brangle{\bigr\rangle}
\def\Brangle{\Bigr\rangle}
\def\bi{\begin{itemize}}
\def\ei{\end{itemize}}
\def\ac{\acute}
\def\pr{\prime}
\def\mgn{\mathfrak{n}}

\def\part{\partial}
\def\ul{\underline}
\def\ol{\overline}
\def\vp{\varpi}
\def\nn{\nonumber}
\def\be{\begin{equation}}
\def\ee{\end{equation}}
\def\bea{\begin{eqnarray}}
\def\eea{\end{eqnarray}}
\def\bg{\boldsymbol}
\def\bull{$\bullet~$}
\def\ex{\mbb{E}}
\def\opb{\wh{\beta}}

\newcommand{\Slash}[1]{{\ooalign{\hfil/\hfil\crcr$#1$}}}
\vspace{-8mm}
\begin{abstract}
In this work, we develop an equilibrium model for price formation  of securities in a market composed of two populations of different types:
the first one consists of  cooperative agents, while the other one consists of non-cooperative agents.
The trading of every cooperative member is assumed to be coordinated by a central planner.
In the large population limit, the problem for the central planner 
is shown to be a conditional extended mean-field control. 
In addition to the convexity assumptions, if the relative size of the cooperative population is small enough, 
then we are able to show the existence of a unique equilibrium for both the finite-agent and the mean-field models.
The strong convergence to the mean-field model is also proved under the same conditions.
\end{abstract}


\section{Introduction}
In this paper, we study the problem of equilibrium price formation by considering, as an example,  a simple model of 
a securities market. The prices of securities
have  always been the center of interest of many people for centuries. 
They inspire people to consider their properties not only for speculative reasons but also 
for true academic interests, such as how they are determined, how they behave in a given environment,
how they are influenced by preferences and the level of risk-averseness of investors, and so on.
In fact,  since the appearance of modern securities markets, the huge amount of price data they provide
has induced the significant developments of  related academic areas, 
such as financial mathematics, financial economics,   statistical inference  etc.
However, despite its importance, we are still quite far from the complete understanding  of a price process
even when we restrict our attention to a standard liquid stock.
The main reason behind this  difficulty is that any price process is inevitably affected by complicated interactions among 
a large number of market participants, which includes various financial firms, individual investors, 
governments and even supernational institutions. 

The recent progress in the mean field game (MFG) theory has started to shed new light on 
the long-standing problem of multi-agent games.
The MFG theory was pioneered by the seminal works of Lasry \& Lions~\cite{Lions-1, Lions-2, Lions-3}
and Huang et al.~\cite{Caines-Huang,  Caines-Huang-2, Caines-Huang-3}, 
which characterizes a Nash equilibrium by a coupled system of Hamilton-Jacobi-Bellman and Kolmogorov equations.
A probabilistic approach to the MFG theory based on forward backward stochastic differential equations (FBSDEs) 
of McKean-Vlasov type was established by Carmona \& Delarue~\cite{Carmona-Delarue-MFG, Carmona-Delarue-MFTC}.
The two volumes of \cite{Carmona-Delarue-1, Carmona-Delarue-2} by the same authors provide 
readers a detailed account of the topic. The method using the concept of relaxed controls (see, for example, \cite{Lacker-1, Lacker-2}), 
which can significantly weaken the regularity assumptions,  has also been intensively studied.
For interested readers, there are an excellent lecture note~\cite{Cardaliaguet-note} and 
many monographs such as \cite{Bensoussan-mono, Gomes-eco, Gomes-reg} on various topics regarding the MFG theory.

Along side of its theoretical developments,  MFGs  have successfully found vast applications in various 
areas. In particular, energy and financial markets have been popular targets for analysis.
As is well known, any price in these free competitive markets is 
determined so that it balances the supply and demand of the asset.
Unfortunately, this {\it market clearing condition} does not fit well to the concept of Nash equilibrium.
In fact, if we shift a control of one agent away from her equilibrium strategy
while keeping the others unchanged, then the market clearing condition would be broken down.
Since the MFG theory has been developed primarily for the analysis of Nash equilibrium,
a standard approach in the literature assumes a specific structure of the price process without
imposing the market clearing condition. A popular way is to hypothesize 
that the price process is decomposed into two parts; one is a so-called fundamental price
whose dynamics is exogenously given
and the other is a price-impact term whose size is affected by actions of agents in a specific manner. 
Another way is to suppose that a price is given by a certain function
of total demand or supply of the asset.  
One can find many interesting examples on optimal executions~\cite{Fu-Horst-1, Fu-Horst-2, Evangelista-execution},
games in a smart grid~\cite{Matoussi, Djehiche-E}, price formation in an electricity market~\cite{Feron, Feron-2},
entry and exit games~\cite{Tankov-stopping}, trading with different beliefs~\cite{Jaimungal},
and trading of exhaustible commodities~\cite{Graber-commodity}, just to name a few.
There are also macroeconomic applications of MFGs, for examples,  
on  growth, inequality and unemployment, and so on~\cite{BM-2,  Aiyagari, Bayraktar, BM-1}.
 
In the above examples, the structure of a price process is assumed exogenously and the emphasis of research is not on the price process itself.
In this paper, on the other hand,  our primary  interest is in the construction of a price process endogenously
from the market clearing condition.
There has also been recent progress in the field of MFGs regarding this direction of research.
Gomes \& Sa\'{u}de~\cite{Gomes-Saude} present a deterministic model of electricity price formation.
Different mathematical approaches and numerical calculation techniques on the same model
are  developed by Ashrafyan et al.~\cite{Gomes-variational, Gomes-potential}.
Its generalization with a random supply function  is provided by Gomes et al.~\cite{Gomes-random-supply}.
Evangelista et al.~\cite{Evangelista-game-liquidity} use the actual high-frequency data 
to give a promising numerical results for the listed stocks in several exchanges.
Shrivats et al.~\cite{Firoozi-1} solve the price formation problem of solar renewable energy certificate (SREC),
 and Firoozi et al.~\cite{Firoozi-2} deal with a principal-agent problem 
in the associated emission market.  Fujii \& Takahashi~\cite{Fujii-Takahashi} solve a stochastic mean-field price model
of securities and prove 
that the market is cleared  in the large population limit.
In \cite{Fujii-mfg-convergence}, the same authors provide the FBSDE characterization of  the market clearing price in the finite-agent market,
and then prove its strong convergence to the mean-field model of \cite{Fujii-Takahashi}.
Its extension to the presence of a major agent is studied in \cite{Fujii-Major}.

In this paper, we build upon the previous works~\cite{Fujii-Takahashi, Fujii-mfg-convergence, Fujii-Major} 
and develop a new model of a securities market composed of {\bf two populations of different types}:
the first one consists of cooperative agents,  while the other one consists of non-cooperative agents.
Every agent is supposed to be a registered financial firm trying to minimize her cost functional by the continuous trading at the common exchange. 
What is new in this paper is the existence of the cooperative population,
which can be a large number of subsidiaries of a giant financial group or a temporary alliance of many financial firms
for a certain purpose, such as liquidation of assets of a defaulted financial firm.  
In addition to the pure academic interests, understanding the market with cooperative agents is of critical importance
for financial regulators as well as policy makers.
The trading of every cooperative member is assumed to be coordinated by a central planner.
We show that, in the mean field limit, this cooperation makes the problem of the central planner 
an {\it extended mean-field control}~\cite{Carmona-E} with common noise,  
where the law of the controls of the cooperative members  enters her cost functional.
Here, let us refer to some works on extended mean-field control problems:
Graber~\cite{Graber} studies a linear-quadratic model for production of a exhaustible resource;
Acciaio et al.~\cite{Carmona-E} present the necessary and sufficient conditions of Pontryagin's maximum principle; 
Nie \& Yan~\cite{Nie-Yan} study the problem under the partial observation;
Motte \& Pham~\cite{Motte-Pham} investigate a mean-field Markov decision problem;
Djete et al.~\cite{Djete-0, Djete-1} and Djete~\cite{Djete-2} provide a dynamic programing principle 
as well as the limit theory by making use of the relaxed control technique and its generalization.

To the author's knowledge, this is the first work studying the equilibrium price formation 
in the co-presence of cooperative and non-cooperative agents.
Using a general convex cost function for the cooperative agents,
we show that there exists a unique market clearing equilibrium for a given duration if the relative size of the 
cooperative  population is small enough. After solving the problem in the finite-agent market, we construct the corresponding MFG model
and finally prove the strong convergence  when the large population limit is taken.
We want to emphasize that,  although we have chosen a  model of a securities market,
the developed methods  (with appropriate modifications) will also be useful
to describe other markets with cooperative producers and/or buyers,
for example,  the crude oil market with OPEC a cooperative association of  exporters.
In addition to the financial economics perspective, the current work also contributes to 
the MFG theory in two ways. Firstly, in addition to the sufficiency condition, we show the existence of a unique 
optimal solution to an extended mean-field control problem with common noise. The existence result 
using the adjoint equations for this type of problem is new except those obtained in the linear-quadratic settings~\cite{Carmona-E, Graber}.
Moreover, due to the co-presence of non-cooperative agents, the problem has a non-standard form of controlled McKean-Vlasov FBSDEs 
instead of controlled McKean-Vlasov SDEs.  Next, although the general convergence results for  extended mean-field control problems
are already studied by \cite{Djete-1,Djete-2},
our  proof  is directly built upon the result for the standard (not extended) mean-field control problem
by Carmona \& Delarue~\cite{Carmona-Delarue-MFTC}  
and the monotonicity technique by Fujii \& Takahashi \cite{Fujii-mfg-convergence}.
In return for our strong convexity assumptions,  the proof is done in a straightforward manner.

The organization of the paper is as follows. After explaining some notations in Section~\ref{sec-notation},
we solve the finite-agent market model in Section~\ref{sec-finite}. The corresponding mean-field model
is built and solved in Section~\ref{sec-mfg}. Finally, in Section~\ref{sec-convergence}, we prove the strong convergence 
to the mean-field limit. Section~\ref{sec-conclusion} concludes the paper.

\section{Notations}
\label{sec-notation}
We first introduce $(1+N_1+N_2)$ complete probability spaces.
\bi
\small
\item $(\Omega^0,\calf^0,\Prb^0)$ is a complete probability space with a complete and right-continuous filtration $\mbb{F}^0:=(\calf_t^0)_{t\geq 0}$
generated by $d_0$-dimensional standard Brownian motion $W^0:=(W_t^0)_{t\geq 0}$.
\item For each $1\leq i\leq N_1$, $(\ol{\Omega}^{1,i},\ol{\calf}^{1,i},\ol{\Prb}^{1,i})$ is a complete probability space with a complete
and right-continuous filtration $\ol{\mbb{F}}^{1,i}:=(\ol{\calf}_t^{1,i})_{t\geq 0}$ generated by $d_1$-dimensional standard Brownian motion $W^i:=(W^i_t)_{t\geq 0}$ and a $\sigma$-algebra $\sigma(\xi^i)$ belonging to $\ol{\calf}^{1,i}_0$,  generated by a square integrable $\mbb{R}^n$-valued random variable $\xi^i$.
The distribution of $\xi^i$ is the same for every $i$.
\item For each $1\leq j\leq N_2$, $(\ol{\Omega}^{2,j},\ol{\calf}^{2,j},\ol{\Prb}^{2,j})$ is a complete probability space with a complete
and right-continuous filtration $\ol{\mbb{F}}^{2,j}:=(\ol{\calf}_t^{2,j})_{t\geq 0}$ generated by $d_2$-dimensional standard Brownian motion
$B^j:=(B_t^j)_{t\geq 0}$
and a $\sigma$-algebra $\sigma(\eta^j)$ belonging to $\ol{\calf}_0^{2,j}$,  generated by a square integrable $\mbb{R}^n$-valued random variable $\eta^j$.
The distribution of $\eta^j$ is the same for every $j$.
\ei
We then introduce probability spaces defined by the Cartesian products of those defined above.
\bi
\small
\item $(\ol{\Omega},\ol{\calf},\ol{\Prb})$ is a probability space defined by the product  $\ol{\Omega}:=\ol{\Omega}^{1,1}\times \ldots \times
\ol{\Omega}^{1,N_1}\times \ol{\Omega}^{2,1}\times \ldots \times \ol{\Omega}^{2,N_2}$ with $(\ol{\calf},\ol{\Prb})$ the completion of $(\otimes_{i=1}^{N_1}\ol{\calf}^{1,i}\otimes_{j=1}^{N_2}\ol{\calf}^{2,j}, 
~\otimes_{i=1}^{N_1} \ol{\Prb}^{1,i}\otimes_{j=1}^{N_2}\ol{\Prb}^{2,j})$. $\ol{\mbb{F}}:=(\ol{\calf}_t)_{t\geq 0}$
denotes the complete and right-continuous augmentation of $(\otimes_{i=1}^{N_1}\ol{\calf}^{1,i}_t\otimes_{j=1}^{N_2}\ol{\calf}^{2,j}_t)_{t\geq 0}$.
\item $(\Omega,\calf,\Prb)$ is a probability space defined by the product  $\Omega:=\Omega^0\times \ol{\Omega}$ 
with $(\calf,\Prb)$ the completion of $(\calf^0\otimes \ol{\calf}, ~\Prb^0\otimes \ol{\Prb})$. $\mbb{F}:=(\calf_t)_{t\geq 0}$
denotes the complete and right-continuous augmentation of $(\calf_t^0\otimes \ol{\calf}_t)_{t\geq 0}$.
The expectation with respect to $\mbb{P}$ is denoted by $\ex[\cdot]$, and a generic element of $\Omega$ is denoted by
$\omega=(\omega^0,\ol{\omega})$ with $\omega^0\in \Omega^0$ and $\ol{\omega}\in \ol{\Omega}$.
\ei
For notational simplicity, we do not distinguish a random variable $X$ defined on a marginal probability space, such as $(\ol{\Omega},\ol{\calf},
\ol{\Prb})$, with its natural extension to $(\Omega,\calf,\mbb{P})$.
Without loss of generality, we may always suppose that the probability spaces are the completions of those countably generated by working on 
the Borel sets of appropriate Polish spaces.

Throughout the work, $T~(>0)$ is a given time horizon, the symbol $L$ denotes a given positive constant, and the symbol $C$ a general positive constant
which may change line by line. For any measurable space $(\Omega,\calg)$ with filtration $\mbb{G}:=(\calg_t)_{t\geq 0}$, 
we use the following notations for frequently encountered spaces: 
\bi
\small
\item $\cals^n_+$ denotes the space of $n\times n$ strictly positive definite symmetric matrices. 
\item $\mbb{L}^2(\calg;\mbb{R}^d)$ denotes the set of $\mbb{R}^d$-valued $\calg$-measurable random variables $X$ satisfying
\be
\begin{split}
\|X\|_{\mbb{L}^2}:=\ex[|X|^2]^\frac{1}{2}<\infty. \nn
\end{split}
\ee
\item $\mbb{S}^2 (\mbb{G};\mbb{R}^d)$ is the set of $\mbb{R}^d$-valued $\mbb{G}$-adapted continuous processes $X$ satisfying
\be
\small
\begin{split}
\|X\|_{\mbb{S}^2}:=\ex\bigl[\sup_{t\in[0,T]}|X_t|^2\bigr]^\frac{1}{2}<\infty. \nn
\end{split}
\ee
\item $\mbb{H}^2(\mbb{G};\mbb{R}^d)$ is the set of $\mbb{R}^d$-valued $\mbb{G}$-progressively measurable processes $Z$ satisfying
\be
\small
\begin{split}
\|Z\|_{\mbb{H}^2}:=\ex\Bigl[\Bigl(\int_0^T |Z_t|^2dt\Bigr)\Bigr]^\frac{1}{2}<\infty. \nn
\end{split}
\ee
\item $\call(X)$ denotes the law of a random variable $X$.
\item $\calp(\mbb{R}^d)$ is the set of probability measures on $(\mbb{R}^d,\calb(\mbb{R}^d))$.
\item $\calp_p(\mbb{R}^d)$ with $p\geq 1$ is the subset of $\calp(\mbb{R}^d)$ with finite $p$th moment, i.e. the set of 
$\mu\in \calp(\mbb{R}^d)$ satisfying
\be
\small
\begin{split}
M_p(\mu):=\Bigl(\int_{\mbb{R}^d}|x|^p\mu(dx)\Bigr)^\frac{1}{p}<\infty. \nn
\end{split}
\ee
We always assign $\calp_p(\mbb{R}^d)$ the $p$-Wasserstein distance $W_p$, which makes $\calp_p(\mbb{R}^d)$ a Polish space.
It is defined by, for any $\mu, \nu\in \calp_p(\mbb{R}^d)$, 
\be
\small
\begin{split}
W_p(\mu,\nu):=\inf_{\pi\in \Pi_p(\mu,\nu)}\Bigl[\Bigl(\int_{\mbb{R}^d\times \mbb{R}^d}|x-y|^p \pi(dx,dy)\Bigr)^\frac{1}{p}\Bigr], \nn
\end{split}
\ee
whre $\Pi_p(\mu,\nu)\subset \calp_p(\mbb{R}^d\times \mbb{R}^d)$ denotes the set of probability measures
with marginals $\mu$ and $\nu$. 
\item For $k=1,2$, the empirical mean of $N_k$ variables $(x^i)_{i=1}^{N_k}$ is denoted by
$
\mg{m}_k(x):=\frac{1}{N_k}\sum_{i=1}^{N_k}x^i. \nn
$
\ei
We often omit the arguments such as $(\mbb{G}, \mbb{R}^d)$ when they are clear from the context.
\section{Price formation in a finite-agent market}
\label{sec-finite}
In this section, we consider the price formation problem in a finite-agent market composed of two populations:
the first  one consists of $N_1$ cooperative agents, and the second one consists of $N_2$ non-cooperative agents.
We often use  $i$ (respectively, $j$) to index an agent in the first (respectively, second) population.
Each agent represents a registered financial firm trading  at a common securities exchange.
She is supposed to have many individual clients who cannot directly access to the exchange,  and she provides
financial services by using her own inventory to answer the sale and purchase requests from those clients. This is the so-called
over-the-counter (OTC) market. Therefore, every agent faces the stochastic order flows from her  OTC clients
in addition to idiosyncratic as well as common market shocks. Under such an environment,
every agent carries  out the optimal continuous trading at the exchange to minimize her cost functional. 
For the members of the first population, we assume that every trading is coordinated by a central planner.

We suppose that there are $n$ tradable securities in the market, whose prices are denoted by $(\vp_t^k)_{k=1}^n$.
The goal of the paper is to determine the price process $\vp:=(\vp_t)_{t\in[0,T]}$ endogenously so that it clears the market, i.e. 
matches the supply and demand of securities at the exchange.
Unless otherwise stated, we will work on the filtered probability space $(\Omega,\calf,\mbb{F},\Prb)$ defined in the last section.
By construction, $(\xi^i)_{i=1}^{N_1}$ and also $(\eta^j)_{j=1}^{N_2}$ are independently  and identically distributed (i.i.d.).
Let us introduce the following processes to describe the various types of information:
\bi
\small 
\item $c^0:=(c_t^0)_{t\in[0,T]}$ belongs to $\mbb{H}^2(\mbb{F}^0;\mbb{R}^m)$ with $\sup_{t\in[0,T]}\ex[|c_t^0|^2]<\infty$,
which is used to represent common market information including the cash flows from the securities. 
\item For $1\leq i\leq N_1$, $c^i:=(c_t^i)_{t\in[0,T]}$ belongs to $\mbb{H}^2(\ol{\mbb{F}}^{1,i};\mbb{R}^m)$ with
$\sup_{t\in[0,T]}\ex[|c_t^i|^2]<\infty$. They are used to represent idiosyncratic shocks for the first population, and
$(c^i)_{i=1}^{N_1}$ are  i.i.d.
\item For $1\leq j\leq N_2$, $\mg{c}^j:=(\mg{c}_t^j)_{t\in[0,T]}$ belongs to $\mbb{H}^2(\ol{\mbb{F}}^{2,j};\mbb{R}^m)$
with $\sup_{t\in[0,T]}\ex[|\mg{c}_t^j|^2]<\infty$. They are used to represent idiosyncratic shocks for the second population,
and $(\mg{c}^j)_{j=1}^{N_2}$ are  i.i.d.
\ei
Here, the dimension $m\in \mbb{N}$ is commonly chosen just for notational simplicity.
\\

Before going to study the details of the optimization problem faced by each  population separately, 
let us summarize the important features of each problem here.
\bi
	\item The optimization problem for the second population $1\leq j\leq N_2$:
		\bi \small
			\item Each agent $j$ tries to minimize her cost functional by continuous trading at the exchange (see $(\ref{non-c-problem})$).
			\item Her cost functional (see $(\ref{2nd-cost-functional})$) depends on her trading speed, her inventory size, the price of securities
				  as well as the shocks induced by $(c^0,\mg{c}^j)$.
			\item She is just a price taker (see Definition~\ref{A-price-taker}) and hence treats the price of securities $\vp$ as an externally given 					 stochastic process.
			\item  Each agent $1\leq j\leq N_2$ solves her own problem independently. In particular, she does not care about the actions of the 						 other agents.
		\ei
	\item The optimization problem for the first population $1\leq i\leq N_1$:
		\bi \small
			\item The cost functional (see $(\ref{1st-cost-functional})$) of each agent $i$ depends on her trading speed, her inventory size, 
				 the empirical distribution of  the inventories for the first population, the price of securities 
				 as wall as the shocks induced by $(c^0, c^i)$.
			\item They try to minimize the {\it sum} of the cost functionals of the whole population $1\leq i\leq N_1$ by continuous trading
				at the exchange, which is fully coordinated by a central planner (see $(\ref{central-problem})$).
 			\item The first population as a whole has non-negligible market share and hence 
				the actions of the central planner can influence the price of securities $\vp$ (see $(\ref{vp-expression-1})$), 
				which in turn affects the actions of the second population.
			\item Assuming the agents in the second population rationally react to the price change, 
				the central planner solves the cost minimization problem with those reactions taken into account. 
		\ei
\ei
\normalsize 
It is  immediate to see that the problem of each agent in the second population is a usual stand-alone optimization.
Hence it makes perfect sense to start our investigation from solving this problem.
The optimization problem for the central planner of the first population becomes much more complicated.
We will see that the market clearing condition (see Definition~\ref{def-market-clearing}) connects her trading actions
and the price impacts. Through the reactions to the price change, the central planner must control 
not only the actions of the first population but also those of the second population indirectly.
This results in stochastic control of a large system of FBSDEs instead of an SDE.
\subsection{Problem for the non-cooperative agents}
\subsubsection{Problem description}
We first study the problem for the non-cooperative agents in the second population. 
Here, we basically follow the technique developed in \cite{Fujii-Takahashi, Fujii-mfg-convergence}.
Let us introduce three $\mbb{F}^0$-progressively measurable bounded processes, $b:=(b_t)_{t\in[0,T]}$, $\rho_2:=(\rho_2(t))_{t\in[0,T]}$
and $\Lambda=(\Lambda_t)_{t\in[0,T]}$, where $b_t$ and $\rho_2(t)$ are non-negative,  and $\Lambda_t$ is $\cals^n_+$-valued.
We introduce the following measurable functions;
$l_2:[0,T]\times (\mbb{R}^m)^2\rightarrow \mbb{R}^n$, 
$(\sigma_2^0,\sigma_2):[0,T]\times (\mbb{R}^m)^2\rightarrow (\mbb{R}^{n\times d_0},\mbb{R}^{n\times d_2})$,
$\ol{f}_2:[0,T]\times \mbb{R}^n\times (\mbb{R}^m)^2\rightarrow \mbb{R}$ and
$g_2:\mbb{R}^n\times (\mbb{R}^m)^2\rightarrow \mbb{R}$.
For each $t\in [0,T], x,\vp, \alpha  \in \mbb{R}^n, c^0,\mg{c}\in \mbb{R}^m$, we also set
\be
\small
\begin{split}
f_2(t,x,\vp,\alpha,c^0,\mg{c}):=\langle \alpha,\vp\rangle+\frac{1}{2}\langle \alpha, \Lambda_t \alpha\rangle
-b_t\langle x,\vp\rangle+\ol{f}_2(t,x,c^0,\mg{c}).  \nn
\end{split}
\ee
Here, $\langle \cdot, \cdot\rangle $ denotes an inner product of $n$-dimensional vectors.
For given stochastic processes  $(c^0_t, (\mg{c}^j_t)_{j=1}^{N_2},$ $b_t,\Lambda_t)_{t\in[0,T]}$, we use the following abbreviations
to simplify the notations: for $1\leq j\leq N_2$ and $(t,x,\vp,\alpha)\in[0,T]\times (\mbb{R}^n)^3$, 
\be
\small
\begin{split}
&l_2^j(t):=l_2(t,c_t^0,\mg{c}_t^j), \quad (\sigma_2^{j,0}(t), \sigma_2^j(t)):=(\sigma_2^0, \sigma_2)(t,c_t^0,\mg{c}_t^j), \\
&\ol{f}_2^j(t,x):=\ol{f}_2(t,x, c_t^0,\mg{c}_t^j), \quad g_2^j(x):=g_2(x,c_T^0,\mg{c}_T^j), \\
&f_2^j(t,x,\vp,\alpha):=\langle \alpha,\vp\rangle+\frac{1}{2}\langle \alpha, \Lambda_t \alpha\rangle
-b_t\langle x,\vp\rangle+\ol{f}_2^j(t,x). 
\label{j-convention}
\end{split}
\ee

Before providing various conditions on these functions, let us describe 
the financial problem for each non-cooperative agent indexed by $j=1,\ldots, N_2$.
The following concept is important.
\begin{definition}[price taker] An agent is called a price taker if she behaves under the assumption that 
there is no price impact from her trading. In other words, she always accepts the market price as it is
without trying to control it.
\end{definition}
\noindent
The following assumption is used throughout this work.
\begin{assumption}
\label{A-price-taker}
Every non-cooperative agent is a price taker.
\end{assumption}
\noindent
This is a natural assumption if the size of  population $N_2$ is large enough so that 
the market share of every individual agent is negligibly small.
Under Assumption~\ref{A-price-taker}, every non-cooperative agent  considers
the price process $\vp\in \mbb{H}^2(\mbb{F};\mbb{R}^n)$ of the $n$ securities as an exogenously given input.
We suppose that the problem for each non-cooperative agent $1\leq j\leq N_2$ is to
minimize the cost functional 
\be
\inf_{\alpha^j\in \mbb{A}_2}J^j(\alpha^j)
\label{non-c-problem}
\ee
under the dynamic constraint
\be
\small
\begin{split}
dx_t^j&=(\alpha_t^j+\rho_2(t)\vp_t+l_2^j(t))dt+\sigma_2^{j,0}(t)dW_t^0+\sigma_2^j(t) dB_t^j, \quad x_0^j=\eta^j. 
\end{split}
\label{eq-j-state}
\ee
Here, the cost functional is defined by
\be
\small
\begin{split}
J^j(\alpha^j):=\ex\Bigl[\int_0^T f_2^j(t,x_t^j,\vp_t,\alpha_t^j)dt+g_2^j(x_T^j)\Bigr], 
\label{2nd-cost-functional}
\end{split}
\ee
and $\mbb{A}_2:=\mbb{H}^2(\mbb{F};\mbb{R}^n)$ is the space of admissible controls.

Let us provide an economic interpretation of each term;
$x^j=(x_t^j)_{t\in[0,T]}$ is an $\mbb{R}^n$-valued process giving the time evolution of the position size of the $n$ securities for the $j$th agent,
and $\eta^j$ denotes its initial value. Any negative component of $(x_t^j)\in \mbb{R}^n$ represents a short position at time $t$.
$\alpha^j=(\alpha^j_t)_{t\in[0,T]} \in \mbb{A}_2$ is a control of the $j$th agent denoting her trading rate of the securities. Here, 
each component of $\alpha_t^j dt~(\in\mbb{R}^n)$ gives the amount of the corresponding security 
bought (or sold if negative) at the exchange within a time interval $[t,t+dt]$. 
In addition to the direct trading via the exchange, the level of her inventory is affected by 
the stochastic order flows from her OTC clients, which are described by 
$(\rho_2(t)\vp_t+l_2^j(t))dt$ and $(\sigma_2^{j,0}(t)dW_t^0, \sigma_2^j(t)dB_t^j)$.
In particular, the term $\rho_2(t)\vp_t$ represents a reaction of her clients to the price level.
Since it is naturally expected that the OTC clients do not have any information about the agent's securities position,
we assume the terms $(\rho_2, l_2^j, \sigma_2^{j,0}, \sigma_2^j)$, which describe the OTC order flows,  to be independent from $x^j$.
$f_2^j$ and $g_2^j$ denote the running and terminal cost functions, respectively;
$\langle \alpha,\vp\rangle$ in $f_2^j$ represents the direct trading cost of the securities of price $\vp$
with trading rate $\alpha$;  $\frac{1}{2}\langle \alpha, \Lambda_t \alpha\rangle$ an internal trading cost or a trading fee to be paid to 
the exchange; $-b_t\langle x,\vp\rangle$ the reduction of the cost based on the 
mark-to-market value of the inventory with a stochastic discount factor $b_t$; 
$\ol{f}_2^j$ the other contributions to the running cost depending on the inventory level,  informations and cash flows $(c^0,\mg{c}^j)$.
The terminal cost function $g_2^j$ represents the penalty on the open position at the closing time $T$.
\begin{remark}
\label{remark-non-c}
From the problem setup and $(\ref{j-convention})$, we see that all the heterogeneities among the non-cooperative agents  come 
from i.i.d.~inputs only; idiosyncratic informations
$(\mg{c}^j)_{j=1}^{N_2}$, initial conditions $(\eta^j)_{j=1}^{N_2}$,  and the Brownian motions $(B^j)_{j=1}^{N_2}$. 
\end{remark}

\begin{remark}
One may wonder why we do not allow $\vp_T$-dependence in $g_2^j$ as in the work~\cite{Fujii-Takahashi}.
This is due to the existence of the first population. Since they have non-negligible market share, their last-minute trading at $\{T\}$ 
can influence $\vp_T$ despite the fact that it is on the Lebesgue-null set.
In other words, if there exists $\vp_T$ dependence in the terminal cost function, 
the first population can drastically change it without changing their securities position at all.
This easily makes the system ill-defined.
Because of this reason, we assume that the terminal cost functions of the both population (see $(\ref{2nd-cost-functional})$ 
and $(\ref{1st-cost-functional})$ below) are independent from $\vp_T$.~\footnote{The other possibility is to forbid the last-minute trading as in \cite{Fujii-Major}.}
This is  not a big restriction from the modeling perspective, since the cost functionals 
can still depend on the time integral of $(\vp_t)_{t\in[0,T]}$.
\end{remark}

\begin{remark}
\label{remark-info}
The definition of the admissible space $\mbb{A}_2=\mbb{H}^2(\mbb{F};\mbb{R}^n)$ used above
implies that each agent is supposed to know, in addition to the common information $\mbb{F}^0$, 
all the idiosyncratic information $\ol{\mbb{F}}$ of the other agents.  (The perfect information  is to be assumed also for the first population.) 
Ideally, we would like to restrict the information set available to each agent $j$ to the filtration $\mbb{F}^0\otimes \ol{\mbb{F}}^{2,j}$.
Unfortunately however, to the best of the author's knowledge, there is no existing work to
successfully achieve the market clearing equilibrium in the finite agent market without assuming the perfect information.
Interestingly, though, we shall observe that the realistic information setup is recovered in the large population limit.
See the discussion in Remark~\ref{remark-info-structure} and the one following Theorem~\ref{th-convergence}.
\end{remark}
\subsubsection{Solving the individual problem for $j=1,\ldots, N_2$}
In order to solve the above problem, let us introduce the following conditions.
\begin{assumption}
\label{A-non-C}
{\rm (i)} $\Lambda=(\Lambda_t)_{t\in[0,T]}$ is an $\mbb{F}^0$-progressively measurable $\cals_+^n$-valued process such that
both $\Lambda_t$ and $\ol{\Lambda}_t~(:=\Lambda_t^{-1})$ are bounded by $L$ for every $t\in[0,T]$.\\
{\rm (ii)} For every $(t,c^0,\mg{c})\in [0,T]\times (\mbb{R}^m)^2$, 
$
|l_2(t,c^0,\mg{c})|+|\sigma_2^0(t,c^0,\mg{c})|+|\sigma_2(t,c^0,\mg{c})|\leq L(1+|c^0|+|\mg{c}|). \nn 
$
\\
{\rm (iii)} For every $(t,x,c^0,\mg{c})\in[0,T]\times \mbb{R}^n\times (\mbb{R}^m)^2$, 
$
|\ol{f}_2(t,x,c^0,\mg{c})|+|g_2(x,c^0,\mg{c})|\leq L(1+|x|^2+|c^0|^2+|\mg{c}|^2). \nn
$
\\
{\rm (iv)} For every $(t,c^0,\mg{c})\in[0,T]\times (\mbb{R}^m)^2$, $\ol{f}_2$ and $g_2$ are once continuously differentiable in $x$,
and their derivatives have the affine form in $x$:
\be
\small
\begin{split}
&\part_x\ol{f}_2(t,x,c^0,\mg{c})=c_f(t,c^0,\mg{c})x+h_f(t,c^0,\mg{c}), \quad \part_x g_2(x,c^0,\mg{c})=c_g(c^0,\mg{c})x+h_g(c^0,\mg{c}). \nn
\end{split}
\ee
Here, $(c_f,h_f):[0,T]\times (\mbb{R}^m)^2\rightarrow (\cals^n_+, ~\mbb{R}^n)$ and $(c_g, h_g):(\mbb{R}^m)^2\rightarrow (\cals^n_+, ~\mbb{R}^n)$ 
are measurable functions that satisfy, with some positive constants $\gamma_2^f, \gamma_2^g>0$, 
\be
\small
\begin{split}
&|h_f(t,c^0,\mg{c})|+|h_g(c^0,\mg{c})|\leq L(1+|c^0|+|\mg{c}|), \quad |c_f(t,c^0,\mg{c})|+|c_g(c^0,\mg{c})|\leq L, \\
&\langle \theta,c_f(t,c^0,\mg{c})\theta\rangle\geq \gamma_2^f |\theta|^2, ~
\langle \theta,c_g(c^0,\mg{c})\theta\rangle\geq \gamma_2^g |\theta|^2, ~\forall \theta\in \mbb{R}^n. \nn
\end{split}
\ee
{\rm (v)} $b:=(b_t)_{t\in[0,T]}$ and $\rho_2:=(\rho_2(t))_{t\in[0,T]}$ are $\mbb{F}^0$-progressively measurable bounded processes 
satisfying
\be
\small
\begin{split}
0\leq b_t\leq L, \quad \frac{b_t^2}{2\gamma_2^f}\leq \rho_2(t)\leq L, \quad \forall t\in[0,T]. \nn
\end{split}
\ee
\end{assumption}

\begin{remark}
The benefit (i.e.~negative cost) due to the continuous payoff (respectively, lump-sum payoff at $T$) from the securities
can be included in $h_f$ (respectively, $h_g$). 
The convexity of cost functions (i.e.~risk averseness of agents) is determined by $c_f(t,c_t^0,\mg{c}_t^j)$ and $c_g(c_T^0,\mg{c}_T^j)$,
which are stochastic and also heterogeneous among the agents by the factors $(\mg{c}^j)_{j=1}^{N_2}$.
The affine structures in {\rm (iv)} will be used in later sections to guarantee the convexity of Hamiltonian for the ``cooperative" agents.
\end{remark}

With the adjoint variable $y\in \mbb{R}^n$, the Hamiltonian for $(\ref{non-c-problem})$ of the $j$th agent is given by
\be
\small
\begin{split}
H_2^j(t,x,y,\vp,\alpha):=\langle y,\alpha+\rho_2(t)\vp+l_2^j(t)\rangle+\langle \alpha,\vp\rangle+\frac{1}{2}\langle \alpha, \Lambda_t \alpha\rangle-b_t\langle x,\vp\rangle
+\ol{f}_2^j(t,x). \nn
\end{split}
\ee
For each $\omega\in \Omega$,  this gives a map from $[0,T]\times (\mbb{R}^n)^4$ to $\mbb{R}$.
Under Assumption~\ref{A-non-C}, for a given adjoint variable $y$, $H_2^j$ is jointly and strictly convex in $(x,\alpha)$.
Its unique minimizer $\wh{\alpha}$ is given by
$
\wh{\alpha}=-\ol{\Lambda}_t (y+\vp). \nn
$
Hence, for a given $\vp\in \mbb{H}^2(\mbb{F};\mbb{R}^n)$, 
the system of the state and adjoint equations from the maximum principle is given by
\be
\small
\begin{split}
&dx_t^j=(-\ol{\Lambda}_t(y_t^j+\vp_t)+\rho_2(t)\vp_t+l_2^j(t))dt+\sigma_2^{j,0}(t)dW_t^0+\sigma_2^j(t)dB_t^j, \\
&dy_t^j=-(\part_x \ol{f}_2^j(t,x_t^j)-b_t\vp_t)dt+z_t^{j,0}dW_t^0+\sum_{k=1}^{N_1}\mg{z}_t^{j,k}dW_t^k+\sum_{k=1}^{N_2}z_t^{j,k}dB_t^k
\end{split}
\label{fbsde-non-C}
\ee
with $x_0^j=\eta^j$ and $y_T^j=\part_x g_2^j(x_T^j)$, for each agent $1\leq j\leq N_2$.
\begin{theorem}
\label{th-non-c-individual}
Let $\vp\in \mbb{H}^2(\mbb{F};\mbb{R}^n)$ be given. Under Assumption~\ref{A-non-C}, for each $j=1,\ldots,N_2$, 
there exists a unique square integrable solution 
with $(x^j,y^j)\in \mbb{S}^2(\mbb{F};\mbb{R}^n)\times \mbb{S}^2(\mbb{F};\mbb{R}^n)$
to the FBSDE $(\ref{fbsde-non-C})$.~\footnote{Obviously, every integrand of Brownian 
motions is square integrable, such as $z^{j,0}\in \mbb{H}^2(\mbb{F};\mbb{R}^{n\times d_0})$. Since all we need is this square integrability, 
we drop the details to save the space. }
Moreover, the problem $(\ref{non-c-problem})$ has a unique optimal solution
$\wh{\alpha}^j:=(\wh{\alpha}_t^j)_{t\in[0,T]}$ given by $\wh{\alpha}_t^j=-\ol{\Lambda}_t(y_t^j+\vp_t)$.
\begin{proof}
This is a simple adaptation of \cite[Theorem 3.1]{Fujii-mfg-convergence}.
By the convexity and the differentiability of the Hamiltonian, it is standard to check the sufficiency of the Pontryagin's maximum principle.
The uniqueness of the optimal solution follows from the strict convexity of $H_2^j$ in $\alpha$.
Thus it only remains to prove the existence of the unique solution to the FBSDE $(\ref{fbsde-non-C})$.
For each $(t,\omega)\in[0,T]\times \Omega$, let us define the maps from  state variables $u:=(x^j,y^j)\in \mbb{R}^n\times \mbb{R}^n$
to $\mbb{R}^n$ by
\be
\small
\begin{split}
&B_{x^j}(t,u):=-\ol{\Lambda}_t(y^j+\vp_t)+\rho_2(t)\vp_t+l_2^j(t), \\
&F_{y^j}(t,u):=-\part_x \ol{f}_2^j(t,x^j)+b_t \vp_t, \quad G_{y^j}(u):=\part_x g_2^j(x^j). \nn
\end{split}
\ee
Under Assumption~\ref{A-non-C},  they are uniformly Lipschitz continuous in $u$, and $(B_{x^j}(t,u), F_{y^j}(t,u), G_{y^j}(u))$
satisfy the square integrability for a given $u$. Hence it suffices to check the Peng-Wu's monotonicity~\cite[(H2.3)]{Peng-Wu}.
For two inputs $(u,\ac{u})$, let us set $\Del x^j:=x^j-\ac{x}^j, \Del y^j:=y^j-\ac{y}^j$, $\Del B_{x^j}(t):=B_{x^j}(t,u)-B_{x^j}(t,\ac{u})$, 
$\Del F_{y^j}(t):=F_{y^j}(t,u)-F_{y^j}(t,\ac{u})$ and $\Del G_{y^j}:=G_{y^j}(u)-G_{y^j}(\ac{u})$.
Then, from condition (iv), it is straightforward to see
$\langle \Del B_{x^j}(t),\Del y^j\rangle+\langle \Del F_{y^j}(t), \Del x^j\rangle\leq -\gamma_2^f |\Del x^j|^2$
and $\langle \Del G_{y^j}, \Del x^j\rangle \geq \gamma_2^g |\Del x^j|^2$. 
Thus the existence of a unique solution to $(\ref{fbsde-non-C})$ follows from \cite[Theorem~2.6]{Peng-Wu}
with $\mu_1=\gamma_2^g$, $\beta_1=\gamma_2^f$ with the identity matrix $G=I_{n\times n}$.
\end{proof}
\end{theorem}

\subsubsection{Market clearing equilibrium with given order flows from the first population}
\label{sec-given-beta}
Let us denote by $\beta^i:=(\beta_t^i)_{t\in[0,T]}$ 
the trading rate of each agent $i=1,\ldots, N_1$ in the first  population at the securities exchange.
\begin{definition}
\label{def-market-clearing}
We say that the market clearing condition is satisfied if the next equality holds:
\be
\small
\begin{split}
\sum_{j=1}^{N_2}\alpha_t^j+\sum_{i=1}^{N_1}\beta_t^i=0, \quad dt\otimes d\mbb{P}\text{-a.e.} \nn
\end{split}
\ee
\end{definition}

\begin{definition}
With given order flows $(\beta^i\in \mbb{H}^2(\mbb{F};\mbb{R}^n), i=1,\ldots, N_1)$ from the first population,
if there exists a price process $\vp\in \mbb{H}^2(\mbb{F};\mbb{R}^n)$ and the set of optimal solutions $(\wh{\alpha}^j)_{j=1}^{N_2}$
to the problem $(\ref{non-c-problem})$ that satisfies the market clearing condition, 
then we say that the market clearing equilibrium exists
with the equilibrium price process $\vp$ for the given order flows $(\beta^i)_{i=1}^{N_1}$.
\end{definition}

Since we know $\wh{\alpha}_t^j=-\ol{\Lambda}_t(y_t^j+\vp_t)$ from Theorem~\ref{th-non-c-individual},
the market clearing condition implies that the price process $\vp_t$ must be equal to $\vp_t(y_t,\beta_t)$
defined by
\be
\small
\vp_t(y,\beta):=-\mg{m}_2(y)+\deln\Lambda_t \mg{m}_1(\beta), 
\label{vp-expression-1}
\ee
where 
\be
\small
\deln:=N_1/N_2 \nn
\ee 
denotes the ratio of the two populations.
Recall that $\mg{m}_2(y)$ and $\mg{m}_1(\beta)$ give the empirical means of $(y^j)_{j=1}^{N_2}$
and $(\beta^i)_{i=1}^{N_1}$, respectively. As in \cite{Fujii-Takahashi, Fujii-mfg-convergence}, the relation $(\ref{vp-expression-1})$
leads us to consider the FBSDEs;
\be
\small
\begin{split}
&dx_t^j=\bigl(-\ol{\Lambda}_t(y_t^j+\vp_t(y_t,\beta_t))+\rho_2(t)\vp_t(y_t,\beta_t)+l_2^j(t)\bigr)dt+\sigma_2^{j,0}(t)dW_t^0+
\sigma_2^j(t)dB_t^j, \\
&dy_t^j=-(\part_x \ol{f}_2^j(t,x_t^j)-b_t\vp_t(y_t,\beta_t))dt+z_t^{j,0}dW_t^0+\sum_{k=1}^{N_1}\mg{z}_t^{j,k}dW_t^k+\sum_{k=1}^{N_2}z_t^{j,k}dB_t^k
\end{split}
\label{fbsde-eq-non-C}
\ee
with $x_0^j=\eta^j$ and $y_T^j=\part_x g_2^j(x_T^j)$, $1\leq j\leq N_2$. It is a coupled system
by the interaction in  $\vp_t(y,\beta)$. The relevance of $(\ref{fbsde-eq-non-C})$ may be understood from the next theorem.

\begin{theorem}
\label{th-non-c-existence}
Let $(\beta^i)_{i=1}^{N_1}$ with $\beta^i\in \mbb{H}^2(\mbb{F};\mbb{R}^n)$ be given.
Under Assumption~\ref{A-non-C}, there exists a unique square integrable solution 
with $(x,y)=((x^j)_{j=1}^{N_2,}, (y^j)_{j=1}^{N_2}) \in \mbb{S}^2(\mbb{F};\mbb{R}^n)^{N_2}\times \mbb{S}^2(\mbb{F};\mbb{R}^n)^{N_2}$
to the system of FBSDEs $(\ref{fbsde-eq-non-C})$. 
Moreover, there exists a unique market clearing equilibrium with the equilibrium price process $(\vp_t=\vp_t(y_t,\beta_t))_{t\in[0,T]}$
for the given order flows $(\beta^i)_{i=1}^{N_1}$.
\begin{proof}
This is an adaptation of \cite[Theorems~3.2 and 3.3]{Fujii-mfg-convergence}.
Firstly, the existence of the solution to $(\ref{fbsde-eq-non-C})$  is 
a necessary condition so that $(\wh{\alpha}^j)_{j=1}^{N_2}$ clears the market with the given order flows $(\beta^i)_{i=1}^{N_1}$.
Conversely, if there exists a square integrable solution to the system $(\ref{fbsde-eq-non-C})$, then one can check that the price process 
$(\vp_t(y_t,\beta_t))_{t\geq 0}$
defined by its solution $y$ actually clears the market when it is used as an input  price process $\vp$ for the problem $(\ref{non-c-problem})$.
In fact, due to the uniqueness of the solution to $(\ref{fbsde-non-C})$ by Theorem~\ref{th-non-c-individual}, 
the backward solutions $y^j$  of the two equations coincide.

Hence it suffices to show that there exists a unique square integrable solution to $(\ref{fbsde-eq-non-C})$. As before, this is done by 
Peng-Wu's continuation method. 
For each $(t,\omega)$, we define the maps from
$u:=((x^j)_{j=1}^{N_2},(y^j)_{j=1}^{N_2}) \in (\mbb{R}^n)^{N_2}\times (\mbb{R}^n)^{N_2}$ to $\mbb{R}^n$ by
\be
\small
\begin{split}
&B_{x^j}(t,u):=-\ol{\Lambda}_t(y^j+\vp_t(y,\beta_t))+\rho_2(t)\vp_t(y,\beta_t)+l_2^j(t), \\
&F_{y^j}(t,u):=-\part_x \ol{f}_2^j(t,x^j)+b_t\vp_t(y,\beta_t), \quad 
G_{y^j}(u):=\part_x g_2^j(x^j) \nn
\end{split}
\ee
with $1\leq j\leq N_2$. By Assumption~\ref{A-non-C}, they are uniformly Lipschitz continuous with respect to $u$.
For each $u$, it is easy to confirm the square integrability; $B_{x^j}(\cdot, u), F_{x^j}(\cdot, u) \in \mbb{H}^2(\mbb{F};\mbb{R}^n)$, and
$G_{y^j}(u) \in \mbb{L}^2(\calf_T;\mbb{R}^n)$. 
Now, it only remains to confirm the Peng-Wu's monotonicity conditions.
For two inputs $(u,\ac{u})$, let us set $\Del x^j:=x^j-\ac{x}^j$, $\Del y^j=y^j-\ac{y}^j$,
$\Del B_{x^j}(t):=B_{x^j}(t,u)-B_{x^j}(t,\ac{u})$, $\Del F_{y^j}(t):=F_{y^j}(t,u)-F_{y^j}(t,\ac{u})$ and $\Del G_{y^j}:=G_{y^j}(u)-G_{y^j}(\ac{u})$.
We have
\be
\small
\begin{split}
&\sum_{j=1}^{N_2}\langle \Del B_{x^j}(t),\Del y^j\rangle+\sum_{j=1}^{N_2}\langle \Del F_{y^j}(t),\Del x^j\rangle \\
&=-\sum_{j=1}^{N_2}\langle \ol{\Lambda}_t(\Del y^j-\mg{m}_2(\Del y))+\rho_2(t)\mg{m}_2(\Del y),\Del y^j\rangle
-\sum_{j=1}^{N_2}\langle c^{f}(t,c_t^0,\mg{c}_t^j)\Del x^j+b_t\mg{m}_2(\Del y), \Del x^j\rangle\\
&\leq -N_2\rho_2(t)|\mg{m}_2(\Del y)|^2-\gamma_2^f\sum_{j=1}^{N_2}|\Del x^j|^2+ b_t\sum_{j=1}^{N_2}|\langle \Del x^j,\mg{m}_2(\Del y)\rangle| 
\leq -\frac{\gamma_2^f}{2}\sum_{j=1}^{N_2}|\Del x^j|^2. \nn
\end{split}
\ee
In the first inequality, we have used the condition (iv) and the fact that $\sum_{j=1}^{N_2}\langle \ol{\Lambda}_t \Del y^j,\Del y^j\rangle 
\geq N_2\langle \ol{\Lambda}_t \mg{m}_2(\Del y),\mg{m}_2(\Del y)\rangle$.
The Young's inequality and the assumption (v) give the last one.
From the condition (iv), we also have
$
\sum_{j=1}^{N_2}\langle \Del G_{y^j}, \Del x^j\rangle\geq \gamma_2^g \sum_{j=1}^{N_2}|\Del x^j|^2. \nn
$
Hence,  we can apply \cite[Theorem~2.6]{Peng-Wu} with $(\beta_1,\mu_1)=(\gamma^f_2/2, \gamma_2^g)$ and the identity matrix $G$.
\end{proof}
\end{theorem}

\begin{remark}
It is interesting to observe an economic role played by the term $\rho_2(t)\vp_t$.
Since $\rho_2$ is positive, it increases the inflow of securities to the agents' inventory 
when the price $\vp$ rises. This is equivalent to the increase of the sales from the OTC clients.
Since agents (i.e.~financial firms) are trying to maintain the optimal inventory level, this
reduces the demand of the securities among the financial firms. Assumption~\ref{A-non-C} (v) guarantees 
that this is stronger than the opposite effect caused by the mark-to-market term $-b_t\langle x, \vp\rangle$.
By making the demand for securities react in an opposite way to the price level, the term $\rho_2(t)\vp_t$
seems to prevent market bubbles/crashes from happening and allow the equilibrium to exist for general duration $T$.
\end{remark}

\subsection{Problem for the cooperative agents}
\subsubsection{Problem description}
\label{sec-coop-finite}
We now study the problem for the first population that consists of cooperative agents.
In contrast to the previous case, the agents are no longer price takers.
Although the market share of each agent may be negligibly small, their coordinated actions in total can influence the behavior of the price process.
In order to describe their problem in a concrete manner,
let us first introduce the following measurable functions;
$l_1:[0,T]\times (\mbb{R}^m)^2\rightarrow \mbb{R}^n$, $(\sigma_1^0, \sigma_1):[0,T]\times (\mbb{R}^m)^2
\rightarrow (\mbb{R}^{n\times d_0},\mbb{R}^{n\times d_1})$, $f_1:[0,T]\times \mbb{R}^n\times \calp_2(\mbb{R}^n)\times \mbb{R}^n\times A_1
\times (\mbb{R}^m)^2\rightarrow \mbb{R}$, and $g_1:\mbb{R}^n\times \calp_2(\mbb{R}^n)\times (\mbb{R}^m)^2\rightarrow \mbb{R}$.
Here, $A_1$ is a closed convex subset of $\mbb{R}^n$. We also introduce $\mbb{R}$-valued $\mbb{F}^0$-progressively measurable process $\rho_1:=(\rho_1(t))_{t\in[0,T]}$.
As in $(\ref{j-convention})$, for given stochastic processes $(c^0_t, c_t^i)_{t\in[0,T]}$,
we often use the following abbreviations
\be
\small
\begin{split}
&l_1^i(t):=l_1(t,c_t^0,c_t^i), \quad (\sigma_1^{i,0}(t), \sigma_1^i(t)):=(\sigma_1^0, \sigma_1)(t,c_t^0,c_t^i), \\
&f_1^i(t,x,\mu,\vp,\beta):=f_1(t,x,\mu,\vp,\beta,c_t^0,c_t^i), \quad
g_1^i(x,\mu):=g_1(x,\mu,c_T^0,c_T^i)
\label{i-convention}
\end{split}
\ee
for  $(t,x,\mu,\vp,\beta)\in[0,T]\times \mbb{R}^n\times \calp_2(\mbb{R}^n)\times \mbb{R}^n\times A_1$ and $1\leq i\leq N_1$. 
We also use, for  $1\leq j\leq N_2$, 
\be 
\small
\begin{split}
c_f^j(t):=c_f^j(t,c_t^0,\mg{c}_t^j), \quad c_g^j:=c_g(c_T^0,\mg{c}_T^j). \nn
\end{split}
\ee

Let us denote by $\beta:=(\beta^i)_{i=1}^{N_1}$ with $\beta^i:=(\beta_t^i)_{t\in[0,T]}$ the trading rate of the cooperative agents at the exchange. 
$X^i=(X_t^i)_{t\in[0,T]}$ is an $\mbb{R}^n$-valued process describing the evolution of the position size of the $n$ securities
of the $i$th agent. We  suppose that it obeys the dynamics given by
\be
\small
\begin{split}
&dX_t^i=(\beta_t^i+\rho_1(t)\vp_t(y_t,\beta_t)+l_1^i(t))dt+\sigma_1^{i,0}(t)dW_t^0+\sigma_1^i(t)dW_t^i, \quad X_0^i=\xi^i, \nn
\end{split}
\ee
$1\leq i\leq N_1$. Here, the term $\vp_t(y_t,\beta_t)$ is defined by $(\ref{vp-expression-1})$ with $y=(y^j)_{j=1}^{N_2}$ being 
the solution to the system of FBSDEs $(\ref{fbsde-eq-non-C})$. The economic interpretation
is similar to the previous case; each component of $\beta_t^i dt~(\in \mbb{R}^n)$ gives the amount
of the corresponding security bought (or sold if negative) by the $i$th agent at the exchange within a time interval $[t,t+dt]$;
$(\rho_1(t)\vp_t(y_t,\beta_t)+l_1^i(t))dt$ and $(\sigma_1^{i,0}(t)dW_t^0, \sigma_1^i(t)dW_t^i)$ give the stochastic order flows
from her OTC clients.
We suppose that the cost functional for each agent $i$ is given by
\be
\small
\begin{split}
J_1^i(\beta):=\ex\Bigl[\int_0^T f_1^i(t,X_t^i, \nu_t^{N_1},\vp_t(y_t,\beta_t),\beta_t^i)dt+g_1^i(X_T^i,\nu_T^{N_1})\Bigr], 
\label{1st-cost-functional}
\end{split}
\ee
where $\nu^{N_1}_t$ is the time-$t$ empirical distribution of the securities position among the 
cooperative agents, i.e. $\nu_t^{N_1}:=\frac{1}{N_1}\sum_{i=1}^{N_1}\del_{X_t^i}$.
$f_1^i$ and $g_1^i$ denote the running and terminal costs, respectively.
Since the agents are cooperative, it is natural to suppose that each of them takes care of 
not only her own inventory but also its distribution across the members. 
This is represented by the dependence on the empirical distribution $\nu^{N_1}$ in the cost functions.
The running cost $f_1^i$ is allowed to have general convex dependence on  $\vp$
to describe the agent's preference  on the price level.  

We suppose that all the trading actions  in the first population are  coordinated by the central planner,
whose  problem is defined by 
\be
\small
\begin{split}
\inf_{\beta^i\in \mbb{A}_1,~ i=1,\ldots, N_1}\sum_{i=1}^{N_1}J_1^i(\beta). 
\end{split}
\label{central-problem}
\ee
Here, $\mbb{A}_1:=\mbb{H}^2(\mbb{F};A_1)$ is the admissible space for each $\beta^i$. Thus, at every time $t\in[0,T]$,
each control $\beta_t^i$ has to take values in a given closed convex set $A_1$.
The central planner is responsible to control every $\beta^i, i=1,\ldots, N_1$ within $\mbb{A}_1$ to minimize the total cost
of the whole population.  
Due to the reactions from the non-cooperative agents,
she needs to deal with  complicated price impacts from her own trading.
Notice that, in addition to the direct impact from $\mg{m}_1(\beta)$-term in $\vp_t(\cdot)$, 
she also receives implicit feedbacks from $\mg{m}_2(y)$-term via $(\ref{fbsde-eq-non-C})$.
In fact, the coupling with $(\ref{fbsde-eq-non-C})$ makes the problem for the central planner 
an optimization with respect to the system of controlled-FBSDEs instead of controlled-SDEs.
See the works by J. Yong~\cite{Yong-1, Yong-2} for general discussions on the problems of controlled-FBSDEs.

\begin{remark}
\label{remark-c}
Similarly to the second population, we see that all the heterogeneities  among the cooperative agents
come from i.i.d.~inputs only; idiosyncratic informations $(c^i)_{i=1}^{N_1}$, initial conditions $(\xi^i)_{i=1}^{N_1}$,
and the Brownian motions $(W^i)_{i=1}^{N_1}$.
\end{remark}

Before giving the set of main assumptions, let us mention about the concept of differentiability of functions defined on the 
space of probability measures. In this work, we adopt the notion of $L$-differentiability used in \cite{Carmona-Delarue-1, Carmona-Delarue-2},
which was first introduced by P.L.Lions in his lecture at College de France.
Here, the differentiation is based on the lifting of function $\calp_2(\mbb{R}^n)\ni \mu\mapsto u(\mu)\in \mbb{R}$ to 
a function $\wt{u}$ defined on the Hilbert space $\mbb{L}^2(\Omega,\calf,\mbb{P};\mbb{R}^n)$ by $\wt{u}(X):=u(\call(X))$
with $X\in \mbb{L}^2(\Omega,\calf,\mbb{P};\mbb{R}^n)$. Here, the probability space $(\Omega,\calf,\mbb{P})$ is chosen
arbitrarily with a Polish set $\Omega$ and an atomless probability measure $\mbb{P}$ to support the random variable
$X$ with $\call(X)=\mu$. The $L$-derivative of $u(\mu)$ is defined as the Frechet derivative of the lifted function $\wt{u}(X)$.

\begin{definition}[Definition 5.22 in \cite{Carmona-Delarue-1}]
A function $u$ on $\calp_2(\mbb{R}^n)$ is said to be $L$-differentiable at $\mu_0\in \calp_2(\mbb{R}^n)$ if there exists
a random variable $X$ with law $\mu_0$ such that the lifted function $\wt{u}$ is Frechet differentiable at $X$.
\end{definition}
We distinguish the Frechet derivative by using the symbol $D$. 
By \cite[Proposition 5.24]{Carmona-Delarue-1}, if $u$ is $L$-differentiable at $\mu_0$, the $\wt{u}$ is 
Frechet differentiable at any $\ac{X}$ with $\call(\ac{X})=\mu_0$ and the law of the pair $(\ac{X}, D\wt{u}(\ac{X}))$
is independent of the choice of the random variable $\ac{X}$. Thus, the $L$-derivative may be denoted by
$\part_\mu u(\mu_0)(\cdot):\mbb{R}^n\ni x\mapsto \part_\mu u(\mu_0)(x)\in \mbb{R}^n$, which is uniquely defined  $\mu_0$-a.e. on $\mbb{R}^n$.
By its definition, the $L$-derivative satisfies
\be
\small
\begin{split}
u(\mu)=u(\mu_0)+\ex\bigl[\langle X-X_0, \part_\mu u(\mu_0)(X_0)\rangle\bigr]+o(\|X-X_0\|_{\mbb{L}^2}), \nn
\end{split}
\ee
for any random variables $X$ and $X_0$ with $\call(X)=\mu$ and $\call(X_0)=\mu_0$.
For example, if the function $u$ is of the form $u(\mu):=\int_{\mbb{R}^n}h(x)\mu(dx)$ for some function 
$h:\mbb{R}^n\rightarrow \mbb{R}$,  we have $\wt{u}(X)=\ex[h(X)]$ with $\call(X)=\mu$. 
If the function $h$ is differentiable, then the definition of $L$-derivative implies that $\part_\mu u(\mu)(\cdot)=\part_x h(\cdot)$. 
See \cite{Carmona-Delarue-1, Carmona-Delarue-2} for details and many more examples.

In our analysis, in particular, functions on empirical distributions play a special role.
Suppose that a function $u:\calp_2(\mbb{R}^n)\rightarrow \mbb{R}$ is given.
By considering the map $(\mbb{R}^n)^N \ni (x^i)_{i=1}^N \mapsto u\Bigr(\frac{1}{N}\sum_{i=1}^N \del_{x^i}\Bigr)\in \mbb{R}$,
$u$ can also be seen as a function on $(\mbb{R}^n)^N$.
If $u:\calp_2(\mbb{R}^n)\rightarrow \mbb{R}$ is $L$-differentiable, 
\cite[Proposition 5.35]{Carmona-Delarue-1} tells that the function $u$, when seen as a function on $(\mbb{R}^n)^N$,  is differentiable for each $x^i$
and satisfies
\be
\small
\begin{split}
\part_{x^i}u(x^1,\ldots, x^N)=\frac{1}{N}\part_\mu u\Bigl(\frac{1}{N}\sum_{i=1}^N \del_{x^i}\Bigr)(x^i).
\end{split}
\label{L-d-empirical}
\ee
Here, with slight abuse of notations, we used the same symbol $u$ to denote the two maps.

Now,  our assumptions on the cooperative agents are given as follows.
\begin{assumption}
\label{A-C} 
{\rm (0)} The population sizes $(N_1, N_2)$ satisfy $\deln\leq L$, where $\deln:=N_1/N_2$. 
\\
{\rm (i)} For every $(t,c^0,c)\in [0,T]\times (\mbb{R}^m)^2$, 
$
|l_1(t,c^0,c)|+|\sigma_1^0(t,c^0,c)|+|\sigma_1(t,c^0,c)|\leq L(1+|c^0|+|c|). 
$
\\
{\rm (ii)} For every $(t,x,\mu,\vp,\beta,c^0,c )\in[0,T]\times \mbb{R}^n\times \calp_2(\mbb{R}^n)\times \mbb{R}^n\times A_1
\times(\mbb{R}^m)^2$,
the cost functions satisfy
\be
\small
\begin{split}
&|f_1(t,x,\mu,\vp,\beta,c^0,c)|\leq L(1+|x|^2+M_2(\mu)^2+|\vp|^2+|\beta|^2+|c^0|^2+|c|^2), \\
&|g_1(x,\mu,c^0,c)|\leq L(1+|x|^2+M_2(\mu)^2+|c^0|^2+|c|^2). \nn
\end{split}
\ee
and,  with another inputs $(\ac{x},\ac{\mu},\ac{\vp},\ac{\beta})$, they also satisfy the local Lipschitz continuity;
\be
\small
\begin{split}
&|f_1(t,x,\mu,\vp,\beta,c^0,c)-f_1(t,\ac{x},\ac{\mu},\ac{\vp},\ac{\beta},c^0,c)|\\
&\quad \leq L\bigl(1+|x|+|\ac{x}|+M_2(\mu)+M_2(\ac{\mu})
+|\vp|+|\ac{\vp}|+|c^0|+|c|\bigr)(|x-\ac{x}|+W_2(\mu,\ac{\mu})+|\vp-\ac{\vp}|+|\beta-\ac{\beta}|), \\
&|g_1(x,\mu,c^0,c)-g_1(\ac{x},\ac{\mu},c^0,c)|\leq L(1+|x|+|\ac{x}|+M_2(\mu)+M_2(\ac{\mu})+|c^0|+|c|)(|x-\ac{x}|+W_2(\mu,\ac{\mu})). \nn
\end{split}
\ee
{\rm (iii)} The cost functions $f_1$ and $g_1$ are
once continuously differentiable in $(x,\mu,\vp,\beta)$ and $(x,\mu)$ respectively with
uniformly Lipschitz continuous derivatives in the sense that
\be
\small
\begin{split}
&|[(\part_x , \part_\vp , \part_\beta)f_1](t,x,\mu,\vp,\beta,c^0,c)-[(\part_x ,\part_\vp,\part_\beta)f_1](t,\ac{x},\ac{\mu},\ac{\vp},\ac{\beta},c^0,c)| \leq L(|x-\ac{x}|+W_2(\mu,\ac{\mu})+|\vp-\ac{\vp}|+|\beta-\ac{\beta}|), \\ 
&|\part_\mu f_1(t,x,\mu,\vp,\beta,c^0,c)(v)-\part_\mu f_1(t,\ac{x},\ac{\mu},\ac{\vp},\ac{\beta},c^0,c)(\ac{v})|
\leq L(|x-\ac{x}|+W_2(\mu,\ac{\mu})+|\vp-\ac{\vp}|+|\beta-\ac{\beta}|+|v-\ac{v}|), \\
&|\part_x g_1(x,\mu, c^0,c)-\part_x g_1(\ac{x},\ac{\mu}, c^0,c)|\leq L(|x-\ac{x}|+W_2(\mu,\ac{\mu})), \\
&|\part_\mu g_1(x,\mu,c^0,c)(v)-\part_\mu g_1(\ac{x},\ac{\mu},c^0,c)(\ac{v})|\leq L(|x-\ac{x}|+W_2(\mu,\ac{\mu})+|v-\ac{v}|), \nn
\end{split}
\ee
hold for every $(t,c^0,c)\in[0,T]\times (\mbb{R}^m)^2$ and $(x,\mu,\vp,\beta,v), (\ac{x},\ac{\mu},\ac{\vp},\ac{\beta},\ac{v})\in \mbb{R}^n
\times \calp_2(\mbb{R}^n)\times \mbb{R}^n\times A_1\times \mbb{R}^n$.
Moreover, the derivatives satisfy the linear growth property:
\be
\small
\begin{split}
&|[(\part_x,\part_\vp ,\part_\beta )f_1](t,x,\mu,\vp,\beta,c^0,c)|+
|\part_\mu f_1(t,x,\mu,\vp,\beta,c^0,c)(v)|\leq L(1+|x|+M_2(\mu)+|\vp|+|\beta|+|v|+|c^0|+|c|), \\
&|\part_x g_1(x,\mu, c^0,c)| +|\part_\mu g_1(x,\mu,c^0,c)(v)|  \leq L(1+|x|+M_2(\mu)+|v|+|c^0|+|c|). \nn
\end{split}
\ee
{\rm (iv)} For any $(t,c^0,c)\in [0,T]\times (\mbb{R}^m)^2$, 
$f_1$ is jointly convex in $(x,\mu,\vp,\beta)$
in the sense that
\be
\small
\begin{split}
&f_1(t,\ac{x},\ac{\mu},\ac{\vp},\ac{\beta},c^0,c)-f_1(t,x,\mu,\vp,\beta,c^0,c)-\blangle \part_{(x,\vp,\beta)}f_1(t,x,\mu,\vp,\beta,c^0,c),
(\ac{x}-x,\ac{\vp}-\vp,\ac{\beta}-\beta)\brangle \\
&-\ex\bigl[\langle \part_\mu f_1(t,x,\mu,\vp,\beta,c^0,c)(X), \ac{X}-X\rangle\bigr]
\geq \frac{\gamma_1^f}{2}|\ac{x}-x|^2+\frac{\lambda_\beta}{2}|\ac{\beta}-\beta|^2+\frac{\lambda_\vp}{2}|\ac{\vp}-\vp|^2 \nn
\end{split}
\ee
holds for any pair of inputs $(x,\mu,\vp,\beta), (\ac{x},\ac{\mu},\ac{\vp},\ac{\beta})\in \mbb{R}^n\times \calp_2(\mbb{R}^n)\times \mbb{R}^n
\times A_1$ and random variables $X, \ac{X}$ with $\call(X)=\mu, \call(\ac{X})=\ac{\mu}$.
Here,  $\gamma_1^f, \lambda_\beta, \lambda_\vp>0$ are some positive constants, and $\part_{(x,\vp,\beta)} f_1$ stands for the gradient
of $f_1$ in the joint variables $(x,\vp,\beta)$.
\\
{\rm (v)} For any $(c^0,c)\in (\mbb{R}^m)^2$, $g_1$ is jointly convex in $(x,\mu)$ with some positive constant $\gamma_1^g>0$, 
\be
\small
\begin{split}
&g_1(\ac{x},\ac{\mu},c^0,c)-g_1(x,\mu,c^0,c)-\langle \part_x g_1(x,\mu,c^0,c),\ac{x}-x\rangle
-\ex\bigl[\langle \part_\mu g_1(x,\mu,c^0,c)(X),\ac{X}-X\rangle\bigr] \geq \frac{\gamma_1^g}{2}|\ac{x}-x|^2 \nn
\end{split}
\ee
holds for any pair of inputs $(x,\mu), (\ac{x},\ac{\mu})\in \mbb{R}^n\times \calp_2(\mbb{R}^n)$
and random variables $X, \ac{X}$ with $\call(X)=\mu, \call(\ac{X})=\ac{\mu}$. 
\\
{\rm (vi)} $\rho_1:=(\rho_1(t))_{t\in[0,T]}$ is an $\mbb{R}$-valued  $\mbb{F}^0$-progressively measurable bounded process 
with $|\rho_1(t)|\leq L, ~\forall t\in[0,T]$. Moreover, $\rho_2$ satisfies 
\be
\small
\begin{split}
\max\bigl(\frac{1}{L}, \frac{2 b_t^2}{\gamma_2^f}\bigr)\leq \rho_2(t)\leq L, \quad \forall t\in[0,T]. \nn
\end{split}
\ee
\end{assumption}
\begin{remark}
Note that, $f_1$ can have  much more general form than $f_2$. 
In particular, $f_1$ having a linear quadratic form in $(x, \vp, \beta)$ with appropriate convexity is a special example satisfying the above assumptions. 
\end{remark}
\subsubsection{Adjoint equations}

Let us first introduce the notations $X, P\in (\mbb{R}^n)^{N_1}, \beta\in A_1^{N_1}$  and $x,y,p,r\in (\mbb{R}^n)^{N_2}$
in the following way;
$X:=(X^i)_{i=1}^{N_1}$, $P:=(P^i)_{i=1}^{N_1}$, $\beta:=(\beta^i)_{i=1}^{N_1}$,
$x:=(x^j)_{j=1}^{N_2}$,
$y:=(y^j)_{j=1}^{N_2}$,  $p:=(p^j)_{j=1}^{N_2}$ and
$r:=(r^j)_{j=1}^{N_2}$. 

With adjoint variables $(P, p, r)$, the Hamiltonian $H$ for the problem $(\ref{central-problem})$ is defined by
\be
\small
\begin{split}
&H(t,X,x,y,P,p,r,\beta)\\
&:=\sum_{i=1}^{N_1}\bigl\{\langle P^i, \beta^i+\rho_1(t)\vp_t(y,\beta)+l_1^i(t)\rangle+f_1^i(t,X^i,\nu^{N_1},\vp_t(y,\beta),\beta^i)\bigr\}\\
&\quad +\sum_{j=1}^{N_2}\bigl\{\langle p^j, -\ol{\Lambda}_t(y^j+\vp_t(y,\beta))+\rho_2(t)\vp_t(y,\beta)+l_2^j(t)\rangle
+\langle r^j, -\part_x \ol{f}_2^j(t,x^j)+b_t \vp_t(y,\beta)\rangle\bigr\},  
\label{H-1-finite}
\end{split}
\ee
which gives, for each $(t,\omega)$, a map from $(\mbb{R}^n)^{N_1}\times (\mbb{R}^n)^{N_2}\times  (\mbb{R}^n)^{N_2}
\times (\mbb{R}^n)^{N_1}\times (\mbb{R}^n)^{N_2}\times  (\mbb{R}^n)^{N_2}\times A_1^{N_1}$ to $\mbb{R}$.
Here, $\vp_t(y,\beta)$ is defined by $(\ref{vp-expression-1})$ and $\nu^{N_1}:=\frac{1}{N_1}\sum_{i=1}^{N_1}\del_{X^i}$.
With slight abuse of notation, let us also write the above Hamiltonian 
as $H(t,X,x,y,P,p,r,\vp,\beta)$ when we treat $\vp \in \mbb{R}^n$ as a separate argument.
From the number of arguments involved in the expression, one may  understand which definition is used without confusion.
Note that, in the both definitions, the dependence on $\nu^{N_1}$ is treated as a function of $X=(X^i)_{i=1}^{N_1}$.
For each $u:=(X,x,y,P,p,r)$, we wan to find a minimizer:
\be
\small
\begin{split}
\opb(t,u):={\rm argmin}\bigl(H(t,X,x,y,P,p,r,\beta):\beta=(\beta^i)_{i=1}^{N_1} \in A_1^{N_1}\bigr).\nn
\end{split}
\ee
We often use the following derivatives: with $\vp:=\vp_t(y,\beta)$ and $f_1^i(t,u,\beta^i):=f_1^i(t,X^i,\nu^{N_1},\vp,\beta^i)$,
\be
\small
\begin{split}
&\part_{\beta^i}H(t,u,\beta)=\part_{\beta^i}H(t,u,\vp,\beta)+\frac{\Lambda_t}{N_2}\part_\vp H(t,u,\vp,\beta), \\
&=P^i+\part_{\beta^i}f_1^i(t,u,\beta^i)+\frac{\Lambda_t}{N_2}\sum_{k=1}^{N_1} 
\part_\vp f_1^k(t,u,\beta^k)
+\deln \rho_1(t)\Lambda_t\mg{m}_1(P)+ (-I+\rho_2(t)\Lambda_t)\mg{m}_2(p)+ b_t\Lambda_t \mg{m}_2(r), \\
&\part_{X^i} H(t,u,\beta)=\part_x f_1^i(t,u,\beta^i)+\frac{1}{N_1}\sum_{k=1}^{N_1}\part_\mu f_1^k(t,u,\beta^k)(X^i), \\
&\part_{x^j}H(t,u,\beta)=-c_f^j(t)r^j, \\
&\part_{y^j}H(t,u,\beta)=-\frac{1}{N_2}\sum_{k=1}^{N_1}\part_\vp f_1^k(t,u,\beta^k)-\deln \rho_1(t)\mg{m}_1(P)-b_t \mg{m}_2(r)
+\ol{\Lambda}_t(\mg{m}_2(p)-p^j)-\rho_2(t)\mg{m}_2(p). 
\end{split}
\label{eq-H1-beta}
\ee
Here, we have used $(\ref{L-d-empirical})$ to get $\part_{X^i}H(t,u,\beta)$.
\begin{lemma}
\label{lemma-H-1-finite}
Under Assumptions~\ref{A-non-C}-\ref{A-C}, for each $(t,\omega)$ and a given set of adjoint variables 
$(P,p,r)\in (\mbb{R}^n)^{N_1}\times (\mbb{R}^n)^{N_2}\times (\mbb{R}^n)^{N_2}$, the Hamiltonian $H$ defined by $(\ref{H-1-finite})$ is jointly 
convex in $(X,x,y,\beta)\in  (\mbb{R}^n)^{N_1}\times (\mbb{R}^n)^{N_2}\times (\mbb{R}^n)^{N_2}\times A_1^{N_1}$ 
and strictly so in $(X,\beta)$. In particular,  for any $t\in[0,T]$ and input $u:=(X,x,y,P,p,r)$, there exists a 
unique minimizer $(\opb_t^i=\opb^i(t,u))_{i=1}^{N_1}$ with $\opb_t^i\in A_1$ that satisfies the following linear growth and
the Lipschitz continuity properties:
\be
\small
\begin{split}
&\sum_{i=1}^{N_1}|\opb^i(t,u)|^2\leq C\sum_{i=1}^{N_1}\bigr[1+|X^i|^2+|P^i|^2+|\mg{m}_2(y)|^2+|\mg{m}_2(p)|^2+|\mg{m}_2(r)|^2+|c^0_t|^2+|c^i_t|^2\bigr],\\
&\sum_{i=1}^{N_1}|\opb^i(t,u)-\opb^i(t,\ac{u})|^2
\leq C\sum_{i=1}^{N_1}\bigl[|X^i-\ac{X}^i|^2+|P^i-\ac{P}^i|^2+|\mg{m}_2(y-\ac{y})|^2+|\mg{m}_2(p-\ac{p})|^2+|\mg{m}_2(r-\ac{r})|^2\bigr], \nn
\end{split}
\ee
where $\ac{u}:=(\ac{X},\ac{x},\ac{y},\ac{P},\ac{p},\ac{r})$ is another input and $C$ is some positive constant independent of the population sizes.
Moreover,  the minimizer $\opb(t,u)$ is coupled to the second population in a symmetric way;
it is independent from $x$ and depends on $y,p,r$ only through their empirical means.
\begin{proof}
First, we check the relevant convexity of $H$. For a given  $(P,p,r)$, we consider two arbitrary sets of state variables 
$(X,x,y), (\ac{X},\ac{x},\ac{y})$ and controls $\beta,\ac{\beta}$.
With $u:=(X,x,y,P,p,r)$ and $\ac{u}:=(\ac{X},\ac{x},\ac{y},P,p,r)$, we compare $H(t,u,\beta)$ and $H(t,\ac{u},\ac{\beta})$.
For notational simplicity, let us set  $\nu^{N_1}:=\frac{1}{N_1}\sum_{i=1}^{N_1}\del_{X^i}, \ac{\nu}^{N_1}:=\frac{1}{N_1}\sum_{i=1}^{N_1}\del_{\ac{X}^i}$,
$\vp:=\vp_t(y,\beta), \ac{\vp}:=\vp_t(\ac{y},\ac{\beta})$, $\Del X^i:=\ac{X}^i-X^i$, $\Del \beta^i:=\ac{\beta}^i-\beta^i$,
$\Del \vp:=\ac{\vp}-\vp=-\mg{m}_2(\Del y)+\deln \Lambda_t\mg{m}_1(\Del \beta)$, etc.
We also put $f_1^i(t,u,\beta^i):=f_1^i(t,X^i,\nu^{N_1},\vp, \beta^i), f_1^i(t,\ac{u},\ac{\beta}^i):=f_1^i(t,\ac{X}^i,\ac{\nu}^{N_1},\ac{\vp},\ac{\beta}^i)$.
Since the linear terms in $H$ cancel,  we obtain by direct calculation using $(\ref{eq-H1-beta})$,
\be
\small
\begin{split}
&H(t,\ac{u},\ac{\beta})-H(t,u,\beta)\\
&-\sum_{i=1}^{N_1}\bigl[\langle \part_{\beta^i}H(t,u,\beta),\Del \beta^i\rangle+
\langle \part_{X^i}H(t,u,\beta),\Del X^i\rangle\bigr] -\sum_{j=1}^{N_2}\bigl[\langle \part_{x^j}H(t,u,\beta), \Del x^j\rangle+\langle \part_{y^j}H(t,u,\beta),\Del y^j\rangle\bigr]\\
&=\sum_{i=1}^{N_1}\bigl[f_1^i(t,\ac{u},\ac{\beta}^i)-f_1^i(t,u,\beta^i)-\langle \part_{\beta} f_1^i(t,u,\beta^i),\Del \beta^i\rangle
-\langle \part_\vp f_1^i(t,u,\beta^i),\Del\vp\rangle\\
&-\langle \part_x f_1^i(t,u,\beta^i), \Del X^i\rangle-\ex^\theta[\langle \part_\mu f_1^i(t,u,\beta^i)(X^\theta),\Del X^\theta\rangle]\bigr]
\geq \frac{1}{2}\sum_{i=1}^{N_1}\bigl(\gamma_1^f|\Del X^i|^2+\lambda_\beta |\Del \beta^i|^2+\lambda_\vp |\Del \vp|^2\bigr),
\label{H1-joint-convexity}
\end{split}
\ee
which gives the desired convexity. Since $A_1$ is convex and closed, it guarantees the existence of the unique minimizer.
Here, we have introduced a random variable $\theta$ on a probability space $(\Omega^\theta,\calf^\theta,\mbb{P}^\theta)$, which  is distributed 
uniformly on the set $\{1,2,\ldots, N_1\}$. We use $\ex^\theta[\cdot]$ to denote the expectation with respect to $\mbb{P}^\theta$. 
In particular, 
\be
\small
\begin{split}
\frac{1}{N_1}\sum_{i=1}^{N_1}\sum_{k=1}^{N_1}\langle \part_\mu f_1^k(t,u,\beta^k)(X^i),\Del X^i\rangle
=\sum_{k=1}^{N_1}\ex^\theta\bigl[\langle \part_\mu f_1^k(t,u,\beta^k)(X^\theta),\Del X^\theta\rangle\bigr].
\end{split}
\label{theta-technique}
\ee
Since the distribution of the random variables $X^\theta, \ac{X}^\theta$
are equal to $\nu^{N_1}$ and $\ac{\nu}^{N_1}$ respectively, Assumption~\ref{A-C} (iv) gives the inequality $(\ref{H1-joint-convexity})$.

We now prove the latter claims.
The convexity $(\ref{H1-joint-convexity})$ implies that,  for any fixed $(\beta_0^i\in A_1)_{i=1}^{N_1}$, 
\be
\small
\begin{split}
H(t,u,\beta_0)\geq H(t,u,\opb(t,u)) \geq H(t,u,\beta_0)+\sum_{i=1}^{N_1}\langle \opb_t^i-\beta_0^i,\part_{\beta^i}H(t,u,\beta_0)\rangle+\frac{\lambda_\beta}{2}\sum_{i=1}^{N_1}|\opb_t^i-\beta_0^i|^2, \nn
\end{split}
\ee
which gives, by Young's inequality, 
$\sum_{i=1}^{N_1}|\opb_t^i-\beta_0^i|^2\leq C\sum_{i=1}^{N_1}|\part_{\beta^i}H(t,u,\beta_0)|^2$,
where $C$ depends on $\lambda_{\beta}$. Hence $(\ref{eq-H1-beta})$, the conditions (0) and (iii)
give the desired growth property for $\opb$. 
Finally, let us put $\opb_t^i=\opb^i(t,u)$ and $\opb_t^{\prime, i}=\opb^i(t,\ac{u})$ the minimizers of $H$ with two inputs $u=(X,x,y,P,p,r)$ and $\ac{u}=(\ac{X},\ac{x},\ac{y},\ac{P},\ac{p},\ac{r})$, respectively.
Using the optimality condition, we have
$
\sum_{i=1}^{N_1} \langle \opb_t^{\prime,i}-\opb_t^i, \part_{\beta^i}H(t,\ac{u},\opb_t^{\prime})-\part_{\beta^i}H(t,u,\opb_t)\rangle \leq 0. \nn
$
By rearrangement,
\be
\small
\begin{split}
\sum_{i=1}^{N_1}\langle \opb_t^{\prime,i}-\opb_t^i, \part_{\beta^i}H(t,u,\opb_t^\prime)-\part_{\beta^i}H(t,u,\opb_t)\rangle
\leq \sum_{i=1}^{N_1}\langle \opb_t^{\prime,i}-\opb_t^i,  \part_{\beta^i}H(t,u,\opb_t^\prime)-\part_{\beta^i}H(t,\ac{u},\opb_t^\prime)
\rangle. \nn
\end{split}
\ee
The convexity of the left hand side in  $\beta$ yields
$
\lambda_\beta \sum_{i=1}^{N_1}|\opb_t^{\prime,i}-\opb_t^i|^2\leq \sum_{i=1}^{N_1}\langle \opb_t^{\prime,i}-\opb_t^i,  \part_{\beta^i}H(t,u,\opb_t^\prime)-\part_{\beta^i}H(t,\ac{u},\opb_t^\prime)\rangle
$
and hence, an application of Young's inequality gives 
\be
\small
\begin{split}
\sum_{i=1}^{N_1}|\opb_t^{\prime,i}-\opb_t^i|^2\leq C\sum_{i=1}^{N_1}|\part_{\beta^i}H(t,u,\opb_t^{\prime})-\part_{\beta^i}H(t,\ac{u},\opb_t^\prime)|^2. \nn
\end{split}
\ee
From $(\ref{eq-H1-beta})$, the conditions (0) and (iii) give the desired Lipschitz continuity. 
The last statement on the coupling of $\opb(t,u)$  is a direct consequence of the form of this Lipschitz continuity.
\end{proof}
\end{lemma}

The system of coupled equations $1\leq i\leq N_1, 1\leq j\leq N_2$ that characterizes the optimal solution to the problem $(\ref{central-problem})$ 
is given as follows: 
\be
\small
\begin{split}
dX_t^i&=(\opb_t^i+\rho_1(t)\vp_t(y_t,\opb_t)+l_1^i(t))dt+\sigma_1^{i,0}(t)dW_t^0+\sigma^i_1(t)dW_t^i, \\
dx_t^j&=(-\ol{\Lambda}_t(y_t^j+\vp_t(y_t,\opb_t))+\rho_2(t)\vp_t(y_t,\opb_t)+l_2^j(t))dt+\sigma_2^{j,0}(t)dW_t^0+\sigma_2^j(t)dB_t^j, \\
dy_t^j&=-(\part_x \ol{f}_2^j(t,x_t^j)-b_t \vp_t(y_t,\opb_t))dt+z_t^{j,0}dW_t^0+\sum_{k=1}^{N_1}\mg{z}_t^{j,k}dW_t^k+\sum_{k=1}^{N_2}z_t^{j,k}dB_t^k, \\
dP_t^i&=-\Bigl\{\part_x f_1^i(t,X_t^i,\nu_t^{N_1},\vp_t(y_t,\opb_t),\opb_t^i)+\frac{1}{N_1}\sum_{k=1}^{N_1}\part_\mu f_1^k(t,X_t^k,\nu_t^{N_1},
\vp_t(y_t,\opb_t),\opb_t^k)(X_t^i)\Bigr\}dt\\
&\qquad+Q_t^{i,0}dW_t^0+\sum_{k=1}^{N_1}Q_t^{i,k}dW_t^k+\sum_{k=1}^{N_2}\mg{Q}_t^{i,k}dB_t^k, \\
dp_t^j&=c_f^j(t)r_t^jdt+q_t^{j,0}dW_t^0+\sum_{k=1}^{N_1}\mg{q}_t^{j,k}dW_t^k+\sum_{k=1}^{N_2}q_t^{j,k}dB_t^k, \\
dr^j_t&=-\Bigl\{\ol{\Lambda}_t(\mg{m}_2(p_t)-p_t^j)-\rho_2(t)\mg{m}_2(p_t)-\deln \rho_1(t) \mg{m}_1(P_t)-b_t \mg{m}_2(r_t) \\
&\qquad -\frac{1}{N_2}\sum_{k=1}^{N_1}\part_\vp f_1^k(t,X_t^k, \nu_t^{N_1},\vp_t(y_t,\opb_t),\opb_t^k)\Bigr\}dt
\end{split}
\label{fbsde-full-finite}
\ee
with the boundary conditions; $X_0^i=\xi^i$, $x_0^j=\eta^j$, $r_0^j=0$ and
\be
\small
\begin{split}
y_T^j=\part_x g_2^j(x_T^j), \quad P^i_T=\part_x g_1^i(X_T^i, \nu_T^{N_1})+\frac{1}{N_1}\sum_{k=1}^{N_1}\part_\mu g_1^k(X_T^k,\nu_T^{N_1})(X_T^i), \
\quad p_T^j=-c_g^j r_T^j.  \nn
\end{split}
\ee
Note that the two populations are coupled only through the empirical means contained in $\vp_t(y,\beta)$.
The relevance of  $(\ref{fbsde-full-finite})$ is based on the next theorem.
\begin{remark}
From $(\ref{eq-H1-beta})$, one can check that the drift terms of adjoint variables $(P^i, p^j, r^j)$ are given by
$-\part_{X^i}H(t,u,\opb)$, $-\part_{x^j}H(t,u,\opb)$ and $-\part_{y^j}H(t,u,\opb)$, respectively.
\end{remark}

\begin{theorem}
\label{th-sufficiency-finite}
Let Assumption~\ref{A-non-C}-\ref{A-C} be in force. Suppose that there exists a unique square integrable 
solution to the system of equations $(\ref{fbsde-full-finite})$ with
$u:=(X,x,y,P,p,r)\in \mbb{S}^2(\mbb{F};\mbb{R}^n)^{N_1}\times 
\mbb{S}^2(\mbb{F};\mbb{R}^n)^{N_2}\times \mbb{S}^2(\mbb{F};\mbb{R}^n)^{N_2}\times \mbb{S}^2(\mbb{F};\mbb{R}^n)^{N_1}\times 
\mbb{S}^2(\mbb{F};\mbb{R}^n)^{N_2}\times \mbb{S}^2(\mbb{F};\mbb{R}^n)^{N_2}$  
and $\opb_t:=(\opb^i(t,u_t))_{i=1}^{N_1}$ the minimizer of  $H(t,u_t,\cdot)$
for $dt\otimes d\mbb{P}$-a.e. Then, $\opb_t,t\in[0,T]$ is the unique optimal solution to $(\ref{central-problem})$.
\begin{proof}
Let $u_t:=(X_t,x_t,y_t,P_t,p_t,r_t), t\in[0,T]$ denote the solution to $(\ref{fbsde-full-finite})$
with $\opb_t$ the minimizer of $H(t,u_t,\cdot)$, $dt\otimes d\mbb{P}$-a.e.
By Lemma~\ref{lemma-H-1-finite} and the properties of $(c^0, c^i)$ given at the beginning of Section~\ref{sec-finite}, 
$\opb$ is square integrable and hence admissible.
We denote by  $(\ac{X}_t,\ac{x}_t,\ac{y}_t), t\in[0,T]$ the state processes corresponding to another control $\ac{\beta}=(\ac{\beta}^i)_{i=1}^{N_1}
\in \mbb{A}_1^{N_1}$ and set $\ac{u}_t:=(\ac{X}_t,\ac{x}_t,\ac{y}_t, P_t, p_t, r_t), t\in[0,T]$. 
Let us put $\nu^{N_1}_t:=\frac{1}{N_1}\sum_{i=1}^{N_1}\del_{X_t^i}, \ac{\nu}_t^{N_1}:=\frac{1}{N_1}\sum_{i=1}^{N_1}\del_{\ac{X}_t^i}$,
$\ac{\vp}_t:=\vp_t(\ac{y}_t,\ac{\beta}_t)$, $\vp_t:=\vp_t(y_t,\opb_t)$, 
$\Del X:=\ac{X}-X$, $\Del x:=\ac{x}-x$, $\Del \beta=\ac{\beta}-\opb$ and similarly for the others. 
For notational ease, we also use $f^i_1(t,u_t,\opb^i_t):=f_1^i(t,X_t^i,\nu_t^{N_1},\vp_t(y_t,\opb_t),\opb_t^i)$
and $f_1^i(t,\ac{u}_t,\ac{\beta}_t^i):=f_1^i(t,\ac{X}_t^i,\ac{\nu}_t^{N_1},\vp_t(\ac{y}_t,\ac{\beta}_t),\ac{\beta}_t^i)$.

From Assumption~\ref{A-C} (v) and the technique used in $(\ref{theta-technique})$,  we have
\be
\small
\begin{split}
\sum_{i=1}^{N_1} \bigl(g_1^i(\ac{X}_T^i,\ac{\nu}_T^{N_1})-g_1^i(X_T^i,\nu_T^{N_1})\bigr)&\geq \sum_{i=1}^{N_1} \langle P_T^i, \Del X_T^i\rangle
=\sum_{i=1}^{N_1} \langle P_T^i, \Del X_T^i\rangle+\sum_{j=1}^{N_2}\bigl(\langle p_T^j,\Del x_T^j\rangle+\langle r_T^j, \Del y_T^j\rangle\bigr).  \nn
\end{split}
\ee
In the latter equality, we used the fact  $\langle p_T^j, \Del x_T^j\rangle+\langle r_T^j, \Del y_T^j\rangle=0$.
By the initial condition $\Del X^i_0=\Del x^i_0=r^j_0=0$ and the optimality $\sum_i \langle \part_{\beta^i}H(t,u_t,\opb_t),\Del\beta^i_t \rangle \geq 0$
 $dt\otimes d\mbb{P}$-a.e., an application of It\^o formula yields
\be
\small
\begin{split}
\sum_{i=1}^{N_1}J_1^i(\ac{\beta})-\sum_{i=1}^{N_1}J_1^i(\opb) &\geq \ex \Bigl[ \sum_{i=1}^{N_1} \langle P_T^i, \Del X_T^i\rangle+\sum_{j=1}^{N_2}\bigl(\langle p_T^j,\Del x_T^j\rangle+\langle r_T^j, \Del y_T^j\rangle\bigr)
+\int_0^T \sum_{i=1}^{N_1}\bigl[f_1^i(t,\ac{u}_t,\ac{\beta}_t^i)-f_1^i(t,u_t,\opb_t^i)\bigr]dt\Bigr]\\
&\geq \ex\int_0^T \bigl[H(t,\ac{u}_t,\ac{\beta}_t)-H(t,u_t,\opb_t)
-\sum_{i=1}^{N_1}\bigl(\langle \part_{X^i}H(t,u_t,\opb_t), \Del X_t^i \rangle+\langle
 \part_{\beta^i}H(t,u_t,\opb_t),\Del \beta_t^i \rangle\bigr) \\
&\quad -\sum_{j=1}^{N_2}\bigl(\langle  \part_{x^j}H(t,u_t,\opb_t),\Del x_t^j \rangle +\langle \part_{y^j}H(t,u_t,\opb_t), \Del y_t^j \rangle\bigr)\bigr]dt. \nn
\end{split}
\ee
The conclusion follows from the joint convexity $(\ref{H1-joint-convexity})$ in  Lemma~\ref{lemma-H-1-finite}.
\end{proof}
\end{theorem}

\subsubsection{Existence of the optimal solution}
\label{sec-existence-finite}
The aim of this section is to prove the existence of a strong solution to $(\ref{fbsde-full-finite})$,
which then will be combined with Theorem~\ref{th-sufficiency-finite} to solve the optimization problem for the first population.
We use the following maps to study the monotonicity~\cite[(H2.3)]{Peng-Wu}.
For each $(t,\omega)$, we define the maps on $(\mbb{R}^n)^{N_1}\times (\mbb{R}^n)^{N_2}\times(\mbb{R}^n)^{N_2}
\times (\mbb{R}^n)^{N_1}\times (\mbb{R}^n)^{N_2}\times(\mbb{R}^n)^{N_2}\ni u:=(X,x,y,P,p,r)\mapsto \mbb{R}^n$ by
\be
\small
\begin{split}
&B_{X^i}(t,u):=\opb^i+\rho_1(t)\vp_t(y,\opb)+l_1^i(t), \\
&B_{x^j}(t,u):=-\ol{\Lambda}_t(y^j+\vp_t(y,\opb))+\rho_2(t)\vp_t(y,\opb)+l_2^j(t), \\
&B_{r^j}(t,u):=-\bigl\{\ol{\Lambda}_t(\mg{m}_2(p)-p^j)-\rho_2(t)\mg{m}_2(p)-\deln \rho_1(t)\mg{m}_1(P)-b_t \mg{m}_2(r)-\frac{1}{N_2}\sum_{k=1}^{N_1}\part_\vp f_1^k(t,X^k,\nu^{N_1},\vp_t(y,\opb),\opb^k)\bigr\}, \\
&F_{P^i}(t,u):=-\bigl\{\part_x f_1^i(t,X^i,\nu^{N_1},\vp_t(y,\opb),\opb^i)+\frac{1}{N_1}\sum_{k=1}^{N_1}\part_\mu f_1^k(t,X^k,\nu^{N_1},
\vp_t(y,\opb),\opb^k)(X^i)\bigr\}, \\
&F_{y^j}(t,u):=-(\part_x \ol{f}_2^j(t,x^j)-b_t \vp_t(y,\opb)), \quad
F_{p^j}(t,u):=c_f^j(t)r^j, \nn
\end{split}
\ee
and 
\be
\small
\begin{split}
G_{P^i}(u):=\part_x g_1^i(X^i,\nu^{N_1})+\frac{1}{N_1}\sum_{k=1}^{N_1}\part_\mu g_1^k(X^k,\nu^{N_1})(X^i), \quad
G_{y^j}(u):=\part_x g_2^j(x^j), \quad
G_{p^j}(u):=-c_g^j r^j,  \nn
\end{split}
\ee
with $1\leq i\leq N_1, 1\leq j\leq N_2$. Here,  $\nu^{N_1}:=\frac{1}{N_1}\sum_{i=1}^{N_1}\del_{X^i}$ and $(\opb^i:=\opb^i(t,u))_{i=1}^{N_1}$ is the minimizer of the Hamiltonian $H(t,u,\cdot)$.
Let $u:=(X,x,y, P,p,r), \ac{u}:=(\ac{X},\ac{x},\ac{y},\ac{P},\ac{p},\ac{r})\in (\mbb{R}^n)^{2N_1+4N_2}$ 
be two arbitrary inputs. We use  abbreviations, such as, 
$\opb^i=\opb^i(t,u)$, $\opb^{\prime,i}:=\opb^i(t,\ac{u})$, $\vp:=\vp_t(y,\opb)$, $\ac{\vp}:=\vp_t(\ac{y},\opb^\prime)$,  
$f_1^i(t,u,\opb^i):=f_1^i(t,X^i,\nu^{N_1},\vp,\opb^i)$, $f_1^i(t,\ac{u},\opb^{\prime,i}):=f_1^i(t,\ac{X}^i,\ac{\nu}^{N_1},\ac{\vp},\opb^{\prime,i})$
and also put
$\Del X^i:=X^i-\ac{X}^i$, $\Del y^j:=y^j-\ac{y}^j$, $\Del \opb^i:=\opb^i-\opb^{\prime, i}$, 
$\Del \vp:=\vp-\ac{\vp}$, and $\Del B_{X^i}(t):=B_{X^i}(t,u)-B_{X^i}(t,\ac{u})$,
$\Del F_{y^j}(t,u):=F_{y^j}(t,u)-F_{y^j}(t,\ac{u})$, $\Del G_{P^i}:=G_{P^i}(u)-G_{P^i}(\ac{u})$, etc.
\begin{lemma}
\label{pw-terminal}
Under Assumptions~\ref{A-non-C}-\ref{A-C}, for any inputs $(u,\ac{u})$, the terminal functions satisfy 
\be
\small
\begin{split}
\sum_{i=1}^{N_1}\langle \Del G_{P^i},\Del X^i\rangle+\sum_{j=1}^{N_2}\bigl(\langle \Del G_{y^j},\Del x^j\rangle+
\langle (-1)\Del G_{p^j},\Del r^j\rangle\bigr)\geq \gamma_1^g \sum_{i=1}^{N_1}|\Del X^i|^2+
\gamma_2^g \sum_{j=1}^{N_2}(|\Del x^j|^2+|\Del r^j|^2). \nn
\end{split}
\ee
\begin{proof}
This is a direct result of Assumption~\ref{A-non-C} (iv) and Assumption~\ref{A-C} (v). 
\end{proof}
\end{lemma}

\noindent Although the proof requires rather lengthy calculation, the following result is crucial. 
It tells that the drift terms {\it do not} satisfy the Peng-Wu's monotonicity conditions. 
However, we shall show that the 
violation of the monotonicity can be controlled by $\deln:=N_1/N_2$, the ratio of the two population sizes.
\begin{lemma}
\label{pw-drift}
Let Assumptions~\ref{A-non-C}-\ref{A-C} hold.
Then, for any $t\in[0,T]$ and 
inputs $(u, \ac{u})$, there exists a positive constant $\del_0^F$ such that,  for any $(N_1,N_2)$ satisfying $\deln:=N_1/N_2\leq \del_0^F$, 
the inequality 
\be
\small
\begin{split}
\cald(t) \leq -\gamma^f \bigl[ \sum_{i=1}^{N_1}|\Del X^i|^2+\sum_{j=1}^{N_2}(|\Del x^j|^2+|\Del r^j|^2)\bigr]+\deln C
\bigl[\sum_{i=1}^{N_1}|\Del P^i|^2 +\sum_{j=1}^{N_2}(|\Del y^j|^2+|\Del p^j|^2)\bigr]
\nn 
\end{split}
\ee
holds with $\gamma^f:=\min\bigl(\frac{\gamma_1^f}{2}, \frac{\gamma_2^f}{3}\bigr)$ and 
some positive constant $C$ independent of the population sizes. Here, 
\be
\small
\begin{split}
\cald(t)&:=\sum_{i=1}^{N_1}\langle \Del B_{X^i}(t),\Del P^i\rangle+\sum_{j=1}^{N_2}\bigl[ \langle \Del B_{x^j}(t),\Del y^j\rangle
+\langle (-1)\Del B_{r^j}(t), \Del p^j\rangle\bigr]\\
&~+\sum_{i=1}^{N_1}\langle \Del F_{P^i}(t),\Del X^i\rangle+\sum_{j=1}^{N_2}\bigl[\langle \Del F_{y^j}(t),\Del x^j\rangle
+\langle (-1)\Del F_{p^j}(t), \Del r^j\rangle\bigr]. \nn
\end{split}
\ee
\begin{proof}
Let us evaluate each term. From $(\ref{eq-H1-beta})$, we can express $\Del P^i$ as
\be
\small
\begin{split}
\Del P^i&=\part_{\beta^i}H(t,u,\opb)-\part_{\beta^i}H(t,\ac{u},\opb^\prime)-(\part_{\beta^i}f_1^i(t,u,\opb^i)
-\part_{\beta^i}f_1^i(t,\ac{u},\opb^{\prime,i}))+(I-\rho_2(t)\Lambda_t)\mg{m}_2(\Del p)\\
&-b_t\Lambda_t \mg{m}_2(\Del r)-\deln \rho_1(t)\Lambda_t\mg{m}_1(\Del P)-\frac{\Lambda_t}{N_2}\sum_{k=1}^{N_1}(\part_\vp f_1^k(t,u,\opb^k)-\part_\vp f_1^k(t,\ac{u},\opb^{\prime,k})). \nn
\end{split}
\ee
It gives, 
together with the optimality condition $\sum_i\langle \part_{\beta^i}H(t,u,\opb)-\part_{\beta^i}H(t,\ac{u},\opb^\prime),\Del \opb^i\rangle\leq 0$, 
\be
\small
\begin{split}
&\sum_{i=1}^{N_1}\langle \Del B_{X^i}(t),\Del P^i\rangle
=\sum_{i=1}^{N_1}\langle \Del \opb^i+\rho_1(t)\Del \vp,\Del P^i\rangle \\
&\leq -\sum_{i=1}^{N_1}\langle \part_{\beta^i}f_1^i(t,u,\opb^i)-\part_{\beta^i}f_1^i(t,\ac{u},\opb^{\prime,i}),\Del \opb^i\rangle
-\sum_{i=1}^{N_1}\langle \part_{\vp}f_1^i(t,u,\opb^i)-\part_{\vp}f_1^i(t,\ac{u},\opb^{\prime,i}),\Del \vp+\mg{m}_2(\Del y)\rangle \\
&\quad+N_1\langle (I-\rho_2(t)\Lambda_t)\mg{m}_2(\Del p)-b_t\Lambda_t \mg{m}_2(\Del r), \mg{m}_1(\Del \opb)\rangle
-N_1\rho_1(t)\langle \mg{m}_2(\Del y),\mg{m}_1(\Del P)\rangle. \nn
\end{split}
\ee
Since $\sum_j \langle \ol{\Lambda}_t (\Del y^j-\mg{m}_2(\Del y)), \Del y^j\rangle \geq 0$
and $\sum_j \langle \ol{\Lambda}_t (\Del p^j-\mg{m}_2(\Del p)), \Del p^j\rangle \geq 0$, 
we have
\be
\small
\begin{split}
&\sum_{j=1}^{N_2}\langle B_{x^j}(t),\Del y^j\rangle \leq -N_2 \rho_2(t)|\mg{m}_2(\Del y)|^2+N_1\langle (\rho_2(t)\Lambda_t-I)\mg{m}_1(\Del \opb), \mg{m}_2(\Del y)\rangle, \\
&\sum_{j=1}^{N_2}\langle (-1)\Del B_{r^j}(t),\Del p^j\rangle \leq -N_2\rho_2(t)|\mg{m}_2(\Del p)|^2-N_2 b_t \langle \mg{m}_2(\Del r),\mg{m}_2(\Del p)\rangle-N_1 \rho_1(t)\langle \mg{m}_1(\Del P),\mg{m}_2(\Del p)\rangle\\
&\qquad-\sum_{i=1}^{N_1}\langle \part_\vp f_1^i(t,u,\opb^i)-\part_\vp f_1^i(t,\ac{u},\opb^{\prime,i}), \mg{m}_2(\Del p)\rangle. \nn
\end{split}
\ee
By the technique used for $(\ref{theta-technique})$ in Lemma~\ref{lemma-H-1-finite}, we get
\be
\small
\begin{split}
&\sum_{i=1}^{N_1}\langle \Del F_{P^i}(t),\Del X^i\rangle
=-\sum_{i=1}^{N_1}\langle \part_x f_1^i(t,u,\opb^i)-\part_x f_1^i(t,\ac{u},\opb^{\prime,i}), \Del X^i\rangle\\
&\qquad -\sum_{i=1}^{N_1}\ex^\theta\bigl[\langle \part_\mu f_1^i(t,u,\opb^i)(X^{\theta})-\part_\mu f_1^i(t,\ac{u},\opb^{\prime,i})(\ac{X}^\theta),\Del X^\theta\rangle\bigr]. \nn
\end{split}
\ee
Finally, as a direct result from Assumption~\ref{A-non-C} (iv), we have
\be
\small
\begin{split}
\sum_{j=1}^{N_2}\langle \Del F_{y^j}(t),\Del x^j\rangle+\sum_{j=1}^{N_2}\langle (-1)\Del F_{p^j}(t),\Del r^j\rangle
\leq -\gamma_2^f \sum_{j=1}^{N_2}(|\Del x^j|^2+|\Del r^j|^2)+N_2b_t \langle \Del \vp, \mg{m}_2(\Del x)\rangle. \nn
\end{split}
\ee
Summing all the terms and using Assumption~\ref{A-C} (iv), we get
\be
\small
\begin{split}
\cald(t)\leq -\gamma_1^f \sum_{i=1}^{N_1}|\Del X^i|^2-\gamma_2^f \sum_{j=1}^{N_2}(|\Del x^j|^2+|\Del r^j|^2)
-\sum_{i=1}^{N_1}(\lambda_\beta |\Del \opb^i|^2+\lambda_\vp |\Del \vp|^2)+\calr(t), 
\label{eq-cald}
\end{split}
\ee
where the residual term $\calr(t)$ is defined by
\be
\small
\begin{split}
\calr(t)&:=-N_2\rho_2(t)(|\mg{m}_2(\Del y)|^2+|\mg{m}_2(\Del p)|^2)-\sum_{i=1}^{N_1}\langle \part_\vp f_1^i(t,u,\opb^i)
-\part_\vp f_1^i(t,\ac{u},\opb^{\prime,i}),\mg{m}_2(\Del y+\Del p)\rangle \\
&\quad+N_1\langle (1-\rho_2(t)\Lambda_t)\mg{m}_2(\Del p-\Del y)-b_t\Lambda_t \mg{m}_2(\Del r), \mg{m}_1(\Del \opb)\rangle 
-N_1\rho_1(t)\langle \mg{m}_1(\Del P),\mg{m}_2(\Del y+\Del p)\rangle\\
&\quad -N_2 b_t \langle \mg{m}_2(\Del r),\mg{m}_2(\Del p)\rangle+N_2 b_t\langle \Del \vp, \mg{m}_2(\Del x)\rangle \nn.
\end{split}
\ee

We get an estimate on $\calr(t)$ from the Lipschitz continuity of $\part_\vp f_1^i$ and Young's inequality. In particular,
\be
\small
\begin{split}
&N_1|\rho_1(t)\langle \mg{m}_1(\Del P), \mg{m}_2(\Del y+\Del p)\rangle| \leq N_2\frac{\rho_2(t)}{2}(|\mg{m}_2(\Del y)|^2+|\mg{m}_2(\Del p)|^2)+\frac{N_1^2 \rho_1(t)^2}{N_2 \rho_2(t)}\Bigl(\frac{1}{N_1}\sum_{i=1}^{N_1}|\Del P^i|^2\Bigr) \\
&\qquad \leq N_2\frac{\rho_2(t)}{2}(|\mg{m}_2(\Del y)|^2+|\mg{m}_2(\Del p)|^2)+C\frac{N_1}{N_2}\sum_{i=1}^{N_1}|\Del P^i|^2, \\
&N_2b_t|\langle \Del \vp, \mg{m}_2(\Del x)\rangle|
\leq \frac{\gamma_2^f}{2}\sum_{j=1}^{N_2}|\Del x^j|^2+\frac{N_2 b_t^2}{\gamma_2^f}|\mg{m}_2(\Del y)|^2+C\frac{N_1}{N_2}\sum_{i=1}^{N_1}|\Del \opb^i|^2 \nn
\end{split}
\ee
with some constant $C$. Similar calculations yield
\be
\small
\begin{split}
\calr(t)\leq& \sum_{i=1}^{N_1}\bigl[\frac{\gamma_1^f}{2}|\Del X^i|^2+\frac{\lambda_\vp}{2}|\Del \vp|^2+\frac{\lambda_\beta}{2}|\Del \opb^i|^2\bigr]
+\frac{\gamma_2^f}{2}\sum_{j=1}^{N_2}(|\Del x^j|^2+|\Del r^j|^2) \\
&+CN_1(|\mg{m}_2(\Del y)|^2+|\mg{m}_2(\Del p)|^2+|\mg{m}_2(\Del r)|^2)+C\frac{N_1}{N_2}\sum_{i=1}^{N_1}(|\Del \opb^i|^2+|\Del P^i|^2)\\
&-N_2\frac{\rho_2(t)}{2}(|\mg{m}_2(\Del y)|^2+|\mg{m}_2(\Del p)|^2)+\frac{N_2b_t^2}{2\gamma_2^f}|\mg{m}_2(\Del p)|^2
+\frac{N_2 b_t^2}{\gamma_2^f}|\mg{m}_2(\Del y)|^2 \\
\leq &\sum_{i=1}^{N_1}\bigl[\frac{\gamma_1^f}{2}|\Del X^i|^2+\frac{\lambda_\vp}{2}|\Del \vp|^2+\frac{\lambda_\beta}{2}|\Del \opb^i|^2\bigr]
+\frac{\gamma_2^f}{2}\sum_{j=1}^{N_2}(|\Del x^j|^2+|\Del r^j|^2) \\
&+CN_1(|\mg{m}_2(\Del y)|^2+|\mg{m}_2(\Del p)|^2+|\mg{m}_2(\Del r)|^2)+C\frac{N_1}{N_2}\sum_{i=1}^{N_1}(|\Del \opb^i|^2+|\Del P^i|^2),
\end{split}
\ee
where we have used Assumption~\ref{A-C} (vi) in the second inequality.
Note that, the constant $C$ depends on $L$ and convexity parameters but not on $(N_1,N_2)$.
Now we choose  $\del_0^F:=\min\bigl(\frac{\lambda_\beta}{2C},\frac{\gamma_2^f}{6C}\bigr)$. 
Then, for any $\deln=N_1/N_2$ satisfying $\deln\leq \del_0^F$, we have 
\be
\small
\begin{split}
\calr(t)\leq& \sum_{i=1}^{N_1}\bigl[\frac{\gamma_1^f}{2}|\Del X^i|^2+\frac{\lambda_\vp}{2}|\Del \vp|^2+\lambda_\beta |\Del \opb^i|^2\bigr]
+\gamma_2^f \sum_{j=1}^{N_2}\bigl[\frac{1}{2}|\Del x^j|^2+\frac{2}{3}|\Del r^j|^2\bigr]\\
&+C\deln \bigl[\sum_{i=1}^{N_1}|\Del P^i|^2+\sum_{j=1}^{N_2}(|\Del y^j|^2+|\Del p^j|^2)\bigr]. \nn
\end{split}
\ee
By inserting into $(\ref{eq-cald})$, we obtain the desired estimate on $\cald(t)$.
\end{proof}
\end{lemma}

We now provide the first main result.
Since Peng-Wu's monotonicity does not hold,  we need in general to make all the Lipschitz constants small enough 
to guarantee the well-posedness of FBSDEs for a given duration.
However,  the result of Lemma~\ref{pw-drift},  which shows the violation of the monotonicity is proportional to  $\deln$, 
suggests that the well-posedness of the system $(\ref{fbsde-full-finite})$ may be available as long as the ratio $\deln=N_1/N_2$ is small enough. 
We shall show that this is actually the case  by the next theorem.

\begin{theorem}
\label{th-existence-finite}
Let Assumptions~\ref{A-non-C}-\ref{A-C} hold. Then, there exists some positive constant $\del_*^F \leq \del_0^F$
such that, for any $(N_1, N_2)$ satisfying $\deln\leq \del_*^F$, there exists a 
unique square integrable solution to the system of  FBSDEs $(\ref{fbsde-full-finite})$ with 
$u:=(X,x,y,P,p,r)\in  \mbb{S}^2(\mbb{F};\mbb{R}^n)^{N_1}\times 
\mbb{S}^2(\mbb{F};\mbb{R}^n)^{N_2}\times \mbb{S}^2(\mbb{F};\mbb{R}^n)^{N_2}\times \mbb{S}^2(\mbb{F};\mbb{R}^n)^{N_1}\times 
\mbb{S}^2(\mbb{F};\mbb{R}^n)^{N_2}\times \mbb{S}^2(\mbb{F};\mbb{R}^n)^{N_2}$.  
\begin{proof}
First, we put $\gamma:=\min\bigl(\frac{\gamma_1^f}{2},\frac{\gamma_2^f}{3}, \gamma_1^g, \gamma_2^g\bigr)$.
Let $I_{X^i}, I_{x^j}, I_{r^j}, I_{P^i}, I_{y^j}, I_{p^j}$ be in $\mbb{H}^2(\mbb{F};\mbb{R}^n)$ and $\theta_{P^i}, \theta_{y^j}, \theta_{p^j}$ be  in $\mbb{L}^2(\calf_T, \mbb{R}^n)$ for every $1\leq i\leq N_1, 1\leq j\leq N_2$.
We hypothesize that, for any choice of  $(I_{X^i},I_{x^j},I_{y^j},I_{P^i},I_{p^j},I_{r^j})$ and $(\theta_{P^i},\theta_{y^j},\theta_{p^j})$,
there exists some $\vr\in[0,1)$ such that
there exists a unique square integrable solution with $\ol{u}=(\ol{X},\ol{x},\ol{y},\ol{P},\ol{p},\ol{r})\in \mbb{S}^2(\mbb{F};\mbb{R}^n)^{2N_1+4N_2}$
to the following system of FBSDEs, $1\leq i\leq N_1, 1\leq j\leq N_2$:
\be
\small
\begin{split}
&d\ol{X}_t^i=(\vr B_{X^i}(t,\ol{u}_t)+I_{X^i}(t))dt+\sigma_1^{i,0}(t)dW_t^0+\sigma_1^i(t)dW_t^i, \\
&d\ol{x}_t^j=(\vr B_{x^j}(t,\ol{u}_t)+I_{x^j}(t))dt+\sigma_2^{j,0}(t)dW_t^0+\sigma_2^j(t)dB_t^j, \\
&d\ol{r}_t^j=(\vr B_{r^j}(t,\ol{u}_t)+I_{r^j}(t))dt,  \\
&d\ol{P}_t^i=-\bigl[(1-\vr)\gamma \ol{X}_t^i-\vr F_{P^i}(t,\ol{u}_t)+I_{P^i}(t)\bigr]dt+\ol{Q}_t^{i,0}dW_t^0+\sum_{k=1}^{N_1}\ol{Q}_t^{i,k}dW_t^k
+\sum_{k=1}^{N_2}\ol{\mg{Q}}_t^{i,k}dB_t^k, \\
&d\ol{y}_t^j=-\bigl[(1-\vr)\gamma\ol{x}_t^j-\vr F_{y^j}(t,\ol{u}_t)+I_{y^j}(t)\bigr]dt+\ol{z}_t^{j,0}dW_t^0+
\sum_{k=1}^{N_1}\ol{\mg{z}}_t^{j,k}dW_t^k+\sum_{k=1}^{N_2}\ol{z}_t^{j,k}dB_t^k, \\
&d\ol{p}_t^j=-\bigl[-(1-\vr)\gamma \ol{r}_t^j-\vr F_{p^j}(t,\ol{u}_t)+I_{p^j}(t)\bigr]dt+\ol{q}_t^{j,0}dW_t^0+
\sum_{k=1}^{N_1}\ol{\mg{q}}_t^{j,k}dW_t^k+\sum_{k=1}^{N_2}\ol{q}_t^{j,k}dB_t^k,  \nn
\end{split}
\ee
with $\ol{X}_0^i=\xi^i,~\ol{x}_0^j=\eta^j,~\ol{r}_0^j=0$ and
$
\ol{P}_T^i=\vr G_{P^i}(\ol{u}_T)+(1-\vr)\ol{X}_T^i+\theta_{P^i}, 
~\ol{y}_T^j=\vr G_{y^j}(\ol{u}_T)+(1-\vr)\ol{x}_T^j+\theta_{y^j},~\ol{p}_T^j=\vr G_{p^j}(\ol{u}_T)-(1-\vr)\ol{r}_T^j+\theta_{p^j}. \nn 
$
The hypothesis clearly holds when $\vr=0$ since then the system becomes the {\it decoupled} Lipschitz FBSDEs.

Now, with a given input process $\ol{u}\in \mbb{S}^2(\mbb{F};\mbb{R}^n)^{2N_1+4N_2}$, we consider the 
perturbed FBSDEs on $u=(X,x,y,P,p,r)$ with $\zeta\in(0,1)$, $1\leq i\leq N_1, 1\leq j\leq N_2$:
\be
\small
\begin{split}
&dX_t^i=\bigl[\vr B_{X^i}(t,u_t)+\zeta B_{X^i}(t,\ol{u}_t)+I_{X^i}(t)\bigr]dt+\sigma_1^{i,0}(t)dW_t^0+\sigma_1^i(t)dW_t^i, \\
&dx_t^j=\bigl[\vr B_{x^j}(t,u_t)+\zeta B_{x^j}(t,\ol{u}_t)+I_{x^j}(t)\bigr]dt+\sigma_2^{j,0}(t)dW_t^0+\sigma_2^j(t)dB_t^j, \\
&dr_t^j=\bigl[\vr B_{r^j}(t,u_t)+\zeta B_{r^j}(t,\ol{u}_t)+I_{r^j}(t)\bigr]dt, \\
&dP_t^i=-\bigl[(1-\vr) \gamma X_t^i-\vr F_{P^i}(t,u_t)+\zeta(-\gamma \ol{X}_t^i-F_{P^i}(t,\ol{u}_t))+I_{P^i}(t)\bigr]dt\\
&\qquad\quad+Q_t^{i,0}dW_t^0+\sum_{k=1}^{N_1}Q_t^{i,k}dW_t^k+\sum_{k=1}^{N_2}\mg{Q}_t^{i,k}dB_t^k, \\
&dy_t^j=-\bigl[(1-\vr)\gamma x_t^j-\vr F_{y^j}(t,u_t)+\zeta(-\gamma \ol{x}_t^j-F_{y^j}(t,\ol{u}_t))+I_{y^j}(t)\bigr]dt\\
&\qquad\quad+z_t^{j,0}dW_t^0+\sum_{k=1}^{N_1}\mg{z}_t^{j,k}dW_t^k+\sum_{k=1}^{N_2}z_t^{j,k}dB_t^k, \\
&dp_t^j=-\bigl[-(1-\vr)\gamma r_t^j-\vr F_{p^j}(t,u_t)+\zeta (\gamma \ol{r}_t^j-F_{p^j}(t,\ol{u}_t))+I_{p^j}(t)\bigr]dt\\
&\qquad\quad+q_t^{j,0}dW_t^0+\sum_{k=1}^{N_1}\mg{q}_t^{j,k}dW_t^k+\sum_{k=1}^{N_2}q_t^{j,k}dB_t^k, 
\label{fbsde-shifted}
\end{split}
\ee 
with the boundary conditions $X_0^i=\xi^i,~x^j_0=\eta^j, ~r_0^j=0$ and
$P_T^i=\vr G_{P^i}(u_T)+(1-\vr)X_T^i+\zeta (G_{P^i}(\ol{u}_T)-\ol{X}_T^i)+\theta_{P^i}$,
$y_T^j=\vr G_{y^j}(u_T)+(1-\vr)x_T^j+\zeta (G_{y^j}(\ol{u}_T)-\ol{x}_T^j)+\theta_{y^j}$
and $p_T^j=\vr G_{p^j}(u_T)-(1-\vr)r_T^j+\zeta (G_{p^j}(\ol{u}_T)+\ol{r}_T^j)+\theta_{p^j}$. 
Note that, by the initial hypothesis, there exists a unique solution $u=(X,x,y,P,p,r)\in\mbb{S}^2(\mbb{F};\mbb{R}^n)^{2N_1+4N_2}$
to the FBSDEs $(\ref{fbsde-shifted})$. This defines a map $\mbb{S}^2(\mbb{F};\mbb{R}^n)^{2N_1+4N_2}\ni \ol{u}\mapsto u \in\mbb{S}^2(\mbb{F};\mbb{R}^n)^{2N_1+4N_2}$. If we show that thie map is a contraction for some $\zeta>0$ independent of  $\vr$, 
then we can extend the hypothesis from $\vr$ to $(\vr+\zeta)$. Then we are done, because $\vr=1$ and $I,\theta=0$ 
give the interested FBSDEs $(\ref{fbsde-full-finite})$. In the following, we shall show that if $\deln\leq \del_*^F$
with some $\del_*^F>0$ determined independently from $\vr$, there exists a sufficiently small $\zeta>0$ independent of $\vr$
such that the above map $\ol{u}\mapsto u$ becomes a  contraction as desired.

\noindent {\bf{First step}}:
Let us denote by $u, \ac{u}\in \mbb{S}^2(\mbb{F};\mbb{R}^n)^{2N_1+4N_2}$ the solutions to $(\ref{fbsde-shifted})$
with two input processes $\ol{u},\ac{\ol{u}}\in \mbb{S}^2(\mbb{F};\mbb{R}^n)^{2N_1+4N_2}$, respectively.
As in Lemmas~\ref{pw-terminal}-\ref{pw-drift}, we use the notations:
$\Del u:=u-\ac{u}$, $\Del \ol{u}:=\ol{u}-\ac{\ol{u}}$, $\Del B_{X^i}(t):=B_{X^i}(t,u_t)-B_{X^i}(t,\ac{u}_t)$,
$\Del \ol{B}_{X^i}(t):=B_{X^i}(t,\ol{u}_t)-B_{X^i}(t,\ac{\ol{u}}_t)$, $\Del F_{P^i}(t):=F_{P^i}(t,u_t)-F_{P^i}(t,\ac{u}_t)$,
$\Del \ol{F}_{P^i}(t):=F_{P^i}(t,\ol{u}_t)-F_{P^i}(t,\ac{\ol{u}}_t)$, $\Del G_{P^i}:=G_{P^i}(u_T)-G_{P^i}(\ac{u}_T)$,
$\Del \ol{G}_{P^i}:=G_{P^i}(\ol{u}_T)-G_{P^i}(\ac{\ol{u}}_T)$, etc.
An application of It\^o formula with the result of Lemma~\ref{pw-drift} gives
\be
\small
\begin{split}
&\sum_{i=1}^{N_1}\ex\bigl[\langle \Del X_T^i, \Del P_T^i\rangle\bigr]+\sum_{j=1}^{N_2}
\ex\bigl[\langle \Del x_T^j, \Del y_T^j\rangle+\langle (-1)\Del r_T^j, \Del p_T^j\rangle\bigr]\\
&\leq -\gamma \ex\int_0^T \bigl[\sum_{i=1}^{N_1}|\Del X_t^i|^2+\sum_{j=1}^{N_2}(|\Del x_t^j|^2+|\Del r_t^j|^2)\bigr]dt \\
&+\deln C \ex\int_0^T \bigl[\sum_{i=1}^{N_1}|\Del P_t^i|^2
+\sum_{j=1}^{N_2}(|\Del y_t^j|^2+|\Del p_t^j|^2)\bigr] dt+\zeta \ex \int_0^T \sum_{i=1}^{N_1}\bigl[\langle \Del \ol{B}_{X^i}(t),\Del P_t^i\rangle+\langle \gamma \Del \ol{X}_t^i+
\Del \ol{F}_{P^i}(t),\Del X_t^i\rangle\bigr]dt \\
& +\zeta\ex\int_0^T \sum_{j=1}^{N_2}\bigl[\langle \Del \ol{B}_{x^j}(t),\Del y_t^j\rangle+\langle \gamma \Del \ol{x}_t^j+\Del \ol{F}_{y^j}(t),\Del x_t^j\rangle
+\langle (-1)\Del \ol{B}_{r^j}(t),\Del p_t^j\rangle+\langle \gamma \Del \ol{r}_t^j-\Del \ol{F}_{p^j}(t),\Del r_t^j\rangle\bigr]dt. \nn
\end{split}
\ee
We also have, from Lemma~\ref{pw-terminal}, 
\be
\small
\begin{split}
&\sum_{i=1}^{N_1}\ex\bigl[\langle \Del X_T^i, \Del P_T^i\rangle\bigr]+\sum_{j=1}^{N_2}
\ex\bigl[\langle \Del x_T^j, \Del y_T^j\rangle+\langle (-1)\Del r_T^j, \Del p_T^j\rangle\bigr]\\
&\geq (\vr \gamma+(1-\vr))\ex\bigl[\sum_{i=1}^{N_1}|\Del X_T^i|^2+\sum_{j=1}^{N_2}(|\Del x_T^j|^2+|\Del r_T^j|^2)\bigr]\\
&\quad +\zeta\ex\bigl[\sum_{i=1}^{N_1}\langle \Del \ol{G}_{P^i}-\Del \ol{X}_T^i, \Del X_T^i\rangle
+\sum_{j=1}^{N_2}(\langle \Del\ol{G}_{y^j}-\Del \ol{x}_T^j, \Del x_T^j\rangle+\langle \Del \ol{G}_{p^j}+\Del \ol{r}_T^j,(-1)\Del r_T^j\rangle\bigr].\nn
\end{split}
\ee
Put $\gamma_c:=\min(1,\gamma)$, which satisfies $0<\gamma_c\leq \vr \gamma+(1-\vr)$, then the above two inequalities yield
\be
\small
\begin{split}
&\gamma_c\ex\bigl[\sum_{i=1}^{N_1}|\Del X_T^i|^2+\sum_{j=1}^{N_2}(|\Del x_T^j|^2+|\Del r_T^j|^2)\Bigr]+
\gamma_c \ex \int_0^T \bigl[\sum_{i=1}^{N_1}|\Del X_t^i|^2+\sum_{j=1}^{N_2}(|\Del x_t^j|^2+|\Del r_t^j|^2)\bigr]dt\\
&\leq \deln C \ex\int_0^T \bigl[\sum_{i=1}^{N_1}|\Del P_t^i|^2+\sum_{j=1}^{N_2}(|\Del y_t^j|^2+|\Del p_t^j|^2)\bigr]dt
+\zeta \ex \int_0^T \sum_{i=1}^{N_1}\bigl[\langle \Del \ol{B}_{X^i}(t),\Del P_t^i\rangle+\langle \gamma \Del \ol{X}_t^i+
\Del \ol{F}_{P^i}(t),\Del X_t^i\rangle\bigr]dt \\
&\quad +\zeta\ex\int_0^T \sum_{j=1}^{N_2}\bigl[\langle \Del \ol{B}_{x^j}(t),\Del y_t^j\rangle+\langle \gamma \Del \ol{x}_t^j+\Del \ol{F}_{y^j}(t),\Del x_t^j\rangle
+\langle (-1)\Del \ol{B}_{r^j}(t),\Del p_t^j\rangle+\langle \gamma \Del \ol{r}_t^j-\Del \ol{F}_{p^j}(t),\Del r_t^j\rangle\bigr]dt. \\
&\quad -\zeta\ex\bigl[\sum_{i=1}^{N_1}\langle \Del \ol{G}_{P^i}-\Del \ol{X}_T^i, \Del X_T^i\rangle
+\sum_{j=1}^{N_2}(\langle \Del\ol{G}_{y^j}-\Del \ol{x}_T^j, \Del x_T^j\rangle+\langle \Del \ol{G}_{p^j}+\Del \ol{r}_T^j,(-1)\Del r_T^j\rangle\bigr].\nn
\end{split}
\ee
We now expand the right hand side by using Lipschitz continuity of functions, Lemma~\ref{lemma-H-1-finite} and Cauchy-Schwarz inequality.
By choosing $\zeta$ small enough to absorb the terms $(|\Del X|^2, |\Del x|^2, |\Del r|^2)$ into the left hand side, 
we obtain 
\be
\small
\begin{split}
&\ex\bigl[\sum_{i=1}^{N_1}|\Del X_T^i|^2+\sum_{j=1}^{N_2}(|\Del x_T^j|^2+|\Del r_T^j|^2)\bigr]+
\ex\int_0^T\bigl[\sum_{i=1}^{N_1}|\Del X_t^i|^2+\sum_{j=1}^{N_2}(|\Del x_t^j|^2+|\Del r_t^j|^2)\bigr]dt\\
&\leq (\zeta+\deln)C\ex\int_0^T \bigl[\sum_{i=1}^{N_1}|\Del P_t^i|^2+ \sum_{j=1}^{N_2}(|\Del y_t^j|^2+|\Del p_t^j|^2)\bigr]dt\\
&\quad +\zeta C\ex\bigl[\sum_{i=1}^{N_1}|\Del \ol{X}_T^i|^2+\sum_{j=1}^{N_2}(|\Del \ol{x}_T^j|^2+|\Del \ol{r}_T^j|^2)\bigr]\\
&\quad +\zeta C\ex \int_0^T \bigl[\sum_{i=1}^{N_1}(|\Del \ol{X}_t^i|^2+|\Del \ol{P}_t^i|^2)+\sum_{j=1}^{N_2}(
|\Del \ol{x}_t^j|^2+|\Del \ol{r}_t^j|^2+|\Del \ol{y}_t^j|^2+|\Del \ol{p}_t^j|^2)\bigr]dt.
\end{split}
\label{first-step-finite}
\ee
Here, the constant $C$ depends only on $L$ and the convexity parameters such as $\gamma, \lambda$. In particular, $C$ 
and the $\zeta$ used in the above transformation are independent of $\vr$ and the population sizes.
\\

\noindent
{\bf Second step}: We need to get  estimates on the backward components $(|\Del P|^2, |\Del y|^2, |\Del p|^2)$ 
in the right hand side of $(\ref{first-step-finite})$.
In addition to $(\ol{u},\ac{\ol{u}})$, let us now treat
$(X, x, r)$ as well as $(\ac{X}, \ac{x}, \ac{r})$ as exogenous input processes.
We consider $(P, y, p)$ as the solution to the BSDEs in $(\ref{fbsde-shifted})$ with the input processes $(\ol{u}, X,x,r)$,
and $(\ac{P},\ac{y},\ac{p})$  with  $(\ac{\ol{u}}, \ac{X}, \ac{x}, \ac{r})$, respectively.
We can then apply the standard stability result for Lipschitz BSDEs (see, for example, \cite[Theorem 4.4.4]{Zhang-book} or
\cite[Theorem 5.21]{Pardoux-Rascanu})
to obtain
\be
\small
\begin{split}
&\ex\int_0^T\bigl[\sum_{i=1}^{N_1}|\Del P_t^i|^2+\sum_{j=1}^{N_2}(|\Del y_t^j|^2+|\Del p_t^j|^2)\bigr]dt\\
&\leq C\ex\bigl[\sum_{i=1}^{N_1}|\Del X_T^i|^2+\sum_{j=1}^{N_2}(|\Del x_T^j|^2+|\Del r_T^j|^2)\bigr]
+C\ex\int_0^T \bigl[\sum_{i=1}^{N_1}|\Del X_t^i|^2+\sum_{j=1}^{N_2}(|\Del x_t^j|^2+|\Del r_t^j|^2)\bigr]dt\\
&\quad +\zeta C \ex\bigl[\sum_{i=1}^{N_1}|\Del \ol{X}_T^i|^2+\sum_{j=1}^{N_2}(|\Del \ol{x}_T^j|^2+|\Del \ol{r}_T^j|^2)\bigr] \\
&\quad +\zeta C \ex\int_0^T \bigl[\sum_{i=1}^{N_1}(|\Del \ol{X}_t^i|^2+|\Del \ol{P}_t^i|^2)+\sum_{j=1}^{N_2}(|\Del \ol{x}_t^j|^2+
|\Del \ol{r}_t^j|^2+|\Del \ol{y}_t^j|^2+|\Del \ol{p}_t^j|^2)\bigr]dt, \nn
\end{split}
\ee
where $C$ depends also on $T$ in addition to the Lipschitz constants. This result combined with $(\ref{first-step-finite})$ then implies
\be
\small
\begin{split}
&\ex\Bigl[\sum_{i=1}^{N_1}|\Del X_T^i|^2+\sum_{j=1}^{N_2}(|\Del x_T^j|^2+|\Del r_T^j|^2) +\int_0^T \bigl[\sum_{i=1}^{N_1}(|\Del X_t^i|^2+|\Del P_t^i|^2)+\sum_{j=1}^{N_2}
(|\Del x_t^j|^2+|\Del r_t^j|^2+|\Del y_t^j|^2+|\Del p_t^j|^2)\bigr]dt\Bigr]\\
&\leq (\zeta+\deln)C\ex\int_0^T \bigl[\sum_{i=1}^{N_1}|\Del P_t^i|^2+\sum_{j=1}^{N_2}(|\Del y_t^j|^2+|\Del p_t^j|^2)\bigr]dt\\
&\quad +\zeta C\ex\bigl[\sum_{i=1}^{N_1}|\Del \ol{X}_T^i|^2+\sum_{j=1}^{N_2}(|\Del \ol{x}_T^j|^2+|\Del \ol{r}_T^j|^2)\bigr]\\
&\quad +\zeta C\ex \int_0^T \bigl[\sum_{i=1}^{N_1}(|\Del \ol{X}_t^i|^2+|\Del \ol{P}_t^i|^2)+\sum_{j=1}^{N_2}(
|\Del \ol{x}_t^j|^2+|\Del \ol{r}_t^j|^2+|\Del \ol{y}_t^j|^2+|\Del \ol{p}_t^j|^2)\bigr]dt.
\end{split}
\label{second-step-finite}
\ee
Choose $\del_*^F\leq \del_0^F$ so that $C\del_*^F<1$.  Then, for any  $\deln=N_1/N_2 \leq \del_*^F$, 
there exists $\zeta>0$ sufficiently small so that
\be
\small
\begin{split}
(\zeta+\deln)C<1, ~\text{and}~\frac{\zeta C}{1-(\zeta+\deln)C}<1. \nn
\end{split}
\ee
For such a $\zeta$, the inequality $(\ref{second-step-finite})$ implies that the map $\ol{u}\mapsto u$ is a contraction 
with respect to the norm $\ex[|u_T|^2+\int_0^T |u_t|^2 dt]^\frac{1}{2}$.  By a simple application of Burkholder-Davis-Gundy (BDG)
inequality, one can also show that it is a contraction map  in the space $\mbb{S}^2(\mbb{F};\mbb{R}^n)^{2N_1+4N_2}$.
\end{proof}
\end{theorem}

By Theorems~\ref{th-sufficiency-finite} and \ref{th-existence-finite}, 
 $(\opb_t^i:=\opb^i(t,u_t), i=1,\ldots,N_1)_{t\in[0,T]}$ and $(\vp_t(y_t,\opb_t))_{t\in[0,T]}$
defined by the solution $u$  of $(\ref{fbsde-full-finite})$ give the unique optimal solution to $(\ref{central-problem})$
and the associated equilibrium price process,  respectively.
As one can see from the form of $\vp_t(\cdot, \cdot)$,
the cooperative agents gain more market power, i.e.~larger price impacts,  as their relative population size $\deln$ grows.
If the power is too strong, it may allow the cooperative agents to make arbitrary amount of profit by  some sort of price manipulation.
In fact, in the literature on the optimal trade executions, it has been known that large price impacts 
may cause a so-called ``hot-potato" problem of strongly oscillating sale and purchase. (See, for example, Schied \& Zhang (2017)~\cite{Schied}.)
It seems that this is one of the economic reasons why we need a small $\deln$ for the well-posedness.

Since  $(\ref{fbsde-full-finite})$ has a strong unique  $\mbb{F}$-adapted solution, 
Yamada-Watanabe Theorem for FBSDEs~\cite[Theorem~1.33]{Carmona-Delarue-2} implies that  
there exists a some measurable function $\Phi$ satisfying
\be
\small
\begin{split}
\bigl((X^i,P^i)_{\{i=1,\ldots, N_1\}}, (x^j,y^j,p^j,r^j)_{\{j=1,\ldots,N_2\}}\bigr)=\Phi\bigl(W^0, (\xi^i, W^i)_{\{i=1,\ldots, N_1\}},
(\eta^j, B^j)_{\{j=1,\ldots, N_2\}}\bigr). \nn
\end{split}
\ee
By the symmetry in the interactions of the agents (see the last statement of Lemma~\ref{lemma-H-1-finite})
with the form of  empirical means, as well as the symmetry in their heterogeneities as  mentioned in Remarks~\ref{remark-non-c} and \ref{remark-c}, the function $\Phi$ does not depend on the order of indexes 
$i$ and $j$. Hence, due to the i.i.d.~property of $(\xi^i, W^i)_{\{i=1,\ldots, N_1\}}$ and
$(\eta^j, B^j)_{\{j=1,\ldots, N_2\}}$, we can see that the joint distributions of $(X^i, P^i, \xi^i, c^i, \opb^i, W^i)_{\{i=1,\ldots, N_1\}}$ as well as  $(x^j, y^j,p^j,r^j,\eta^j,\mg{c}^j,B^j)_{\{j=1,\ldots, N_2\}}$ are invariant under the  permutation of their indexes respectively. In other words, they are the exchangeable random variables.
Since $W^0$ is independent from the other inputs, the exchangeability also holds  $\calf^0$-conditionally. 
Moreover, the permutation of $(x^j, y^j,p^j,r^j,\eta^j,\mg{c}^j,B^j)_{\{j=1,\ldots, N_2\}}$
does not affect the distribution of  $(X^i, P^i, \xi^i, c^i, \opb^i, W^i)_{\{i=1,\ldots, N_1\}}$ and vice versa.

\section{Mean-field equilibrium}
\label{sec-mfg}
\subsection{Notations and preliminary remarks}
\label{sec-mfg-notation}
Before setting up the mean-field model, let us introduce some notations to be used in this section.
Since we will single out an arbitrarily chosen pair of representative agents $(i,j)$ from the two populations, and the other agents are to be decoupled,
it is useful to work on smaller probability spaces.
\bi
\small
\item $(\ol{\Omega},\ol{\calf},\ol{\mbb{P}})$ is a probability space defined by the product $\ol{\Omega}:=\ol{\Omega}^{1,i}\times \ol{\Omega}^{2,j}$ with $(\ol{\calf},\ol{\mbb{P}})$ the completion of $(\ol{\calf}^{1,i}\otimes \ol{\calf}^{2,j},~\ol{\mbb{P}}^{1,i}\otimes \ol{\mbb{P}}^{2,j})$.
$\ol{\mbb{F}}:=(\ol{\calf}_t)_{t\geq 0}$ denotes the complete and right-continuous augmentation of $(\ol{\calf}_t^{1,i}\otimes \ol{\calf}_t^{2,j})_{t\geq 0}$.
\item $(\Omega,\calf,\Prb)$ is a probability space defined by the  product  $\Omega:=\Omega^0\times \ol{\Omega}$ 
with $(\calf,\Prb)$ the completion of $(\calf^0\otimes \ol{\calf}, ~\Prb^0\otimes \ol{\Prb})$. $\mbb{F}:=(\calf_t)_{t\geq 0}$
denotes the complete and right-continuous augmentation of $(\calf_t^0\otimes \ol{\calf}_t)_{t\geq 0}$.
Here, $(\Omega^0,\calf^0,\mbb{P}^0)$ with $\mbb{F}^0$ is defined in the same way as in Section~\ref{sec-notation}.
\ei
\noindent
$\ex[\cdot]$ denotes the expectation with respect to $\mbb{P}$, 
and $\omega=(\omega^0,\ol{\omega})$ with $\omega^0\in \Omega^0$ and $\ol{\omega}\in \ol{\Omega}$
a generic element of $\Omega$. Recall that $\mbb{F}^0$ denotes the filtration generated by the common noise $W^0$. 
We also need the following:
\bi
\small
\item $(\Omega^{1,i},\calf^{1,i},\mbb{P}^{1,i})$ is a probability space defined by the product $\Omega^{1,i}:=\Omega^0\times\ol{\Omega}^{1,i}$
with $(\calf^{1,i},\mbb{P}^{1,i})$ the completion of $(\calf^0\otimes \ol{\calf}^{1,i},\mbb{P}^0\otimes \ol{\mbb{P}}^{1,i})$.
$\mbb{F}^{1,i}:=(\calf^{1,i}_t)_{t\geq 0}$ denotes the complete and right-continuous augmentation of $(\calf^0_t\otimes \ol{\calf}_t^{1,i})_{t\geq 0}$. 
$(\Omega^{2,j},\calf^{2,j},\mbb{P}^{2,j})$ with $\mbb{F}^{2,j}:=(\calf^{2,j}_t)_{t\geq 0}$ is defined similarly as above with
$(\ol{\Omega}^{1,i},\ol{\calf}^{1,i},\ol{\mbb{P}}^{1,i})$ and $\ol{\mbb{F}}^{1,i}$ replaced by $(\ol{\Omega}^{2,j},\ol{\calf}^{2,j},\ol{\mbb{P}}^{2,j})$
and $\ol{\mbb{F}}^{2,j}$.
\ei

Let us give some remarks on the properties of $\calf^0$-conditional distributions and expectations.
By \cite[Lemma~2.4]{Carmona-Delarue-2}, it is shown that
for any $\mbb{R}^n$-valued random variable $X$ on $(\Omega,\calf,\mbb{P})$, $X(\omega^0,\cdot)$
is a random variable on $(\ol{\Omega},\ol{\calf},\ol{\mbb{P}})$ for $\mbb{P}^0$-a.e. $\omega^0\in \Omega^0$.
In particular, by assigning an arbitrary $\calp(\mbb{R}^n)$ value for the exceptional set, 
the map defined by $\omega^0 \mapsto \call^0(X)(\omega^0):=\call(X(\omega^0,\cdot))$
becomes a  $\calp(\mbb{R}^n)$-valued random variable on $(\Omega^0,\calf^0,\mbb{P}^0)$,
and  $\call^0(X)$ provides the conditional law of $X$ given $\calf^0$.
Moreover, if  $X=(X_t)_{t\geq 0}$ is a stochastic process in $\mbb{S}^2(\mbb{F};\mbb{R}^n)$
then \cite[Lemma~2.5]{Carmona-Delarue-2} shows that we can find a version of $(\call^0(X_t))_{t\geq 0}$ such that
the process $(\call^0(X_t))_{t\geq 0}$ is $\mbb{F}^0$-adapted and has continuous paths in $\calp_2(\mbb{R}^n)$.
We denote by $\ex^0[\cdot]$ 
the $\calf^0$-conditional expectation, or equivalently the integration over $\ol{\Omega}$ by $\ol{\mbb{P}}$. 
For any integrable $\mbb{F}$-adapted process $X=(X_t)_{t \geq 0}$, we have
$\ex^0[X_t](\omega^0)=\ex[X_t(\omega^0,\cdot)] =\ex[X_t|\calf_t^0](\omega^0)=\ex[X_t|\calf^0](\omega^0)$
since  $\sigma(W_s^0-W_t^0),~s>t$ is independent of $\sigma(X_t)$. 
In \cite[Section 4.3.1]{Carmona-Delarue-2}, 
it is shown that we can find a version of $(\ex^0[X_t])_{t\geq 0}$ which is $\mbb{F}^0$-progressively measurable
by modifying $dt\otimes d\mbb{P}^0$-null set when the process $X=(X_t)_{t\geq 0}$ is $\mbb{F}$-progressively measurable. 
In the remainder, we always use such a version.

As we have discussed in Subsection~\ref{sec-coop-finite}, we need to lift  a function on probability measures to that on
$\mbb{L}^2$-random variables to define the $L$-derivative. To do this, we need another probability space.
\bi
\small
\item $(\wt{\ol{\Omega}},\wt{\ol{\calf}}, \wt{\ol{\mbb{P}}})$ is a copy of the probability space $(\ol{\Omega},\ol{\calf},\ol{\mbb{P}})$. 
For any random variable $X$ on $\Omega=\Omega^0\times \ol{\Omega}$, 
the symbol $\wt{X}$ is used to denote a random variable defined as a copy of $X$ on the space $\wt{\Omega}:=\Omega^0\times \wt{\ol{\Omega}}$.
We also use the abbreviations such as, $\wt{f}_1^i(t,\cdot):=f_1(t,\cdot,c_t^0,\wt{c}_t^i)$ and 
$\wt{g}_1^i(\cdot):=g_1(\cdot,c_T^0,\wt{c}_T^i)$ to ease the notations.
\ei
In particular, conditionally on $\calf^0$ (i.e.~for each $\omega^0\in \Omega^0, \mbb{P}^0$-a.e.), $X$ and $\wt{X}$ are independent and have the same distribution 
$\call^0(X)=\call^0(\wt{X})$. 
The corresponding expectations $\wt{\ex}[\cdot]$ and $\wt{\ex}^0[\cdot]$ are defined in the same way as
$\ex[\cdot]$ and $\ex^0[\cdot]$, respectively.
When we need to put both $X$ and $\wt{X}$ on the common space, we always assume that they are extended naturally
to the space $\Omega^0\times \ol{\Omega} \times \wt{\ol{\Omega}}$ and still use the same symbols. We have $\ex\wt{\ex}[\cdot]=\ex\wt{\ex}^0[\cdot]=\wt{\ex}\ex^0[\cdot]$.
In particular, if  $F:\mbb{R}^n\times \mbb{R}^n\rightarrow \mbb{R}$ is a measurable function such that $F(X, \wt{X})$ is integrable, 
then Fubini's theorem implies $\ex\wt{\ex}^0[F(X,\wt{X})]=\ex\wt{\ex}^0[F(\wt{X},X)]$.
Throughout this section, we work under Assumptions~\ref{A-non-C}-\ref{A-C}.
\subsection{Mean-field problem}
Firstly, let us provide a formal derivation of  the corresponding mean field problem.
To do so, it is beneficial for us to recall the famous result of De Finetti's theory of 
exchangeable sequence of random variables:
\begin{theorem}[{\cite{Carmona-Delarue-2}[Theorem~2.1]}]
\label{th-De-Finetti}
For any exchangeable sequence of random variables $(X_n)_{n\geq 1}$ with $\mbb{E}[|X_1|]<\infty$, 
it holds, $\mbb{P}$-a.s., 
\be
\lim_{n\rightarrow \infty}\frac{1}{n}\sum_{i=1}^n X_i=\mbb{E}[X_1|\calf_\infty], \nn
\ee
where $\calf_\infty$ is the tail $\sigma$-field $\calf_\infty=\bigcap_{n\geq 1}\sigma\{X_k, k\geq n\}$.
\end{theorem}
\noindent
Suppose that $(y^j)_{j=1,\ldots, N_2}$ and $(X^i,\beta^i)_{i=1,\ldots, N_1}$ are exchangeable sequences of random variables respectively.  
If we take $N_1, N_2\rightarrow \infty$ with $\deln:=N_1/N_2$ kept fixed, then we expect to have, from Theorem~\ref{th-De-Finetti} with $\calf_{\infty}$ replaced by $\calf^0$, 
\be
\vp_t(y_t,\beta_t):=-\mg{m}_2(y_t)+\deln\Lambda_t \mg{m}_1(\beta_t)
\longrightarrow -\ex^0[y_t^1]+\deln\Lambda_t \ex^0[\beta_t^1] ~{\text{$\mbb{P}$-a.s. as $N_1, N_2\rightarrow \infty$}}. \nn
\ee
It is also natural to suppose 
\be
\nu_t^{N_1}:=\frac{1}{N_1}\sum_{i=1}^{N_1}\del_{X^i}\longrightarrow \call^0(X_t^1)~{\text{$\mbb{P}$-a.s. as $N_1 \rightarrow \infty$}}. \nn
\ee

Considering the exchangeability  of the solution to the system of equations $(\ref{fbsde-full-finite})$ as discussed  at
the end of Section~\ref{sec-existence-finite}, the above observation motivates us 
to study the following problem: Solve
\be
\small
\inf_{\beta^i \in \mbb{A}_1^i}\calj_1^i(\beta^i), 
\label{mfg-c-problem}
\ee
with the cost functional defined by
\be
\small
\begin{split}
\calj_1^i(\beta^i):=\ex\Bigl[\int_0^T f_1^i(t,X_t^i,\call^0(X_t^i),\pi_t(y_t^j,\beta_t^i),\beta_t^i)dt+g_1^i(X_T^i,\call^0(X_T^i))\Bigr], \nn
\end{split}
\ee
under the dynamic constraints:
\be
\small
\begin{split}
&dX^i_t=(\beta_t^i+\rho_1(t)\pi_t(y_t^j,\beta_t^i)+l_1^i(t))dt+\sigma_1^{i,0}(t)dW_t^0+\sigma_1^i(t)dW_t^i, \\
&dx_t^j=(-\ol{\Lambda}_t(y_t^j+\pi_t(y_t^j,\beta_t^i))+\rho_2(t)\pi_t(y_t^j,\beta_t^i)+l_2^j(t))dt+\sigma_2^{j,0}(t)dW_t^0+\sigma_2^j(t)dB_t^j, \\
&dy_t^j=-(\part_x \ol{f}_2^j(t,x_t^j)-b_t \pi_t(y_t^j,\beta_t^i))dt+z_t^{j,0}dW_t^0+z_t^{j,j}dB_t^j, 
\end{split}
\label{mfg-c-dynamics}
\ee
with the boundary conditions; $X^i_0=\xi^i, ~x^j_0=\eta^j$ and $y^j_T=\part_x g_2^j(x_T^j)$.
Here, $\mbb{A}_1^i:=\mbb{H}^2(\mbb{F}^{1,i};A_1)$ is the space of admissible controls, and $\pi_t(\cdot,\cdot)$ is defined by
\be
\small
\pi_t(y,\beta):=-\ex^0[y]+\deln \Lambda_t \ex^0[\beta]. 
\label{def-pi}
\ee 
In this problem, we have chosen a pair $\bf{(i,j)}$  as {\bf the  representative agents} from the two populations
and they are kept fixed throughout this section.
The rigorous discussions on the relation between this problem with the finite-agent equilibrium discussed in the previous sections
will be the topic in the next section.

\begin{remark}
\label{remark-given-beta}
Let Assumptions~\ref{A-non-C}-\ref{A-C} hold.
Then, for a given $\beta^i\in \mbb{A}_1^i$,  there exists a unique solution $x^j,y^j \in \mbb{S}^2(\mbb{F}^{2,j};\mbb{R}^n)$ of 
$(\ref{mfg-c-dynamics})$. This can be shown by the Peng-Wu's continuation method  in a completely parallel manner to 
Theorem~\ref{th-non-c-existence}. In fact, this is a special situation handled in \cite[Theorem~4.2]{Fujii-Takahashi}.
Therefore, for each $\beta^i \in \mbb{A}_1^i$, the state processes $(X^i,x^j,y^j)\in \mbb{S}^2(\mbb{F}^{1,i};\mbb{R}^n)\times
(\mbb{S}^2(\mbb{F}^{2,j};\mbb{R}^n))^2$ are defined uniquely.
\end{remark}

\begin{remark}
\label{remark-info-structure}
Note that, the symmetric interaction contained in $\vp_t(\cdot,\cdot)$ as the empirical means does not exist in $\pi_t(\cdot, \cdot)$
anymore. This induces a decoupling among the agents, which can also be observed in the information structure.
Since $X^i$ is adapted to $\mbb{F}^{1,i}$, the 
representative agent $i$ cares only about the common information $\mbb{F}^0$ and her own idiosyncratic 
information $\ol{\mbb{F}}^{1,i}$ (see the definition of $\mbb{A}_1^i$ given above).
The same observation also holds for the agent $j$ of the second population. 
In fact, the optimal control of the agent $j$ is now given by $\wh{\alpha}_t^j=-\ol{\Lambda}_t(y_t^j+\pi_t(y_t^j,\beta_t^i))$,
which is adapted to $\mbb{F}^0\otimes \ol{\mbb{F}}^{2,j}$.
\end{remark}

The involvement of $\ex^0[\beta^i_t]$-term in the cost function makes our problem the conditional extended mean-field control~\cite{Carmona-E}.
For this type of problem, it is convenient to work on the lifted Hamiltonian $\calh$ defined on the $\mbb{L}^2$-space. 
We define the map $\calh:[0,T]\times \mbb{L}^2(\calf_t^{1,i};\mbb{R}^n)\times (\mbb{L}^2(\calf_t^{2,j};\mbb{R}^n))^2\times
\mbb{L}^2(\calf_t^{1,i};\mbb{R}^n)\times (\mbb{L}^2(\calf_t^{2,j};\mbb{R}^n))^2\times \mbb{L}^2(\calf_t^{1,i};A_1)
\ni (t, X^i_t,x^j_t,y^j_t, P^i_t, p^j_t,r^j_t,\beta^i_t)\mapsto \calh(t,X^i_t,x^j_t,y^j_t,P^i_t,p^j_t,r^j_t,\beta^i_t)\in \mbb{R}$ by
\be
\small
\begin{split}
&\calh(t,X^i_t,x^j_t,y^j_t,P^i_t,p^j_t,r^j_t,\beta^i_t)\\
&:=\ex\Bigl[\langle P^i_t,\beta^i_t+\rho_1(t)\pi_t(y^j_t,\beta^i_t)+l_1^i(t)\rangle
+f_1^i(t,X^i_t,\call^0(X^i_t),\pi_t(y^j_t,\beta^i_t),\beta^i_t)\\
&\quad +\oldeln\bigl\{\langle p^j_t, -\ol{\Lambda}_t(y^j_t+\pi_t(y^j_t,\beta^i_t))+\rho_2(t)\pi_t(y^j_t,\beta^i_t)+l_2^j(t)\rangle
+\langle r_t^j, -\part_x\ol{f}_2^j(t,x^j_t)+b_t\pi_t(y_t^j,\beta^i_t)\rangle\bigr\}\Bigr],
\end{split}
\label{lifted-Hamiltonian}
\ee
where $\oldeln:=1/\deln=N_2/N_1$. The dependence on the conditional distribution $\call^0(X^i_t)$
as well as the conditional expectations $(\ex^0[y^j_t], \ex^0[\beta^i_t])$ in $\pi_t(\cdot,\cdot)$
is treated as a function on the random variables $X_t^i$ and $(y_t^j, \beta^i_t)$, respectively.
Since there is no information about the size of populations in the dynamics $(\ref{mfg-c-dynamics})$,
it must be given by hand to the Hamiltonian $\calh$ using the coefficient  $\oldeln$. 
The way in which $\calh$ depends on $\oldeln$ may be guessed from $H$ in $(\ref{H-1-finite})$.
As before, with slight abuse of notation, we write the lifted Hamiltonian as $\calh(t,X_t^i, x^j_t,y^j_t,P^i_t, p^j_t,r^j_t,\pi_t,\beta^i_t)$
when we treat $\pi_t\in \mbb{L}^2(\calf_t^0;\mbb{R}^n)$ as a separate argument. 

For each $u_t:=(X^i_t,x^j_t,y^j_t,P^i_t,p^j_t,r^j_t)$, 
we want to find the minimizer:
\be
\small
\begin{split}
\opb^i(t,u_t):={\rm argmin}\bigl(\calh(t,X^i_t,x^j_t,y^j_t,P^i_t,p^j_t,r^j_t,\beta_t^i); \beta^i_t\in \mbb{L}^2(\calf_t^{1,i};A_1)\bigr). \nn
\end{split}
\ee
The optimization problem in a Hilbert space has been well studied and we can follow the standard technique 
using the weak sequential lower-semicontinuity. See, for example, \cite{Okelo} and a brief monograph \cite{Peypouquet}.
Below, we often use the following results on Frechet derivatives of $\calh$: with $\pi_t:=\pi_t(y^j_t,\beta^i_t)$ and $\mu_t:=\call^0(X^i_t)$,
\be
\small
\begin{split}
D_{\beta^i} \calh(t,u_t,\beta^i_t)&=\part_{\beta^i} \calh(t,u_t,\pi_t,\beta^i_t)+\deln \Lambda_t 
\ex^0[\part_\pi \calh(t,u_t,\pi_t,\beta^i_t)] \\
&=P_t^i+\part_\beta f_1^i(t,X^i_t,\mu_t,\pi_t,\beta^i_t)+\deln \Lambda_t \ex^0[\part_\vp f_1^i(t,X_t^i,\mu_t,
\pi_t,\beta_t^i)] \\
&\quad+\deln \rho_1(t)\Lambda_t \ex^0[P_t^i]+(-I+\rho_2(t)\Lambda_t)\ex^0[p_t^j]+b_t\Lambda_t \ex^0[r^j_t], \\
D_{X^i}\calh(t,u_t,\beta_t^i)&=\part_x f_1^i(t,X^i_t,\mu_t,\pi_t,\beta^i_t)+
\wt{\ex}^0[\part_\mu \wt{f}_1^i(t,\wt{X}^i_t,\mu_t, \pi_t,\wt{\beta}^i_t)(X^i_t)], \\
D_{x^j}\calh(t,u_t,\beta^i_t)&=\part_{x^j} \calh(t,u_t,\beta_t^i)=-c_f^j(t)r^j_t, \\
D_{y^j} \calh(t,u_t,\beta^i_t)&=-\oldeln \ol{\Lambda}_t p^j_t-\ex^0[\part_\pi \calh(t,u_t,\pi_t,\beta^i_t)] \\
&\hspace{-8mm}=-\ex^0[\part_\vp f_1^i(t,X^i_t,\mu_t,\pi_t,\beta^i_t)]-\rho_1(t)\ex^0[P^i_t]+\oldeln \bigl(\ol{\Lambda}_t(\ex^0[p^j_t]-p^j_t)
-\rho_2(t)\ex^0[p^j_t]- b_t\ex^0[r^j_t]\bigr).
\end{split}
\label{H-Frechet-D}
\ee

\begin{lemma}
\label{mfg-H-minimization}
Under Assumptions~\ref{A-non-C}-\ref{A-C}, for each $t\in[0,T]$ and a given set of adjoint variables
$(P_t^i,p^j_t,r^j_t)\in \mbb{L}^2(\calf_t^{1,i};\mbb{R}^n)\times (\mbb{L}^2(\calf_t^{2,j};\mbb{R}^n))^2$,
the lifted Hamiltonian $\calh$ defined by $(\ref{lifted-Hamiltonian})$ is jointly convex in $(X^i_t,x^j_t,y^j_t,\beta^i_t)\in 
\mbb{L}^2(\calf_t^{1,i};\mbb{R}^n)\times (\mbb{L}^2(\calf_t^{2,j};\mbb{R}^n))^2\times \mbb{L}^2(\calf_t^{1,i};A_1)$
and strictly so in $(X^i_t,\beta^i_t)$. In particular, for any $t\in[0,T]$ and input  $u_t:=(X_t^i,x^j_t,y^j_t,P^i_t,p^j_t,r^j_t)$, 
there exists a unique minimizer $\opb_t^i:=\opb^i(t,u_t)\in \mbb{L}^2(\calf_t^{1,i};A_1)$ that satisfies
the following linear growth and the Lipschitz continuity properties:
\be
\small
\begin{split}
&\ex\bigl[|\opb^i(t,u_t)|^2\bigr]\leq C\ex\Bigl[1+|X_t^i|^2+|P_t^i|^2+|\ex^0[y_t^j]|^2+|\ex^0[p_t^j]|^2+|\ex^0[r_t^j]|^2+|c_t^0|^2+|c_t^i|^2\Bigr], \\
&\ex\bigl[|\opb^i(t,u_t)-\opb^i(t,\ac{u}_t)|^2\bigr] \leq C\ex\Bigl[|X^i_t-\ac{X}^i_t|^2+|P_t^i-\ac{P}^i_t|^2+|\ex^0[y^j_t-\ac{y}^j_t]|^2
+|\ex^0[p_t^j-\ac{p}^j_t]|^2+|\ex^0[r_t^j-\ac{r}^j_t]|^2\Bigr]. \nn
\end{split}
\ee
where $\ac{u}_t:=(\ac{X}^i_t,\ac{x}^j_t,\ac{y}^j_t,\ac{P}^i_t,\ac{p}^j_t,\ac{r}^j_t)$ is another input and $C$ is some positive constant independent of 
$\deln$.
\begin{proof}
Firstly,  we check the relevant convexity of $\calh$. For a given set of adjoint variables $(P_t^i,p_t^j,r_t^j)$, let us consider 
two arbitrary inputs $(X^i_t,x^j_t,y^j_t), (\ac{X}^i_t,\ac{x}^j_t,\ac{y}^j_t)\in \mbb{L}^2(\calf^{1,i}_t;\mbb{R}^n)\times (\mbb{L}^2(\calf_t^{2,j};\mbb{R}^n))^2$
and $\beta^i_t,\ac{\beta}^i_t \in \mbb{L}^2(\calf^{1,i}_t;A_1)$.
For notational simplicity, we put $u_t:=(X^i_t,x^j_t,y^j_t,P^i_t,p^j_t,r^j_t)$ and $\ac{u}_t:=(\ac{X}^i_t,\ac{x}^j_t,\ac{y}^j_t,P^i_t,p^j_t,r^j_t)$,
$\mu_t:=\call^0(X^i_t), \ac{\mu}_t:=\call^0(\ac{X}^i_t), \pi_t:=\pi_t(y^j_t,\beta^i_t), \ac{\pi}_t:=\pi_t(\ac{y}^j_t,\ac{\beta}^i_t)$,
$\Del X^i_t:=\ac{X}^i_t-X^i_t, \Del \beta_t^i:=\ac{\beta}^i_t-\beta^i_t$ and  $\Del \pi_t:=\ac{\pi}_t-\pi_t=-\ex^0[\Del y^j_t]+\deln\Lambda_t\ex^0[\Del \beta^i_t]$, etc. Then, the Frechet derivatives  $(\ref{H-Frechet-D})$, Assumption~\ref{A-C} (iv) and Fubini's theorem yield
\be
\small
\begin{split}
&\calh(t,\ac{u}_t,\ac{\beta}^i_t)-\calh(t,u_t,\beta^i_t)-\ex\bigl[\langle D_{X^i} \calh(t,u_t,\beta^i_t), \Del X^i_t \rangle 
+\langle D_{x^j} \calh(t,u_t,\beta^i_t), \Del x^j_t \rangle \\
&\quad+\langle D_{y^j}\calh(t,u_t,\beta^i_t), \Del y^j_t \rangle+\langle D_{\beta^i} \calh(t,u_t,\beta^i_t),\Del \beta^i_t \rangle\bigr]\\
&=\ex\bigl[f_1^i(t,\ac{X}^i_t,\ac{\mu}_t,\ac{\pi}_t,\ac{\beta}^i_t)-f_1^i(t,X^i_t,\mu_t,\pi_t,\beta^i_t)
-\wt{\ex}^0[\langle \part_\mu f_1^i(t,X^i_t,\mu_t,\pi_t,\beta^i_t)(\wt{X}^i_t),\Del \wt{X}^i_t\rangle] \\
&\quad -\langle \part_x f_1^i(t,X_t^i,\mu_t,\pi_t,\beta^i_t),\Del X^i_t \rangle 
-\langle \ex^0[\part_\vp f_1^i(t,X^i_t,\mu_t,\pi_t,\beta^i_t)], \Del \pi_t\rangle-\langle \part_\beta f_1^i(t,X^i_t,\mu_t,\pi_t,\beta^i_t),\Del \beta_t^i\rangle\bigr]
\\
&\geq \frac{1}{2}\ex\bigl[\gamma_1^f|\Del X^i_t|^2+\lambda_\beta |\Del \beta^i_t|^2+\lambda_\vp |\Del \pi_t|^2\bigr]. 
\end{split}
\label{lifted-H-convexity}
\ee
This gives the desired convexity. 
Since the map $\beta^i_t\mapsto \calh(t,u_t,\beta^i_t)$ is coercive due to its strict $\lambda_\beta$-convexity,
we can restrict its minimization on the set $K:=\{\beta^i_t\in \mbb{L}^2(\calf_t^{1,i};A_1), \|\beta^i_t\|_{\mbb{L}^2}\leq M\}$
for some constant $M>0$. Since our probability spaces are assumed to be the completion of those countably generated,
it is known that $H:=\mbb{L}^2(\calf_t^{1,i};\mbb{R}^n)$ is a separable Hilbert space. Since $K$
is a bounded and closed convex subset of $H$, $K$ is weakly sequentially compact by \cite[Proposition~2.6]{Okelo}. 
Since the map $\beta^i_t\mapsto \calh(t,u_t,\beta^i_t)$ is strongly continuous and convex,  it  is 
also weakly sequentially lower semicontinuous by \cite[Lemma~2.7]{Okelo}. Therefore, by \cite[Theorem~3.1]{Okelo},
there exists a minimizer $\opb^i(t,u_t)$ of the map, which is unique by the strict convexity. 


The latter claims can be proved similarly to those in  Lemma~\ref{lemma-H-1-finite}.
Let us fix $\beta^0 \in \mbb{L}^2(\calf_t^{1,i};A_1)$ arbitrarily,
and denote by $\opb_t^i:=\opb^i(t,u_t), \opb_t^{\prime,i}:=\opb^i(t,\ac{u}_t)\in \mbb{L}^2(\calf_t^{1,i};A_1)$ the unique minimizers of $\calh(t,u_t,\cdot)$
and $\calh(t,\ac{u}_t,\cdot)$, respectively.  Here, $u_t:=(X^i_t,x^j_t,y^j_t,P^i_t,p^j_t,r^j_t)$ and 
$\ac{u}_t:=(\ac{X}^i_t,\ac{x}^j_t,\ac{y}^j_t,\ac{P}^i_t,\ac{p}^j_t,\ac{r}^j_t)$
are two arbitrary inputs. Using the convexity $(\ref{lifted-H-convexity}$) and the optimality condition,  we can show that
the following inequalities hold:
\be
\small
\begin{split}
&\frac{\lambda_\beta}{2}\ex[|\opb^i_t-\beta^0|^2]\leq \ex\bigl[\langle \beta^0-\opb^i_t,  D_{\beta^i} \calh(t,u_t,\beta^0)\rangle\bigr], \\
&\lambda_\beta \ex[|\opb_t^{\prime,i}-\opb^i_t|^2]\leq \ex\bigl[\langle \opb_t^{\prime,i}-\opb^i_t, D_{\beta^i}\calh(t,u_t,\opb_t^{\prime,i})
-D_{\beta^i} \calh(t,\ac{u}_t,\opb_t^{\prime,i})\rangle\bigr]. \nn
\end{split}
\ee
Yong's equality and Lipschitz continuity from Assumption~\ref{A-C} (iii) then give the desired result.
\end{proof}
\end{lemma}

The system of state and adjoint equations for the problem $(\ref{mfg-c-problem})$ is given as follows:
\be
\small
\begin{split}
&dX^i_t=(\opb^i_t+\rho_1(t)\pi_t(y^j_t,\opb^i_t)+l_1^i(t))dt+\sigma_1^{i,0}(t)dW_t^0+\sigma_1^i(t)dW_t^i, \\
&dx^j_t=(-\ol{\Lambda}_t(y^j_t+\pi_t(y^j_t,\opb^i_t))+\rho_2(t)\pi_t(y^j_t,\opb^i_t)+l_2^j(t))dt+\sigma_2^{j,0}(t)dW_t^0+\sigma_2^j(t)dB_t^j, \\
&dy^j_t=-(\part_x \ol{f}_2^j(t,x^j_t)-b_t\pi_t(y^j_t,\opb^i_t))dt+z_t^{j,0}dW_t^0+z_t^{j,j}dB_t^j, \\
&dP^i_t=-\bigl\{\part_x f_1^i(t,X^i_t,\call^0(X^i_t),\pi_t(y^j_t,\opb^i_t),\opb^i_t)+\wt{\ex}^0[\part_\mu \wt{f}_1^i(t,\wt{X}^i_t,\call^0(X^i_t),\pi_t(y^j_t,\opb^i_t),
\wt{\opb^i_t})(X^i_t)]\bigr\}dt\\
&\qquad\quad+Q_t^{i,0}dW_t^0+Q_t^{i,i}dW_t^i, \\
&dp^j_t=c_f^j(t)r^j_t dt+q_t^{j,0}dW_t^0+q_t^{j,j}dB_t^j, \\
&dr^j_t=-\bigl\{ \ol{\Lambda}_t(\ex^0[p^j_t]-p^j_t)-\rho_2(t)\ex^0[p^j_t]-b_t\ex^0[r^j_t]\\
&\qquad\quad -\deln \rho_1(t)\ex^0[P^i_t]-\deln \ex^0[\part_\vp f_1^i(t,X^i_t,\call^0(X_t^i),\pi_t(y^j_t,\opb^i_t),\opb^i_t)]\bigr\}dt,
\end{split}
\label{fbsde-mfg}
\ee
with boundary conditions: $X^i_0=\xi^i, x^j_0=\eta^j, r^j_0=0$ and 
\be
\small
\begin{split}
y_T^j=\part_x g_2^j(x^j_T), \quad P^i_T=\part_x g_1^i(X^i_T, \call^0(X^i_T))+\wt{\ex}^0[\part_\mu \wt{g}_1^i(\wt{X}^i_T,\call^0(X^i_T))(X^i_T)],
\quad p_T^j=-c_g^j r^j_T. \nn
\end{split}
\ee
Here, $\forall t\in[0,T]$, $\opb^i_t:=\opb^i(t,u_t)$ is the minimizer of $\calh(t,u_t,\cdot)$
with $u_t:=(X^i_t,x^j_t,y^j_t,P^i_t,p^j_t,r^j_t)$.
The next theorem gives a mean-field counterpart of Theorem~\ref{th-sufficiency-finite}.

\begin{remark}
Using $(\ref{H-Frechet-D})$, one can check that the drift terms of the adjoint processes $(P^i,p^j,r^j)$
are given by $-D_{X^i} \calh(t,u_t,\opb^i_t)$, $-\deln D_{x^j} \calh(t,u_t,\opb^i_t)$
and $-\deln D_{y^j} \calh(t,u_t,\opb^i_t)$, respectively.
\end{remark}

\begin{theorem}
\label{mfg-sufficiency}
Let Assumptions~\ref{A-non-C}-\ref{A-C} be in force. Suppose that there exists a unique square integrable solution to the 
system of equations $(\ref{fbsde-mfg})$ with $(X^i,x^j,y^j,P^i,p^j,r^j)\in \mbb{S}^2(\mbb{F}^{1,i};\mbb{R}^n)\times \mbb{S}^2(\mbb{F}^{2,j};\mbb{R}^n)^2
\times \mbb{S}^2(\mbb{F}^{1,i};\mbb{R}^n)\times \mbb{S}^2(\mbb{F}^{2,j};\mbb{R}^n)^2$ with $\opb^i_t:=\opb^i(t,u_t)$
the minimizer of $\calh(t, X^i_t,x^j_t,y^j_t,P^i_t,p^j_t,r^j_t,\cdot)$ for $dt$-a.e.~$t\in[0,T]$.
Then, $\opb_t^i, t\in[0,T]$ is the unique optimal solution to $(\ref{mfg-c-problem})$.
\begin{proof}
Let $u:=(X^i,x^j,y^j,P^i,p^j,r^j)$ be 
the solution to $(\ref{fbsde-mfg})$ with $\opb^i_t=\opb^i(t,u_t)$ the minimizer of $\calh(t,u_t,\cdot)$,
$dt$-a.e.  We denote by $(\ac{X}^i,\ac{x}^j,\ac{y}^j)$ the sate processes corresponding to another control $\ac{\beta}^i\in \mbb{A}_1^i$
and set $\ac{u}:=(\ac{X}^i,\ac{x}^j,\ac{y}^j,P^i,p^j,r^j)$. 
We put $\mu_t:=\call^0(X_t^i), \ac{\mu}_t:=\call^0(\ac{X}_t^i)$, $\pi_t:=\pi_t(y_t^j,\opb_t^i), \ac{\pi}_t:=\pi_t(\ac{y}_t^j,\ac{\beta}_t^i)$,
$\Del X_t^i=\ac{X}_t^i-X_t^i$, $\Del x_t^j:=\ac{x}_t^j-x_t^j$, $\Del \beta_t^i:=\ac{\beta}_t^i-\opb_t^i$
and $\Del\pi_t:=\ac{\pi}_t-\pi_t$, etc.
From Assumption~\ref{A-C} (v), Fubini's theorem and the equality
$\langle p^j_T,\Del x^j_T\rangle+\langle r^j_T,\Del y^j_T\rangle=0$, we obtain
\be
\small
\begin{split}
\ex\bigl[g_1^i(\ac{X}^i_T,\ac{\mu}_T)-g_1^i(X_T^i,\mu_T)\bigr]\geq \ex\bigl[\langle P^i_T,\Del X^i_T\rangle+
\oldeln(\langle p^j_T,\Del x^j_T\rangle+\langle r^j_T,\Del y^j_T\rangle)\bigr]. \nn
\end{split}
\ee
Thus,  an application of It\^o-formula and the optimality condition $\ex\bigl[\langle D_{\beta^i} \calh(t,u_t,\opb^i_t),\Del \beta^i_t\rangle\bigr]\geq 0$
$dt$-a.e. yield
\be
\small
\begin{split}
\calj_1^i(\ac{\beta}^i)-\calj_1^i(\opb^i)&\geq \int_0^T \Bigl[\calh(t,\ac{u}_t,\ac{\beta}^i_t)-\calh(t,u_t,\opb^i_t)-\ex\bigl[\langle D_{X^i} \calh(t,u_t,\opb^i_t),\Del X^i_t\rangle
+\langle D_{x^j}\calh(t,u_t,\opb^i_t),\Del x^j_t\rangle\\
&\qquad\quad+\langle D_{y^j}\calh(t,u_t,\opb^i_t),\Del y^j_t\rangle+\langle D_{\beta^i} \calh(t,u_t,\opb^i_t),\Del \beta^i_t\rangle\bigr]\Bigr]dt \\
&\geq \frac{1}{2} \int_0^T \ex\bigl[\gamma_1^f|\Del X^i_t|^2+\lambda_\beta |\Del \beta^i_t|^2+\lambda_\vp |\Del \pi_t|^2\bigr]dt, \nn
\end{split}
\ee
which proves the claim.
\end{proof}
\end{theorem}

\subsection{Existence of mean-field equilibrium}
By Theorem~\ref{mfg-sufficiency},  it only remains to show the existence of a solution to 
the FBSDE $(\ref{fbsde-mfg})$ of conditional McKean-Vlasov type. Thanks to the similarity in the form of equations, 
we can proceed in a parallel fashion to Section~\ref{sec-existence-finite}.
For each $t$, let us denote by $u_t:=(X^i_t,x^j_t,y^j_t,P^i_t,p^j_t,r^j_t)$ a generic element in $\mbb{L}^2(\calf^{1,i}_t;\mbb{R}^n)
\times (\mbb{L}^2(\calf^{2,j}_t;\mbb{R}^n))^2\times \mbb{L}^2(\calf^{1,i}_t;\mbb{R}^n)\times 
(\mbb{L}^2(\calf_t^{2,j};\mbb{R}^n))^2$. As we have done before, 
we introduce the following maps:
\be
\small
\begin{split}
B_{X^i}(t,u_t)&:=\opb^i_t+\rho_1(t)\pi_t(y^j_t,\opb^i_t)+l_1^i(t), \\
B_{x^j}(t,u_t)&:=-\ol{\Lambda}_t(y^j_t+\pi_t(y^j_t,\opb^i_t)+\rho_2(t)\pi_t(y^j_t,\opb^i_t)+l_2^j(t), \\
B_{r^j}(t,u_t)&:=-\bigl\{ \ol{\Lambda}_t(\ex^0[p^j_t]-p^j_t)-\rho_2(t)\ex^0[p^j_t]-b_t\ex^0[r^j_t]-\deln \rho_1(t)\ex^0[P^i_t]\\
&\qquad~ -\deln \ex^0[\part_\vp f_1^i(t,X^i_t,\mu_t,\pi_t(y^j_t,\opb^i_t),\opb^i_t)]\bigr\}, \\
F_{P^i}(t,u_t)&:=-\bigl\{\part_x f_1^i(t,X^i_t,\mu_t,\pi_t(y^j_t,\opb^i_t),\opb^i_t)+\wt{\ex}^0[\part_\mu \wt{f}_1^i(t,\wt{X}^i_t,\mu_t,\pi_t(y^j_t,\opb^i_t),
\wt{\opb^i_t})(X^i_t)]\bigr\}, \\
F_{y^j}(t,u_t)&:=-\part_x \ol{f}_2^j(t,x^j_t)+b_t\pi_t(y^j_t,\opb^i_t), \quad
F_{p^j}(t,u_t):=c_f^j(t)r^j_t, 
\end{split}
\label{BFG-mfg}
\ee
and similarly
\be
\small
\begin{split}
G_{P^i}(u_T):=\part_x g_1^i(X^i_T,\mu_T)+\wt{\ex}^0[\part_\mu \wt{g}_1^i(\wt{X}^i_T,\mu_T)(X^i_T)], \quad
G_{y^j}(u_T):=\part_x g_2^j(x^j_T), \quad G_{p^j}(u_T):=-c_g^j r^j_T. \nn
\end{split}
\ee
Here, $\mu_t:=\call^0(X^i_t)$ and $\opb^i_t:=\opb^i(t,u_t)$ is the minimizer of $\calh(t,u_t,\cdot)$.

Let $u_t:=(X^i_t,x^j_t,y^j_t, P^i_t,p^j_t,r^j_t)$ and $\ac{u}_t:=(\ac{X}^i_t,\ac{x}^j_t,\ac{y}^j_t,\ac{P}^i_t,\ac{p}^j_t,\ac{r}^j_t)$ be two arbitrary inputs
in the space  $\mbb{L}^2(\calf^{1,i}_t;\mbb{R}^n)
\times (\mbb{L}^2(\calf^{2,j}_t;\mbb{R}^n))^2\times \mbb{L}^2(\calf^{1,i}_t;\mbb{R}^n)\times 
(\mbb{L}^2(\calf_t^{2,j};\mbb{R}^n))^2$.
We introduce the following abbreviations:
$\ac{\mu}_t:=\call^0(\ac{X}^i_t)$, $\opb_t^{\prime,i}:=\opb^i(t,\ac{u}_t)$, 
$\pi_t:=\pi_t(y^j_t,\opb^i_t), \ac{\pi}_t:=\pi_t(\ac{y}^j_t,\opb_t^{\prime,i})$, 
$f_1^i(t,u_t,\opb^i_t):=f_1^i(t,X^i_t,\mu_t,\pi_t,\opb^i_t)$
and $f_1^i(t,\ac{u}_t,\opb_t^{\prime,i}):=f_1^i(t,\ac{X}^i_t,\ac{\mu}_t,\ac{\pi}_t,\opb_t^{\prime,i})$.
We also put
$\Del u_t:=u_t-\ac{u}_t$, $\Del X^i_t:=X^i_t-\ac{X}^i_t$,
$\Del y^j_t:=y^j_t-\ac{y}^j_t$, $\Del \pi_t:=\pi_t-\ac{\pi}_t$, $\Del \opb^i_t:=\opb^i_t-\opb_t^{\prime,i}$,  
$\Del B_{X^i}(t):=B_{X^i}(t,u_t)-B_{X^i}(t,\ac{u}_t)$, $\Del F_{P^i}(t):=F_{P^i}(t,u_t)-F_{P^i}(t,\ac{u}_t)$,
$\Del G_{P^i}:=G_{P^i}(u_T)-G_{P^i}(\ac{u}_T)$, etc.

\begin{lemma}
\label{pw-terminal-mfg}
Under Assumptions~\ref{A-non-C}-\ref{A-C}, the terminal functions satisfy for any inputs $(u_T,\ac{u}_T)$, 
\be
\small
\begin{split}
\ex\bigl[\langle \Del G_{P^i}, \Del X^i_T\rangle+\oldeln (\langle \Del G_{y^j}, \Del {x}^j_T\rangle+
\langle (-1)\Del G_{p^j}, \Del r^j_T\rangle )\bigr]\geq 
\ex\bigl[\gamma_1^g |\Del X^i_T|^2+\oldeln \gamma_2^g (|\Del x^j_T|^2+|\Del r^j_T|^2)\bigr]. \nn
\end{split}
\ee
\begin{proof}
Using Fubini's theorem, it directly follows from Assumption~\ref{A-non-C} (iv) and Assumption~\ref{A-C} (v).
\end{proof}
\end{lemma}

\begin{lemma}
\label{pw-drift-mfg}
Let Assumptions~\ref{A-non-C}-\ref{A-C} hold. Then, for any $t\in[0,T]$ and inputs $(u_t,\ac{u}_t)$,
there exists a positive constant $\del_0^M$ such that, for any $\deln\leq \del_0^M$, the inequality
\be
\small
\begin{split}
\cald(t)\leq -\gamma^f \ex\bigl[|\Del X^i_t|^2+\oldeln (|\Del x^j_t|^2+|\Del r^j_t|^2)\bigr]+\deln C\ex\bigl[|\Del P^i_t|^2
+\oldeln(|\Del y_t^j|^2+|\Del p_t^j|^2)\bigr] \nn
\end{split}
\ee
holds with $\gamma^f:=\min\bigl(\frac{\gamma_1^f}{2},\frac{\gamma_2^f}{3}\bigr)$ and some positive constant $C$ independent of $\deln$.
Here, 
\be
\small
\begin{split}
\cald(t)&:=\ex\bigl[\langle \Del B_{X^i}(t),\Del P^i_t\rangle+\oldeln (\langle \Del B_{x^j}(t),\Del y^j_t\rangle
+\langle (-1)\Del B_{r^j}(t),\Del p^j_t\rangle)\bigr]\\
&~+\ex\bigl[\langle \Del F_{P^i}(t),\Del X^i_t\rangle+\oldeln (\langle \Del F_{y^j}(t),\Del x^j_t\rangle+
\langle (-1)\Del F_{p^j}(t),\Del r^j_t\rangle)\bigr]. \nn
\end{split}
\ee
\begin{proof}
From $(\ref{H-Frechet-D})$, we now have
\be
\small
\begin{split}
\Del P^i_t&=
D_{\beta^i} \calh(t,u_t,\opb^i_t)-D_{\beta^i} \calh(t,\ac{u}_t,\opb_t^{\prime,i})-(\part_\beta f_1^i(t,u_t,\opb^i_t)-\part_\beta 
f_1^i(t,\ac{u}_t,\opb_t^{\prime,i})
)+(I-\rho_2(t)\Lambda_t)\ex^0[\Del p^j_t]\\
&-b_t \Lambda_t \ex^0[\Del r^j_t]-\deln \rho_1(t)\Lambda_t \ex^0[\Del P^i_t]-\deln \Lambda_t 
\ex^0[\part_\vp f_1^i(t,u_t,\opb^i_t)-\part_\vp f_1^i(t,\ac{u}_t,\opb_t^{\prime,i})]. \nn
\end{split}
\ee 
Using it, with simple replacement of summations to expectations,  we can proceed in almost the same way as  in Lemma~\ref{pw-drift}, hence we 
omit the details.
\end{proof}
\end{lemma}

\begin{theorem}
\label{th-existence-mfg}
Let Assumptions~\ref{A-non-C}-\ref{A-C} be in force. Then, there exists some positive constant $\del_*^M\leq \del_0^M$ such that,
for any $\deln\leq \del_*^M$, there exists a unique square integrable solution to the FBSDE $(\ref{fbsde-mfg})$ of McKean-Vlasov type
with $u:=(X^i,x^j,y^j,P^i,p^j,r^j)\in \mbb{S}^2(\mbb{F}^{1,i};\mbb{R}^n)\times \mbb{S}^2(\mbb{F}^{2,j};\mbb{R}^n)\times \mbb{S}^2(\mbb{F}^{2,j};\mbb{R}^n)
\times \mbb{S}^2(\mbb{F}^{1,i};\mbb{R}^n)\times \mbb{S}^2(\mbb{F}^{2,j};\mbb{R}^n)\times \mbb{S}^2(\mbb{F}^{2,j};\mbb{R}^n)$.
With the solution $u$,  the process $(\opb^i_t:=\opb^i(t,u_t))_{t\in[0,T]}$ gives the unique optimal solution to $(\ref{mfg-c-problem})$.
\begin{proof}
We put $\gamma:=\min\bigl(\frac{\gamma_1^f}{2},\frac{\gamma_2^f}{3},\gamma_1^g, \gamma_2^g\bigr)$
and choose $I_{X^i}, I_{P^i}\in \mbb{H}^2(\mbb{F}^{1,i};\mbb{R}^n)$, $I_{x^j}, I_{r^j}, I_{y^j}, I_{p^j} \in \mbb{H}^2(\mbb{F}^{2,j};\mbb{R}^n)$, 
and $\theta_{P^i}\in \mbb{L}^2(\calf_T^{1,i};\mbb{R}^n)$, $\theta_{y^j}, \theta_{p^j} \in \mbb{L}^2(\calf_T^{2,j};\mbb{R}^n)$
arbitrarily. For $\vr\in[0,1)$ and $\zeta\in(0,1)$, we consider the map
from $\ol{u}:=(\ol{X}^i,\ol{x}^j,\ol{y}^j,\ol{P}^i,\ol{p}^j,\ol{r}^j)\in \mbb{S}^2(\mbb{F}^{1,i};\mbb{R}^n)\times (\mbb{S}^2(\mbb{F}^{2,j};\mbb{R}^n))^2
\times  \mbb{S}^2(\mbb{F}^{1,i};\mbb{R}^n)\times (\mbb{S}^2(\mbb{F}^{2,j};\mbb{R}^n))^2$ to 
$u:=(X^i,x^j,y^j,P^i,p^j,r^j)$ in the same space defined by
\be
\small
\begin{split}
&dX^i_t=\bigl[\vr B_{X^i}(t,u_t)+\zeta B_{X^i}(t,\ol{u}_t)+I_{X^i}(t)\bigr]dt+\sigma_1^{i,0}(t)dW_t^0+\sigma_1^i(t)dW_t^i, \\
&dx^j_t=\bigl[\vr B_{x^j}(t,u_t)+\zeta B_{x^j}(t,\ol{u}_t)+I_{x^j}(t)\bigr]dt+\sigma_2^{j,0}(t)dW_t^0+\sigma_2^j(t)dB_t^j, \\
&dr^j_t=\bigl[\vr B_{r^j}(t,u_t)+\zeta B_{r^j}(t,\ol{u}_t)+I_{r^j}(t)\bigr]dt, \\
&dP_t^i=-\bigl[(1-\vr)\gamma X_t^i-\vr F_{P^i}(t,u_t)+\zeta (-\gamma \ol{X}^i_t-F_{P^i}(t,\ol{u}_t))+I_{P^i}(t)\bigr]dt+Q_t^{i,0}dW_t^0+Q_t^{i,i}dW_t^i, \\
&dy^j_t=-\bigl[(1-\vr)\gamma x^j_t-\vr F_{y^j}(t,u_t)+\zeta(-\gamma \ol{x}^j_t-F_{y^j}(t,\ol{u}_t))+I_{y^j}(t)\bigr]dt+z_t^{j,0}dW_t^0+z_t^{j,j}dB_t^j, \\
&dp^j_t=-\bigl[-(1-\vr)\gamma r^j_t-\vr F_{p^j}(t,u_t)+\zeta (\gamma \ol{r}^j_t-F_{p^j}(t,\ol{u}_t))+I_{p^j}(t)\bigr]dt+q_t^{j,0}dW_t^0+q_t^{j,j}dB_t^j, 
\end{split}
\label{fbsde-shifted-mfg}
\ee
with the boundary conditions $X_0=\xi^i, x_0=\eta^j, r_0=0$ and
$P^i_T=\vr G_{P^i}(u_T)+(1-\vr)X^i_T+\zeta (G_{P^i}(\ol{u}_T)-\ol{X}^i_T)+\theta_{P^i}$, 
$y^j_T=\vr G_{y^j}(u_T)+(1-\vr)x^j_T+\zeta (G_{y^j}(\ol{u}_T)-\ol{x}^j_T)+\theta_{y^j}$
and $p^j_T=\vr G_{p^j}(u_T)-(1-\vr)r^j_T+\zeta (G_{p^j}(\ol{u}_T)+\ol{r}^j_T)+\theta_{p^j}$.
Using  $(\ref{fbsde-shifted-mfg})$ (instead of $(\ref{fbsde-shifted})$), 
 Lemmas~\ref{pw-terminal-mfg}-\ref{pw-drift-mfg}, Fubini's theorem, and the estimate on $|\opb^i(t,u_t)-\opb^i(t,\ac{u}_t)|^2$
from Lemma~\ref{mfg-H-minimization}, 
we can prove the claim in an almost (i.e. replacing the summations by appropriate expectations) the same way
as in the case for Theorem~\ref{th-existence-finite}. 
\end{proof}
\end{theorem}

\section{Convergence to the mean-field limit}
\label{sec-convergence}

Finally, in this section, we are going to prove that the finite-agent equilibrium given in Section~\ref{sec-finite}
converges to the one in Section~\ref{sec-mfg} when we take a large population limit $N_1, N_2\rightarrow \infty$
while keeping the ratio $\deln=N_1/N_2$ unchanged. 
To handle the large population limit,  we make the probability space $(\ol{\Omega},\ol{\calf},\ol{\mbb{P}})$ big enough 
$\ol{\Omega}:=\prod_{i=1}^{\infty}\ol{\Omega}^{1,i}\times \prod_{j=1}^\infty \ol{\Omega}^{2,j}$ to support the relevant random variables.
The other conventions are the same as in Section~\ref{sec-notation} and Subsection~\ref{sec-mfg-notation}.
Throughout this section, we work under Assumptions~\ref{A-non-C}-\ref{A-C} with 
\be
\small
\deln\leq \del_*:=\min(\del_*^F,\del_*^M). \nn
\ee
so that Theorems~\ref{th-existence-finite}  and \ref{th-existence-mfg} hold.

Obviously, the analysis for the mean-field model done in Section~\ref{sec-mfg} is independent from the choice of the representative agents $(i,j)$.
Thanks to the existence of unique strong solution $(X^i,P^i, x^j,y^j,p^j,r^j)\in \mbb{S}^2(\mbb{F}^{1,i};\mbb{R}^n)^2
\times \mbb{S}^2(\mbb{F}^{2,j};\mbb{R}^n)^4$ to $(\ref{fbsde-mfg})$, Yamada-Watanabe Theorem for FBSDEs (see, \cite[Theorem~1.33]{Carmona-Delarue-2}) indicates that
there exists some measurable functions $\Phi_1, \Phi_2$ such that
\be
\small
\begin{split}
&(X^i, P^i)=\Phi_1(W^0, \xi^i, W^i), \quad (x^j,y^j,p^j,r^j)=\Phi_2(W^0, \eta^j,B^j) \nn
\end{split}
\ee
hold, where $\Phi_1,\Phi_2$ are independent from the choice of $(i,j)$ by the symmetry of the model (see, Remarks~\ref{remark-non-c} 
and \ref{remark-c}).
This result implies that $( X^i, P^i, \xi^i, c^i, W^i)_{i=1}^{N_1}$ as well as $( x^j, y^j,p^j,r^j, \eta^j, \mg{c}^j,B^j)_{j=1}^{N_2}$
are $\calf^0$-conditionally i.i.d.~within each group,  and also that the two groups are mutually $\calf^0$-conditionally independent.
Moreover, Lemma~\ref{mfg-H-minimization} tells that $\opb^i$ is independent from $x^j$ and depends on $(y^j,p^j,r^j)$ only through their $\calf^0$-conditional expectations $\ex^0[\cdot]$.
Hence, $(X^i, P^i, \opb^i, \xi^i,c^i, W^i)_{i=1}^{N_1}$ are also $\calf^0$-conditionally i.i.d.
As a result, an $\mbb{F}^0$-adapted process
$\pi_t(y^j_t,\opb^i_t)=-\ex^0[y^j_t]+\deln\Lambda_t \ex^0[\opb_t^i], t\in[0,T]$ is independent from the choice of $(i,j)$,
which is as expected if $\pi$ is the market price process in the large population limit.
This also implies that $(X^i,P^i)$ are independent from the choice of  $j$ in $(x^j,y^j,p^j,r^j)$
and vice versa since the two groups interact only through the price process $\pi$.
Recall that, under the finite-agent setup in Section~\ref{sec-finite}, 
we only have the ($\calf^0$-conditional) exchangeability.

In the remainder of this section, for each $(i,j)$ with $1\leq i\leq N_1, 1\leq j\leq N_2$, let us denote by $u:=(X^i, x^j,y^j, P^i,p^j,r^j)\in \mbb{S}^2(\mbb{F}^{1,i};\mbb{R}^n)
\times (\mbb{S}^2(\mbb{F}^{2,j};\mbb{R}^n))^2\times \mbb{S}^2(\mbb{F}^{1,i};\mbb{R}^n)
\times (\mbb{S}^2(\mbb{F}^{2,j};\mbb{R}^n))^2$ the solution to $(\ref{fbsde-mfg})$, and $\opb^i:=(\opb^i(t,u_t))_{t\in[0,T]}\in \mbb{A}_1^i$
the minimizer of the lifted Hamiltonian $\calh(t,u_t,\cdot)$ in $(\ref{lifted-Hamiltonian})$.
On the other hand, we use the symbols with the superscript $\mgn:=(N_1, N_2)$ when we describe the variables in the finite-agent market, 
such as $u^\mgn:=(X^{\mgn}, x^{\mgn},y^{\mgn}, P^{\mgn},
p^{\mgn},r^{\mgn})\in \mbb{S}^2(\mbb{F};\mbb{R}^n)^{2N_1+4N_2}$ and $\opb^\mgn\in \mbb{A}_1^{N_1}$ 
with $X^{\mgn}:=(X^{\mgn,i})_{i=1}^{N_1}, x^{\mgn}:=(x^{\mgn,j})_{j=1}^{N_2}, y^{\mgn}:=(y^{\mgn,j})_{j=1}^{N_2}$,
$P^{\mgn}:=(P^{\mgn,i})_{i=1}^{N_1}, p^{\mgn}:=(p^{\mgn,j})_{j=1}^{N_2}, r^\mgn:=(r^{\mgn,j})_{j=1}^{N_2}$,
and $\opb^\mgn:=(\opb^{\mgn,i})_{i=1}^{N_1}$. In particular, we denote by $u^\mgn$ the solution to the system of FBSDEs $(\ref{fbsde-full-finite})$
and $\opb_t^{\mgn}$ the minimizer of  Hamiltonian $H(t,u^\mgn_t,\cdot)$ in $(\ref{H-1-finite})$.
Note that,  as long as $\deln=N_1/N_2\leq \del_*$ is kept constant, there exists some positive constant $C$ independent of 
the population sizes $(N_1, N_2)$ such that the inequality
\be
\small
\begin{split}
\sup_{t\in[0,T]}\ex\bigl[&|X_t^{\mgn,i}|^2+|x_t^{\mgn, j}|^2+|y_t^{\mgn,j}|^2+
|P_t^{\mgn,i}|^2+|p_t^{\mgn,j}|^2+|r_t^{\mgn, j}|^2+|\opb_t^{\mgn,i}|^2 \\
&+|X_t^i|^2+|x_t^j|^2+|y_t^j|^2+|P_t^i|^2+|p_t^j|^2+|r_t^j|^2+|\opb_t^i|^2\bigr]\leq C 
\end{split}
\label{uniform-moment}
\ee
holds for every $i$ and $j$. Here,  the estimate for the first line is obtained from  the standard result on the $\mbb{L}^2$-moment 
of the FBSDE solutions normalized by each population size using the exchangeability,
and Lemma~\ref{lemma-H-1-finite}. The one for the second line is obvious from the $\calf^0$-conditional i.i.d. property and Lemma~\ref{mfg-H-minimization}.

What we are going to do is the comparison of the two cost functionals:
\be
\small
\begin{split}
J_1^i (\opb^\mgn)=\ex\Bigl[\int_0^T f_1^i(t,X_t^{\mgn,i},\nu_t^{N_1},\vp_t(y^\mgn_t,\opb^\mgn_t),\opb^{\mgn,i}_t)dt+
g_1^i(X_T^{\mgn,i},\nu_T^{N_1})\Bigr] \nn
\end{split}
\ee
and
\be
\small
\begin{split}
\calj_1^i(\opb^i)=\ex\Bigl[\int_0^T f_1^i(t,X_t^i, \mu_t,\pi_t(y_t^j,\opb_t^j),\opb_t^i)dt+g_1^i(X_T^i,\mu_T)\Bigr], \nn
\end{split}
\ee
where  $\nu_t^{N_1}:=\frac{1}{N_1}\sum_{i=1}^{N_1}\del_{X_t^{\mgn,i}}$ and $\mu_t:=\call^0(X_t^i)=\call^0(X_t^1)$.
Note that both $J^i_1(\opb^\mgn)$ and $\calj_1^i(\opb^i)$ are independent from the choice of $i$ due to the 
exchangeability.
Let us first provide two  lemmas concerning the $\calf^0$-conditional i.i.d.~property of the mean-field solutions.

\begin{lemma}
\label{IID-1}
Let Assumptions~\ref{A-non-C}-\ref{A-C} be in force and the ratio $\deln\leq \del_*$ be kept fixed.
Set $\mu_t^{N_1}:=\frac{1}{N_1}\sum_{i=1}^{N_1}\del_{\psi_t^i}, \mu_t:=\call^0(\psi_t^1)$,
$\nu_t^{N_2}:=\frac{1}{N_2}\sum_{j=1}^{N_2}\del_{\varphi_t^j}$ and $\nu_t:=\call^0(\varphi_t^1)$, where $\psi_t^i$ stands for either  $X_t^i$ or $P_t^i$,
and $\varphi_t^j$ for either $x_t^j, y_t^j, p_t^j$, or $r_t^j$.
Then there exist some sequence $(\ep_{N})_{N\geq 1}$ converging to zero as $N\rightarrow \infty$ 
such that
\be
\small
\begin{split}
\sup_{t\in[0,T]}\ex\bigl[W_2(\mu_t^{N_1},\mu_t)^2\bigr]\leq \ep_{N_1}, \quad \sup_{t\in[0,T]}\ex\bigl[W_2(\nu_t^{N_2},\nu_t)^2\bigr]\leq \ep_{N_2}. \nn
\end{split}
\ee
Moreover,  there exists some positive constant $C$ independent of the population sizes that satisfies
\be
\small
\begin{split}
\sup_{t\in[0,T]}\ex\bigl[|\mg{m}_1(\psi_t)-\ex^0[\psi_t^1]|^2\bigr]\leq C N_1^{-1}, \quad 
\sup_{t\in[0,T]}\ex\bigl[|\mg{m}_2(\varphi_t)-\ex^0[\varphi_t^1]|^2\bigr]\leq C N_2^{-1}.  \nn
\end{split}
\ee
In the latter claim,  $\psi_t^i$ can also stand for $\opb_t^i$ and $\part_\vp f_1^i(t,X_t^i,\call^0(X_t^1),\pi_t(y_t^1,\opb_t^1),\opb_t^i)$ 
in addition to $(X^i, P^i)$.
\begin{proof}
The first claim follows from Lemma~\cite[Lemma~4.1]{Fujii-mfg-convergence}.
Let us prove the second claim.  Since $(\psi^i)_{i\geq 1}$ are  $\calf^0$-conditionally i.i.d., the ladder property of the conditional 
expectation yields
\be
\small
\begin{split}
\ex\bigl[|\mg{m}_1(\psi_t)-\ex^0[\psi_t^1]|^2\bigr]&=\ex\bigl[\bigl|\frac{1}{N_1}\sum_{i=1}^{N_1}(\psi_t^i-\ex^0[\psi_t^1])\bigr|^2\bigr]
=\frac{1}{N_1^2}\sum_{i=1}^{N_1}\ex\bigl[|\psi_t^i-\ex^0[\psi_t^1]|^2\bigr]. \nn
\end{split}
\ee
Since $\sup_{t\in[0,T]}\ex[|\psi_t^1-\ex^0[\psi_t^1]|^2]\leq 2\sup_{t\in[0,T]}\ex[|\psi_t^1|^2]\leq C$, the conclusion follows.
Note also that $\opb_t^i$, $\part_\vp f_1^i(t,X_t^i,\call^0(X_t^1),\pi_t(y_t^1,\opb_t^1),\opb_t^i), i\geq 1$
are also $\calf^0$-conditionally i.i.d. Hence,  from the growth property in Lemma~\ref{mfg-H-minimization}, 
the same estimate holds. The proof for $\varphi$ is the same.
\end{proof}
\end{lemma} 

\begin{remark}
Under some additional integrability conditions,  the dependence of  $\ep_N$ on the population size $N$ 
is known to be available in a more explicit form. For details, see \cite[Theorem~5.8]{Carmona-Delarue-1}.
\end{remark}

\begin{lemma}
\label{IID-2}
Let Assumptions~\ref{A-non-C}-\ref{A-C} be in force and the ratio $\deln\leq \del_*$ be kept fixed.
Let us set $\mu_t:=\call^0(X_t^1), \opb_t^i:=\opb^i(t,u_t)$ and $\pi_t:=\pi_t(y_t^1,\opb_t^1),  t\in[0,T]$. 
Then, there exists some positive constant $C$ independent of the population sizes that satisfies
\be
\small
\begin{split}
\sup_{t\in[0,T]}&\ex\bigl[\bigl|\frac{1}{N_1}\sum_{k=1}^{N_1}\part_\mu f_1^k(t,X_t^k,\mu_t,\pi_t,\opb_t^k)(X_t^i)
-\wt{\ex}^0[\part_\mu \wt{f}_1^i(t,\wt{X}_t^i,\mu_t,\pi_t,\wt{\opb^i_t})(X_t^i)]\bigr|^2\bigr]\leq CN_1^{-1}, \quad \text{and}\\
&\ex\bigl[\bigl|\frac{1}{N_1}\sum_{k=1}^{N_1}\part_\mu g_1^k(X_T^k, \mu_T)(X_T^i)-
\wt{\ex}^0[\part_\mu \wt{g}_1^i(\wt{X}_T^i,\mu_T)(X_T^i)\bigr|^2\bigr]\leq CN_1^{-1}. \nn
\end{split}
\ee
\begin{proof}
Let us  first proves the second one.
Recall that  the notations $\wt{g}_1^i(\cdot):=g_1(\cdot,c_T^0,\wt{c}_T^i)$
and also that $\wt{X}^i$ and $\wt{c}^i$ are the copies of $X^i$ and $c^i$ defined on the space 
$\Omega^0\times \wt{\ol{\Omega}}$ (see, Section~\ref{sec-mfg-notation}.). Since $(\wt{X}^i, \wt{c}^i)_{i\geq 1}, (X^i ,c^i)_{i\geq 1}$ are $\calf^0$-conditionally i.i.d., we have
\be
\small
\begin{split}
\wt{\ex}^0[\part_\mu \wt{g}_1^i(\wt{X}_T^i,\mu_T)(X_T^i)]=\ex^0[\part_\mu g_1^j(X_T^j,\mu_T)(v)]\bigl|_{v=X_T^i}
=\ex^0[\part_\mu g_1^j(X_T^j,\mu_T)(X_T^i)| \sigma(X_T^i)] \nn
\end{split}
\ee
for any $j \neq i$. Therefore,
\be
\small
\begin{split}
&\ex\bigl[\bigl|N_1^{-1}\sum_{k=1}^{N_1}\part_\mu g_1^k(X_T^k,\mu_T)(X_T^i)
-\wt{\ex}^0[\part_\mu \wt{g}_1^i(\wt{X}_T^i,\mu_T)(X_T^i)]\bigr|^2\bigr] \\
&=\ex\bigl[\bigl|N_1^{-1}\sum_{k=1}^{N_1}\Bigl(\part_\mu g_1^k(X_T^k,\mu_T)(X_T^i)
-\ex^0[\part_\mu g_1^j(X_T^j,\mu_T)(X_T^i)|\sigma(X_T^i)]\Bigr)\bigr|^2\bigr]\\
&\leq 2\ex\bigl[\bigl|N_1^{-1}\sum_{k\neq i}^{N_1}\Bigl(\part_\mu g_1^k(X_T^k,\mu_T)(X_T^i)
-\ex^0[\part_\mu g_1^j(X_T^j,\mu_T)(X_T^i)|\sigma(X_T^i)]\Bigr)\bigr|^2\bigr] \\
&\quad +2 N_1^{-2}\ex\bigl[ \bigl|\part_\mu g_1^i(X_T^i,\mu_T)(X_T^i)
-\ex^0[\part_\mu g_1^j(X_T^j,\mu_T)(X_T^i)|\sigma(X_T^i)]\bigr|^2\bigr]. \nn
\end{split}
\ee
Obviously,  $(X^k,c^k)_{k\neq i}$ are $\calf^0\vee \sigma(X^i)$-conditionally i.i.d., 
which makes the cross terms in the first line vanish.  Hence the right hand side of the above inequality is bounded by 
\be
\small
\begin{split}
2N_1^{-2}\sum_{k=1}^{N_1}\ex\bigl[\bigl| \part_\mu g_1^k(X_T^k, \mu_T)(X_T^i)-\ex^0[\part_\mu g_1^j(X_T^j,\mu_T)(X_T^i)|\sigma(X_T^i)]
\bigr|^2\bigr]\leq CN_1^{-1}, \nn
\end{split}
\ee
where we have used the linear growth property $\part_\mu g_1^i$. 
The first inequality can be proved in the same way from the $\calf^0$-conditional i.i.d. property of 
$(X^i,\opb^i,c^i)_{i=1}^{N_1}$ and the growth property of $\opb^i(t,\cdot)$ in Lemma~\ref{mfg-H-minimization}.
\end{proof}
\end{lemma}

Let us define the variable $\vep_{N_1}$  by
\be
\small
\begin{split}
\vep_{N_1}:=\max\Bigl(N_1^{-\frac{1}{2}}, \sup_{t\in[0,T]}\ex\bigl[W_2(\mu_t^{N_1},\mu_t)^2\bigr]^\frac{1}{2}\Bigr) 
\end{split}
\label{vepN1-def}
\ee
with $\mu_t:=\call^0(X_t^1)$ and $\mu_t^{N_1}:=\frac{1}{N_1}\sum_{i=1}^{N_1}\del_{X_t^i}$.
The following result is an extension of \cite[Theorem~6.1]{Carmona-Delarue-MFTC} to our $\calf^0$-conditional extended mean-field type 
control problem.

\begin{proposition}
\label{prop-convergence-1}
Let Assumptions~\ref{A-non-C}-\ref{A-C} be in force and the ratio $\deln\leq \del_*$ be kept fixed.
Then there exists some positive constant $C$ independent of the population sizes that satisfies, for any $1\leq i\leq N_1$, 
\be
\small
\begin{split}
J_1^i(\opb^\mgn)-\calj_1^i(\opb^i)&\geq\ex \int_0^T\Bigl[\frac{\gamma_1^f}{2}|X_t^{\mgn,i}-X_t^i|^2+\frac{\lambda_\beta}{2}|\opb_t^{\mgn,i}-\opb_t^i|^2
+\frac{\lambda_\vp}{2}|\vp_t(y^\mgn_t,\opb_t^\mgn)-\pi_t(y_t^1,\opb_t^1)|^2\Bigr]dt \\
&\quad+\frac{\gamma_1^g}{2}\ex\bigl[|X_T^{\mgn,i}-X_T^i|^2\bigr]-C\vep_{N_1}. \nn
\end{split}
\ee
\begin{proof}
We have $J_1^i(\opb^\mgn)-\calj_1^i(\opb^i)=V_1^i+V_2^i$ with
\be
\small
\begin{split}
&V_1^i:=\ex\Bigl[\langle X_T^{\mgn,i}-X_T^i, P_T^i\rangle+\int_0^T[f_1^i(t,X_t^{\mgn,i},\nu_t^{N_1},\vp_t(y^\mgn_t,\opb_t^\mgn),\opb_t^{\mgn,i})
-f_1^i(t,X_t^i,\mu_t,\pi_t(y_t^j,\opb_t^i),\opb_t^i)]dt\Bigr], \\
&V_2^i:=\ex\Bigl[g_1^i(X_T^{\mgn,i},\nu_T^{N_1})-g_1^i(X_T^i,\mu_T)
-\langle X_T^{\mgn,i}-X_T^i, \part_x g_1^i(X_T^i, \mu_T)+\wt{\ex}^0[\part_\mu \wt{g}_1^i(\wt{X}_T^i,\mu_T)(X_T^i)]\rangle\Bigr].  \nn
\end{split}
\ee
By exchangeability, the values of $V_1^i$ and $V_2^i$ are independent from the choice of $i$.
For notional ease, let us put 
$\Del X_t^i:=X_t^{\mgn,i}-X_t^i$, $\Del \opb_t^i:=\opb_t^{\mgn,i}-\opb_t^i$,
$\vp_t:=\vp_t(y_t^\mgn,\opb_t^\mgn)$, $\pi_t:=\pi_t(y_t^1,\opb_t^1)=\pi_t(y_t^j,\opb_t^i)$ and
$\Del \vp_t:=\vp_t-\pi_t$, etc. 
\\

\noindent
{\bf First step: Estimate on $V_2^i$}\\
Let us separate the terms in $V_2^i$ as follows: $V_2^i=V_{2,1}^i-V_{2,2}^i-V_{2,3}^i$ where
\be
\begin{split}
&V_{2,1}^i:=\ex\bigl[g_1^i(X_T^{\mgn,i},\nu_T^{N_1})-g_1^i(X_T^i,\mu_T)\bigr], \\
&V_{2,2}^i:=\ex\bigl[\langle \Del X_T^i, \part_x g_1^i(X_T^i,\mu_T)\rangle\bigr], \quad V_{2,3}^i:=\ex\bigl[\langle \Del X_T^i, \wt{\ex}^0[\part_\mu \wt{g}_1^i(\wt{X}_T^i,\mu_T)(X_T^i)]\rangle\bigr]. \nn
\end{split}
\ee
By local Lipschitz continuity in Assumption~\ref{A-C} (ii), $(\ref{uniform-moment})$  and Cauchy-Schwarz inequality, we have
\be
\small
\begin{split}
V_{2,1}^i &\geq \ex\bigl[g_1^i(X_T^{\mgn,i},\nu_T^{N_1})-g_1^i(X_T^i, \mu_T^{N_1})\bigr]
-C\ex\bigl[(1+|X_T^i|+M_2(\mu_T)+M_2(\mu_T^{N_1})+|c_T^0|+|c_T^i|)W_2(\mu_T,\mu_T^{N_1})\bigr]\\
&\geq  \ex\bigl[g_1^i(X_T^{\mgn,i},\nu_T^{N_1})-g_1^i(X_T^i, \mu_T^{N_1})\bigr]-C\vep_{N_1}. \nn
\end{split}
\ee
From Assumption~\ref{A-C} (iii), similar calculations give
\be
\small
\begin{split}
V_{2,2}^i\leq \ex\bigl[\langle \Del X_T^i, \part_x g_1^i(X_T^i, \mu_T^{N_1})\rangle\bigr]+C\vep_{N_1}. \nn
\end{split}
\ee
Using  Lemma~\ref{IID-2} and Assumption~\ref{A-C} (iii), we get
\be
\small
\begin{split}
V_{2,3}^i 
&\leq \ex\Bigl[\langle \Del X_T^i, \frac{1}{N_1}\sum_{k=1}^{N_1}\part_\mu g_1^k(X_T^k,\mu^{N_1}_T)(X_T^i)\rangle\Bigr]+C\vep_{N_1}.  \nn
\end{split}
\ee
Note that the value of $V_{2,3}^i$ is independent from $i$. From the same technique used in $(\ref{theta-technique})$, we have
\be
\small
\begin{split}
V_{2,3}^i &=\frac{1}{N_1}\sum_{i=1}^{N_1}V_{2,3}^i 
\leq \frac{1}{N_1^2}\sum_{i,k=1}^{N_1}\ex\bigl[\blangle 
\Del X_T^i, \part_\mu g_1^k(X_T^k,\mu^{N_1}_T)(X_T^i)\brangle\bigr]+C\vep_{N_1} \\
&=\frac{1}{N_1}\sum_{k=1}^{N_1}\ex \ex^\theta \bigl[\langle \Del X_T^\theta, \part_\mu g_1^k(X_T^k, \mu_T^{N_1})(X_T^\theta)
\rangle\bigr]+C\vep_{N_1}
=\ex \ex^\theta \bigl[\langle \Del X_T^\theta, \part_\mu g_1^i(X_T^i, \mu_T^{N_1})(X_T^\theta)
\rangle\bigr]+C\vep_{N_1}. \nn
\end{split}
\ee 
Combined the above estimates, we obtain,  from Assumption~\ref{A-C} (v), 
\be
\small
\begin{split}
V_2^i &\geq \ex\bigl[g_1^i(X_T^{\mgn,i},\nu_T^{N_1})-g_1^i(X_T^i, \mu_T^{N_1})\\
&\qquad-\langle \Del X_T^i, \part_x g_1^i(X_T^i, \mu_T^{N_1})\rangle
-\ex^\theta [\langle \Del X_T^\theta, \part_\mu g_1^i(X_T^i, \mu_T^{N_1})(X_T^\theta)
\rangle]\bigr]-C\vep_{N_1} \\
&\geq \frac{\gamma_1^g}{2}\ex\bigl[|\Del X_T^i|^2\bigr]-C\vep_{N_1}. 
\end{split}
\label{conv-first-step}
\ee

\noindent
{\bf Second step: Estimate on $V_1^i$}\\
Since $\langle \Del x_T^j, p_T^j\rangle+\langle \Del y_T^j, r_T^j\rangle=0$ for any $j$, we have
\be
\small
\begin{split}
V_1^i&=\ex\Bigl[\langle \Del X_T^i, P_T^i\rangle+\oldeln (\langle \Del x_T^j, p_T^j\rangle+\langle \Del y_T^j, r_T^j\rangle)
+\int_0^T [f_1^i(t,X_t^{\mgn,i},\nu_t^{N_1},\vp_t,\opb_t^{\mgn,i})-f_1^i(t,X_t^i,\mu_t,\pi_t, \opb_t^i)]dt\Bigr] \\
&\geq \ex\Bigl[\langle \Del X_T^i, P_T^i\rangle+\oldeln (\langle \Del x_T^j, p_T^j\rangle+\langle \Del y_T^j, r_T^j\rangle)\\
&\quad+ \int_0^T [f_1^i(t,X_t^{\mgn,i},\nu_t^{N_1},\vp_t,\opb_t^{\mgn,i})-f_1^i(t,X_t^i,\mu_t,\pi_t, \opb_t^i)
-\langle \Del \opb_t^i, D_{\beta^i} \calh(t,u_t,\opb_t^i)\rangle]dt\Bigr]. \nn
\end{split}
\ee
Here, in the second line, we have set $u_t:=(X_t^i,x_t^j,y_t^j,P_t^i,p_t^j,r_t^j)$ and used the optimality condition on $\opb_t^i=\opb^i(t,u_t)$
for $\calh(t,u_t,\cdot)$. Direct calculation  of the right hand side using the expression of $D_{\beta^i}\calh(\cdot)$ in $(\ref{H-Frechet-D})$ shows that 
$V_1^i\geq V_{1,1}^i+V_{1,2}^i$ with
\be
\small
\begin{split}
V_{1,1}^i&:=\ex\int_0^T\Bigl[f_1^i(t,X_t^{\mgn,i},\nu_t^{N_1},\vp_t,\opb_t^{\mgn,i})-f_1^i(t,X_t^i,\mu_t,\pi_t,\opb_t^i) 
-\langle \Del \opb_t^i, \part_\beta f_1^i(t,X_t^i,\mu_t,\pi_t,\opb_t^i)\rangle \\
&\qquad-\langle \Del X_t^i, \part_x f_1^i(t,X_t^i,\mu_t,\pi_t,\opb_t^i)
+\wt{\ex}^0[\part_\mu \wt{f}_1^i(t,\wt{X}_t^i,\mu_t,\pi_t,\wt{\opb_t^i})(X_t^i)]\rangle \\
&\qquad-\langle \ex^0[\part_\vp f_1^i(t,X_t^i,\mu_t,\pi_t,\opb_t^i)],  -\Del y_t^j+\deln \Lambda_t \Del\opb_t^i \rangle
\Bigr]dt, \\
V_{1,2}^i&:=\ex\int_0^T \Bigl[\langle \rho_1(t) P_t^i, \Del \vp_t\rangle+
\langle \rho_1(t)\ex^0[P_t^i], \Del y_t^j-\deln \Lambda_t \Del\opb_t^i\rangle\\
&\qquad+\oldeln\bigl(\langle (-\ol{\Lambda}_t+\rho_2(t))p_t^j, \Del \vp_t\rangle+
\langle (-\ol{\Lambda}_t+\rho_2(t))\ex^0[p_t^j], \Del y_t^j-\deln \Lambda_t \Del\opb_t^i\rangle \bigr)\\
&\qquad+\oldeln\bigl( \langle  b_t r_t^j, \Del \vp_t \rangle+\langle b_t\ex^0[r_t^j], \Del y_t^j-\deln \Lambda_t \Del \opb_t^i\rangle\bigr)\Bigr]dt. 
\nn
\end{split}
\ee

\noindent
\bull Analysis for $V_{1,2}^i$ \\
For the first line, from the ($\calf^0$-conditional) exchangeability of $(P^i)$, $(\Del y^j)$ and $(\Del \opb^i)$, we have 
\be
\small
\begin{split}
&\ex\bigl[\langle \rho_1(t) P_t^i, \Del \vp_t\rangle+
\langle \rho_1(t)\ex^0[P_t^i], \Del y_t^j-\deln \Lambda_t \Del\opb_t^i\rangle \bigr]\\
&=\ex\bigl[\langle \rho_1(t) (P_t^i-\ex^0[P_t^i]), \Del \vp_t\rangle
+\langle \rho_1(t)\ex^0[P_t^i], \Del \vp_t+\Del y_t^j-\deln \Lambda_t \Del\opb_t^i\rangle \bigr]\\
&=\ex\bigl[\langle \rho_1(t) (\mg{m}_1(P_t)-\ex^0[P_t^1]), \Del \vp_t\rangle
+\langle \rho_1(t)\ex^0[P_t^1], \Del \vp_t+\mg{m}_2(\Del y_t)-\deln \Lambda_t \mg{m}_1(\Del\opb_t)\rangle \bigr].\nn
\end{split}
\ee
Here, the second term is zero since
\be
\small
\begin{split}
&\ex^0[\Del \vp_t+\mg{m}_2(\Del y_t)-\deln \Lambda_t \mg{m}_1(\Del\opb_t)] \\
&=\ex^0[-\mg{m}_2(y_t^\mgn)+\ex^0[y_t^1]+\deln \Lambda_t (\mg{m}_1(\opb_t^\mgn)-\ex^0[\opb_t^1])+\mg{m}_2(\Del y_t)-\deln\Lambda_t 
\mg{m}_1(\Del \opb_t)]=0.
\label{delvp-zero}
\end{split}
\ee
Applying Lemma~\ref{IID-1} and Cauchy-Schwarz inequality to the first term, we obtain, 
\be
\small
\begin{split}
\ex\bigl[\langle \rho_1(t) P_t^i, \Del \vp_t\rangle+
\langle \rho_1(t)\ex^0[P_t^i], \Del y_t^j-\deln \Lambda_t \Del\opb_t^i\rangle \bigr] \geq -C/\sqrt{N_1}\geq -C\vep_{N_1}. \nn
\end{split}
\ee

\noindent
By the same technique and $(\ref{delvp-zero})$, the second and third lines can be estimated as 
\be
\small
\begin{split}
&\ex\bigl[\langle (-\ol{\Lambda}_t+\rho_2(t))p_t^j, \Del \vp_t\rangle+
\langle (-\ol{\Lambda}_t+\rho_2(t))\ex^0[p_t^j], \Del y_t^j-\deln \Lambda_t \Del\opb_t^i\rangle\bigr]\\
&\quad +\ex\bigl[\langle  b_t r_t^j, \Del \vp_t \rangle+\langle b_t\ex^0[r_t^j], \Del y_t^j-\deln \Lambda_t \Del \opb_t^i\rangle\bigr]\\
&=\ex\bigl[\langle (-\ol{\Lambda}_t+\rho_2(t))(\mg{m}_2(p_t)-\ex^0[p_t^j]), \Del \vp_t\rangle
+\langle (-\ol{\Lambda}_t+\rho_2(t))\ex^0[p_t^j], \Del \vp_t+\mg{m}_2(\Del y_t)-\deln \Lambda_t \mg{m}_1(\Del\opb_t)\rangle\bigr]\\
&\quad +\ex\bigl[\langle b_t(\mg{m}_2(r_t)-\ex^0[r_t^j]),\Del \vp_t\rangle+\langle b_t\ex^0[r_t^j], 
\Del\vp_t+\mg{m}_2(\Del y_t)-\deln \Lambda_t \mg{m}_1(\Del \opb_t) \rangle\bigr]\\
&\geq -C/\sqrt{N_2}\geq -C\vep_{N_1}. \nn
\end{split}
\ee
Since every estimate is uniform in $t\in[0,T]$ by $(\ref{uniform-moment})$, we get
\be
\small
V_{1,2}^i\geq -C\vep_{N_1}.
\label{result-V12}
\ee

\noindent
\bull Analysis for $V_{1,1}^i$ \\
From $(\ref{uniform-moment})$ and the (local) Lipschitz continuity, it is easy to see, for the first line,
\be
\small
\begin{split}
&\ex\bigl[f_1^i(t,X_t^{\mgn,i},\nu_t^{N_1},\vp_t,\opb_t^{\mgn,i})-f_1^i(t,X_t^i,\mu_t,\pi_t,\opb_t^i)-\langle \Del \opb_t^i, \part_\beta f_1^i(t,X_t^i,\mu_t,\pi_t,\opb_t^i)\rangle\bigr]\\
&\geq \ex\bigl[f_1^i(t,X_t^{\mgn,i},\nu_t^{N_1},\vp_t,\opb_t^{\mgn,i})-f_1^i(t,X_t^i,\mu_t^{N_1},\pi_t,\opb_t^i)
-\langle \Del \opb_t^i, \part_\beta f_1^i(t,X_t^i,\mu_t^{N_1},\pi_t,\opb_t^i)\rangle
\bigr]-C\vep_{N_1}. \nn
\end{split}
\ee
Using Lemma~\ref{IID-2},  exchangeability and the technique in $(\ref{theta-technique})$,  we have
\be
\small
\begin{split}
&\ex\bigl[\langle \Del X_t^i, \wt{\ex}^0[\part_\mu \wt{f}_1^i(t,\wt{X}_t^i,\mu_t,\pi_t,\wt{\opb_t^i})(X_t^i)]\rangle\bigr]
\leq \frac{1}{N_1}\sum_{i=1}^{N_1}\ex\Bigl[\blangle \Del X_t^i, \frac{1}{N_1}\sum_{k=1}^{N_1}\part_\mu f_1^k(t,X_t^k,\mu_t,\pi_t,\opb_t^k)(X_t^i)\brangle\Bigr]+C /\sqrt{N_1}  \\
&=\ex\ex^\theta\bigl[\langle \Del X_t^\theta, \part_\mu f_1^i(X_t^i,\mu_t,\pi_t,\opb_t^i)(X_t^\theta)\rangle\bigr]
+C/\sqrt{N_1}. \nn
\end{split}
\ee
Thus, from the Lipschitz continuity of $(\part_x f_1, \part_\mu f_1)$, we see that the second line of $V_{1,1}^i$ satisfies
\be
\small
\begin{split}
&\ex\bigl[\langle \Del X_t^i, \part_x f_1^i(t,X_t^i,\mu_t,\pi_t,\opb_t^i)
+\wt{\ex}^0[\part_\mu \wt{f}_1^i(t,\wt{X}_t^i,\mu_t,\pi_t,\wt{\opb_t^i})(X_t^i)]\rangle\bigr]\\
&\leq \ex\bigl[\langle \Del X_t^i, \part_x f_1^i(t,X_t^i,\mu_t^{N_1},\pi_t,\opb_t^i)\rangle+
\ex^\theta[\langle \Del X_t^\theta, \part_\mu f_1^i(X_t^i,\mu_t^{N_1},\pi_t,\opb_t^i)(X_t^\theta)\rangle]\bigr]+C\vep_{N_1}.\nn
\end{split}
\ee
Lastly, the third line of $V_{1,1}^i$ can be estimated as
\be
\small
\begin{split}
&\ex\bigl[\langle \ex^0[\part_\vp f_1^i(t,X_t^i,\mu_t,\pi_t,\opb_t^i)],  -\Del y_t^j+\deln \Lambda_t \Del\opb_t^i \rangle\bigr]
=\ex\bigl[\langle \ex^0[\part_\vp f_1^i(t,X_t^i,\mu_t,\pi_t,\opb_t^i)],\Del \vp_t\rangle\bigr] \\
&\leq  \frac{1}{N_1}\sum_{k=1}^{N_1}\ex\bigl[\langle \part_\vp f_1^k(t,X_t^k,\mu_t,\pi_t,\opb_t^k), \Del \vp_t\rangle\bigr]+C/\sqrt{N_1}\\
&\leq \ex\bigl[\langle \part_\vp f_1^i(t,X_t^i,\mu_t^{N_1},\pi_t,\opb_t^i), \Del \vp_t\rangle\bigr]+C\vep_{N_1}. \nn
\end{split}
\ee
Here, we have used $(\ref{delvp-zero})$ in the first equality, Lemma~\ref{IID-1} in the second line, 
and finally the Lipschitz continuity of $\part_\vp f_1$ in the last line.
Combined with $(\ref{result-V12})$, the joint convexity of $f_1$ in Assumption~\ref{A-C} (iv) gives
\be
\small
\begin{split}
V_1^i &\geq \ex\int_0^T \Bigl[f_1^i(t,X_t^{\mgn,i},\nu_t^{N_1},\vp_t,\opb_t^{\mgn,i})-f_1^i(t,X_t^i,\mu_t^{N_1},\pi_t,\opb_t^i)
-\langle \Del \opb_t^i, \part_\beta f_1^i(t,X_t^i,\mu_t^{N_1},\pi_t,\opb_t^i)\rangle\\
&\qquad -\langle \Del X_t^i, \part_x f_1^i(t,X_t^i,\mu_t^{N_1},\pi_t,\opb_t^i)\rangle
-\ex^\theta[\langle \Del X_t^\theta, \part_\mu f_1^i(t,X_t^i,\mu_t^{N_1},\pi_t,\opb_t^i)(X_t^\theta)\rangle]\\
&\qquad -\langle \Del \vp_t, \part_\vp f_1^i(t,X_t^i,\mu_t^{N_1},\pi_t,\opb_t^i)\rangle\Bigr]dt-C\vep_{N_1}\\
&\geq \ex\int_0^T\Bigl[\frac{\gamma_1^f}{2}|\Del X_t^i|^2+\frac{\lambda_\vp}{2}|\Del \opb_t^i|^2+
\frac{\lambda_\vp}{2}|\Del \vp_t|^2\Bigr]dt-C\vep_{N_1}. 
\end{split}
\label{conv-second-step}
\ee
The conclusion follows from $(\ref{conv-first-step})$ and $(\ref{conv-second-step})$.
\end{proof}
\end{proposition}

We now consider the state process of the finite-agent market 
$\ul{u}^\mgn:=(\ul{X}^{\mgn}, \ul{x}^\mgn, \ul{y}^\mgn)\in \mbb{S}^2(\mbb{F};\mbb{R}^n)^{N_1}\times (\mbb{S}^2(\mbb{F};\mbb{R}^n))^{2N_2}$
with $\ul{X}^\mgn:=(\ul{X}^{\mgn,i})_{i=1}^{N_1}$,
$\ul{x}^{\mgn}:=(\ul{x}^{\mgn,j})_{j=1}^{N_2}$, and $\ul{y}^{\mgn}:=(\ul{y}^{\mgn,j})_{j=1}^{N_2}$ defined as the unique solution to
the system of FBSDEs: $1\leq i\leq N_1, 1\leq j\leq N_2$, 
\be
\small
\begin{split}
&d\ul{X}_t^{\mgn,i}=(\opb_t^i+\rho_1(t)\vp_t(\ul{y}^\mgn,\opb_t)+l_1^i(t))dt+\sigma_1^{i,0}(t)dW_t^0+\sigma_1^i(t)dW_t^i, \\
&d\ul{x}_t^{\mgn,j}=(-\ol{\Lambda}_t(\ul{y}_t^{\mgn,j}+\vp_t(\ul{y}^\mgn_t,\opb_t))+\rho_2(t)\vp_t(\ul{y}^\mgn_t,\opb_t)+l_2^j(t))
dt+\sigma_2^{j,0}(t)dW_t^0+\sigma_2^j(t)dB_t^j,\\
&d\ul{y}_t^{\mgn,j}=-(\part_x \ol{f}_2^j(t,\ul{x}_t^{\mgn,j})-b_t \vp_t(\ul{y}^\mgn_t,\opb_t))dt+\ul{z}_t^{\mgn,j,0}dW_t^0+
\sum_{k=1}^{N_1}\ul{\mg{z}}_t^{\mgn,j,k}dW_t^k+\sum_{k=1}^{N_2}\ul{z}_t^{\mgn,j,k}dB_t^k, 
\label{fbsde-approx}
\end{split}
\ee
with the boundary conditions; $\ul{X}_0^{\mgn,i}=\xi^i, \ul{x}_0^{\mgn,j}=\eta^j$
and $\ul{y}_T^{\mgn,j}=\part_x g_2^j(\ul{x}_T^{\mgn,j})$. As in $(\ref{vp-expression-1})$, we have 
$\vp_t(\ul{y}^\mgn_t,\opb_t)=-\mg{m}_2(\ul{y}^\mgn)+\deln \Lambda_t \mg{m}_1(\opb_t)$.
Recall that $\opb^i$ denotes the optimal solution to the mean-field setup for the representative agents $i$ in the first population
and (arbitrary) $j$ in the second population, which is specified by the solution of $(\ref{fbsde-mfg})$.
In other words, the system of equations $(\ref{fbsde-approx})$ describes the situation 
where each agent $i$ in the first population behaves as if she is a representative agent 
and adopts $\opb^i\in \mbb{A}_1^i=\mbb{H}^2(\mbb{F}^{1,i};A_1)$ as an approximation for the true optimal control $\opb^{\mgn,i}\in \mbb{A}_1:=
\mbb{H}^2(\mbb{F};A_1)$. Moreover,  $(\ul{x}^{\mgn}, \ul{y}^\mgn)$ describes the market clearing
state process of the second population under the given order flow $\opb~(:=(\opb^i)_{i=1}^{N_1})$. The unique existence 
of such a process $(\ul{x}^\mgn,\ul{y}^\mgn)$ is guaranteed 
by Theorem~\ref{th-non-c-existence} as discussed in Subsection~\ref{sec-given-beta}. 
Note that the ($\calf^0$-conditional) exchangeability among $(\ul{X}^{\mgn,i},\opb^i, \xi^i,c^i,W^i)_{i=1}^{N_1}$
and $(\ul{x}^{\mgn,j}, \ul{y}^{\mgn,j}, \eta^j,\mg{c}^j,B^j)_{j=1}^{N_2}$ holds in this case, too. 
The cost functional for the member $i$ in the first population under this setup is given by
\be
\small
\begin{split}
J_1^i(\opb)=\ex\Bigl[\int_0^T f_1^i(t,\ul{X}_t^{\mgn,i},\ul{\nu}_t^{N_1},\vp_t(\ul{y}_t^{\mgn},\opb_t),\opb_t^i)dt+g_1^i(\ul{X}_T^{\mgn,i},
\ul{\nu}_T^{N_1})\Bigr], \nn
\end{split}
\ee
where $\ul{\nu}_t^{N_1}:=\frac{1}{N_1}\sum_{i=1}^{N_1}\del_{\ul{X}_t^{\mgn,i}}$.

The following proof is inspired by the monotonicity technique used in \cite[Theorem~4.2]{Fujii-mfg-convergence}.
\begin{proposition}
\label{prop-convergence-2}
Let Assumptions~\ref{A-non-C}-\ref{A-C} be in force and the ratio $\deln\leq \del_*$ be kept fixed.
Then there exists some positive constant $C$ independent of the population sizes that satisfies, for any $1\leq i\leq N_1$,
\be
\small
\begin{split}
\bigl|J_1^i(\opb)-\calj_1^i(\opb^i)\bigr| \leq C\vep_{N_1}, \nn
\end{split}
\ee
where $\vep_{N_1}$ is defined by $(\ref{vepN1-def})$. 
\begin{proof}
Let us denote by $\ul{u}^\mgn:=(\ul{X}^\mgn, \ul{x}^\mgn,\ul{y}^\mgn)$ and $u:=(X^i,x^j,y^j)$
the unique solution to $(\ref{fbsde-approx})$ and $(\ref{fbsde-mfg})$, respectively.
For given processes $\ul{u}^\mgn, u$ and $\opb$,  we just put
\be
\small
\begin{split}
&B_{\ul{X}^i}(t,\ul{u}_t^{\mgn}):=\opb_t^i+\rho_1(t)\vp_t(\ul{y}^\mgn_t,\opb_t)+l_1^i(t), \\
&B_{\ul{x}^j}(t,\ul{u}_t^\mgn):=-\ol{\Lambda}_t(\ul{y}_t^{\mgn,j}+\vp_t(\ul{y}^\mgn_t,\opb_t))+\rho_2(t)\vp_t(\ul{y}_t^\mgn, \opb_t)+l_2^j(t), \\
&F_{\ul{y}^j}(t,\ul{u}_t^\mgn):=-\part_x\ol{f}_2^j(t,\ul{x}_t^{\mgn,j})+b_t\vp_t(\ul{y}^\mgn_t,\opb_t), \quad G_{\ul{y}^j}(\ul{u}_T^\mgn):=\part_x g_2^j(\ul{x}_T^{\mgn,j}), \nn
\end{split}
\ee
and also $(B_{X^i}(t,u_t),B_{x^j}(t,u_t),F_{y^j}(t,u_t),G_{y^j}(u_T))$ as in $(\ref{BFG-mfg})$.
For notational ease, we shall use 
$\Del X_t^i:=\ul{X}_t^{\mgn,i}-X_t^i,  \Del x_t^j:=\ul{x}_t^{\mgn,j}-x_t^j, \Del y_t^j:=\ul{y}_t^{\mgn,j}-y_t^j$,
$\Del B_{x^j}(t):=B_{\ul{x}^j}(t,\ul{u}^{\mgn}_t)-B_{x^j}(t,u_t)$, $\Del F_{y^j}(t):=F_{\ul{y}^j}(t,\ul{u}^\mgn_t)-F_{y^j}(t,u_t)$, 
$\Del G_{y^j}:=G_{\ul{y}^j}(\ul{u}_T^\mgn)-G_{y^j}(u_T)$, $\Del \vp_t:=\vp_t(\ul{y}^\mgn,\opb_t)-\pi_t(y^j_t,\opb^i_t)$.
We have the following expansion for $\Del \vp_t$:
\be
\small
\begin{split}
\Del \vp_t
&=-\mg{m}_2(\Del y_t)-(\mg{m}_2(y_t)-\ex^0[y_t^1])+\deln\Lambda_t(\mg{m}_1(\opb_t)-\ex^0[\opb_t^1]). 
\label{delvp-expansion}
\end{split}
\ee

Using the exchangeability, $(\ref{delvp-expansion})$ and the fact that $\sum_{j=1}^{N_2}\ex\bigl[\langle \ol{\Lambda}_t(\Del y_t^j-\mg{m}_2(\Del y_t)), \Del y_t^j\rangle
\bigr]\geq 0$,  we have
\be
\small
\begin{split}
&\ex\bigl[\langle \Del B_{x^j}(t), \Del y_t^j\rangle\bigr]=\frac{1}{N_2}\sum_{j=1}^{N_2}\ex\bigl[\langle -\ol{\Lambda}_t(\Del y^j_t+\Del \vp_t)
+\rho_2(t)\Del \vp_t,\Del y_t^j\rangle\bigr] \\
&\leq -\ex\big[\rho_2(t)|\mg{m}_2(\Del y_t)|^2\bigr]+\ex\bigl[\langle -(\mg{m}_2(y_t)-\ex^0[y_t^1])+\deln\Lambda_t(\mg{m}_1(\opb_t)-\ex^0[\opb_t^1]),
(\rho_2(t)-\ol{\Lambda}_t)\mg{m}_2(\Del y_t)\rangle\bigr]. \nn
\end{split}
\ee
Lemma~\ref{IID-1} and Cauchy-Schwarz inequality give
\be
\small
\begin{split}
\ex\bigl[\langle \Del B_{x^j}(t), \Del y_t^j\rangle\bigr]\leq -\ex[\rho_2(t)|\mg{m}_2(\Del y_t)|^2]+C\vep_{N_1}
\ex\bigl[|\mg{m}_2(\Del y_t)|^2\bigr]^\frac{1}{2}. \nn
\end{split}
\ee 
From Assumption~\ref{A-non-C} (iv), $(\ref{delvp-expansion})$ and Young's inequality, we also have
\be
\small
\begin{split}
\ex\bigl[\langle \Del F_{y^j}(t),\Del x_t^j\rangle\bigr]&
\leq \ex\bigl[-\gamma_2^f|\Del x_t^j|^2+b_t\langle -\mg{m}_2(\Del y_t)-(\mg{m}_2(y_t)-\ex^0[y_t^1])+\deln \Lambda_t (\mg{m}_1(\opb_t)
-\ex^0[\opb_t^1]), \Del x_t^j\rangle\bigr]\\
&\leq \ex\Bigl[-\frac{\gamma_2^f}{2}|\Del x_t^j|^2+\frac{b_t^2}{\gamma_2^f}|\mg{m}_2(\Del y_t)|^2|\Bigr]+C\vep_{N_1}^2. \nn
\end{split}
\ee
Since $\frac{b_t^2}{\gamma_2^f}\leq \frac{\rho_2(t)}{2}$ by Assumption~\ref{A-C} (vi), we obtain
\be
\small
\ex\bigl[\langle \Del B_{x^j}(t),\Del y_t^j\rangle+\langle \Del F_{y^j}(t),\Del x_t^j\rangle\bigr]
\leq -\frac{\gamma_2^f}{2}\ex[|\Del x_t^j|^2]+C\bigl(\vep_{N_1}^2+\vep_{N_1}\ex\bigl[|\mg{m}_2(\Del y_t)|^2\bigr]^\frac{1}{2}\bigr).
\label{drift-conv}
\ee
We also have, as a direct result of Assumption~\ref{A-non-C} (iv),
\be
\small
\begin{split}
\ex\bigl[\langle \Del G_{y^j}, \Del x_T^j\rangle\bigr]\geq \gamma_2^g \ex\bigl[|\Del x_T^j|^2].
\end{split}
\label{terminal-conv}
\ee
On the other hand,  It\^o formula applied to $\langle \Del y^j_t, \Del x_t^j\rangle$ gives an equality
\be
\small
\begin{split}
\bigl[\langle \Del G_{y^j}, \Del x_T^j\rangle\bigr]=\ex\int_0^T\bigl[\langle \Del B_{x^j}(t),\Del y_t^j\rangle
+\langle \Del F_{y^j}(t),\Del x_t^j\rangle\bigr]dt.
\nn
\end{split}
\ee
Applying $(\ref{drift-conv})$ and $(\ref{terminal-conv})$ to the above equality, we obtain, with some constant $C$,
\be
\small
\begin{split}
\ex\Bigl[|\Del x_T^j|^2+\int_0^T |\Del x_t^j|^2dt\Bigr]\leq C\Bigl(\vep_{N_1}^2+\vep_{N_1}\int_0^T \ex\bigl[|\mg{m}_2(\Del y_t)|^2\bigr]^\frac{1}{2}dt
\Bigr). 
\end{split}
\label{del-second-pop}
\ee

Next, a simple application of It\^o formula to $|\Del y_t^j|^2$ gives
\be
\small
\begin{split}
\ex\bigl[|\Del y_t^j|^2\bigr] &\leq \ex\Bigl[|\Del G_{y^j}|^2-2\int_t^T \langle \Del F_{y^j}(s), \Del y_s^j\rangle ds\Bigr]
\leq C\ex\Bigl[|\Del x_T^j|^2+\int_t^T (|\Del x_s^j|+|\Del \vp_s|)|\Del y_s^j|ds\Bigr] \\
&\leq C\ex\Bigl[|\Del x_T^j|^2+\int_0^T|\Del x_t^j|^2 dt\Bigr]+C\vep_{N_1}^2+C\int_t^T\ex\bigl[|\Del y_s^j|^2\bigr]ds, \nn
\end{split}
\ee
where, in the last line, we used exchangeability, $(\ref{delvp-expansion})$ and Lemma~\ref{IID-1}.
Hence, from the backward Gronwall's inequality, exchangeability and $(\ref{del-second-pop})$, we get
\be
\small
\begin{split}
\sup_{t\in[0,T]}\ex\bigl[|\Del y_t^j|^2\bigr] 
&\leq C\vep_{N_1}^2+C\vep_{N_1}\int_0^T \ex[|\Del y_t^j|^2]^\frac{1}{2}dt, \nn
\end{split}
\ee
which gives $\sup_{t\in[0,T]}\ex\bigl[|\Del y_t^j|^2\bigr]\leq C\vep_{N_1}^2$ by Young's inequality.
This in turn implies, by $(\ref{del-second-pop})$,
\be
\small
\begin{split}
\sup_{t\in[0,T]}\ex\bigl[|\Del x_t^j|^2+|\Del y_t^j|^2+|\Del \vp_t|^2\bigr]\leq C\vep_{N_1}^2. \nn
\end{split}
\ee
Finally, since $\Del X_t^i=\int_0^t \rho_1(t)\Del \vp_t dt$, we conclude, for any $1\leq i\leq N_1, 1\leq j\leq N_2$,
\be
\small
\begin{split}
\sup_{t\in[0,T]}\ex\bigl[|\Del X_t^i|^2+|\Del x_t^j|^2+|\Del y_t^j|^2+|\Del \vp_t|^2\bigr]\leq C\vep_{N_1}^2. 
\end{split}
\label{conv-prop2-stability}
\ee

Form $(\ref{conv-prop2-stability})$ and the local Lipschitz continuity of $g_1$,  we have
\be
\small
\begin{split}
&\bigl|\ex\bigl[g_1^i(\ul{X}_T^{\mgn,i},\ul{\nu}_T^{N_1})-g_1^i(X_T^i,\mu_T)\bigr]\bigr|\\
&\leq C\ex\bigl[(1+|\ul{X}_T^{\mgn,i}|+|X_T^i|+M_2(\ul{\nu}_T^{N_1})+M_2(\mu_T)+|c_T^0|+|c_T^i|)(|\Del X_T^i|+W_2(\ul{\nu}_T^{N_1},\mu_T))\bigr]\\
&\leq C\ex\bigl[|\Del X_T^i|^2+W_2(\mu_T^{N_1}, \mu_T)^2\bigr]^\frac{1}{2}\leq C\vep_{N_1}. 
\end{split}
\label{conv-prop2-terminal}
\ee
Here, the triangle inequality $W_2(\ul{\nu}_T^{N_1},\mu_T)\leq W_2(\ul{\nu}_T^{N_1}, \mu_T^{N_1})+W_2(\mu_T^{N_1},\mu_T)$
and the fact 
\be
\small
\begin{split}
\ex\bigl[W_2(\ul{\nu}_T^{N_1},\mu_T^{N_1})^2\bigr]\leq \frac{1}{N_1}\sum_{k=1}^{N_1}\ex\bigl[|\ul{X}_T^{\mgn,k}-X_T^k|^2\bigr]
=\ex[|\Del X_T^i|^2] \nn
\end{split}
\ee
were used. Similar calculation using the local Lipschitz continuity also yields
\be
\small
\begin{split}
&\Bigl|\ex\int_0^T\Bigl[f_1^i(t,\ul{X}_t^{\mgn,i},\ul{\nu}_t^{N_1},\vp_t(\ul{y}_t^\mgn,\opb_t),\opb_t^i)
-f_1^i(t,X_t^i,\mu_t,\pi_t(y_t^j,\opb_t^i),\opb_t^i)\Bigr]dt\Bigr|\\
&\leq C\sup_{t\in[0,T]}\ex\bigl[|\Del X_t^i|^2+|\Del \vp_t|^2+W_2(\mu_t^{N_1},\mu_t)^2\bigr]^\frac{1}{2}\leq C\vep_{N_1}.
\end{split}
\label{conv-prop2-running}
\ee
The conclusion follows from inequalities $(\ref{conv-prop2-terminal})$ and $(\ref{conv-prop2-running})$.
\end{proof}
\end{proposition}

We arrived the last main result of this paper.
\begin{theorem}
\label{th-convergence}
Let Assumptions~\ref{A-non-C}-\ref{A-C} be in force  and the ratio $\del_n\leq \del_*$ be kept fixed.
Then the equilibrium of the finite-agent market strongly converges to the mean-field equilibrium in the large population limit of $N_1$~(and hence $N_2$) 
$\rightarrow \infty$ in the sense that, for any $N_1$, there exists some positive constant $C$ independent of population sizes that satisfies,
with $\vep_{N_1}$ defined by $(\ref{vepN1-def})$, 
\be
\small
\begin{split}
{\rm (i):}~&|J_1^i(\opb^\mgn)-\calj^i(\opb^i)| \leq C\vep_{N_1}, \\
{\rm (ii):}~&\ex\int_0^T \Bigl[|\opb_t^{\mgn,i}-\opb_t^i|^2+|\vp_t(y^{\mgn}_t,\opb^\mgn_t)-\pi_t(y_t^1,\opb_t^1)|^2\Bigr]dt \leq C\vep_{N_1}, \\
{\rm (iii):}~&\ex\Bigl[\sup_{t\in[0,T]} \bigl(|X_t^{\mgn,i}-X_t^i|^2+|x_t^{\mgn,j}-x_t^j|^2+|y_t^{\mgn,j}-y_t^j|^2\bigr)\Bigr]\leq C\vep_{N_1} \nn
\end{split}
\ee
for any $1\leq i\leq N_1$ and $1\leq j\leq N_2$.
\begin{proof}
Let us use the same conventions as in Proposition~\ref{prop-convergence-1}, i.e.,
$\Del X_t^i:=X_t^{\mgn,i}-X_t^i$, $\Del \opb_t^i:=\opb_t^{\mgn,i}-\opb_t^i$,
$\vp_t:=\vp_t(y_t^\mgn,\opb_t^\mgn)$, $\pi_t:=\pi_t(y_t^1,\opb_t^1)=\pi_t(y_t^j,\opb_t^i)$ and
$\Del X_t^i:=X_t^{\mgn,i}-X_t^i$, $\Del y_t^j:=y_t^{\mgn, j}-y_t^j$, $\Del \opb_t^i:=\opb_t^{\mgn,i}-\opb_t^i$,
$\Del \vp_t:=\vp_t-\pi_t$, etc. 
Note that,  from the results of  Propositions~\ref{prop-convergence-1} and \ref{prop-convergence-2}, we have
\be
\small
\begin{split}
&\calj_1^i(\opb^i)+\ex\int_0^T \Bigl[\frac{\gamma_1^f}{2}|\Del X_t^i|^2+\frac{\lambda_\beta}{2}|\Del \opb_t^i|^2+
\frac{\lambda_\vp}{2}|\Del \vp_t|^2\Bigr]dt+\frac{\gamma_1^g}{2}\ex\bigl[|\Del X_T^i|^2\bigr]-C\vep_{N_1}\\
&\leq J_1^i(\opb^{\mgn})\leq J_1^i(\opb) \leq \calj_1^i(\opb^i)+C\vep_{N_1}. \nn
\end{split}
\ee
This gives (i) and (ii).
Since 
$
\Del X_t^i=\int_0^t \bigl[\Del \opb_t^i+\rho_1(t)\Del \vp_t\bigr]dt, \nn
$
it is easy to obtain
\be
\small
\begin{split}
\ex\bigl[\sup_{t\in[0,T]}|\Del X_t^i|^2\bigr]\leq C\ex\int_0^T\bigl[|\Del \opb_t^i|^2+|\Del \vp_t|^2\bigr]dt\leq C\vep_{N_1}. \nn
\end{split}
\ee
Moreover,  applying the convexity of $g_2, \ol{f}_2$ to the equality
\be
\small
\begin{split}
\ex\bigl[\langle \Del y_T^j, \Del x_T^j\rangle\bigr]&=\ex\int_0^T \Bigl[\langle -\ol{\Lambda}_t\Del y_t^j+(\rho_2(t)-\ol{\Lambda}_t)\Del \vp_t,
\Del y_t^j\rangle +\langle -(\part_x \ol{f}_2^j(t,x_t^{\mgn,j})-\part_x \ol{f}_2^j(t,x_t^j))+b_t\Del \vp_t, \Del x_t^j\rangle\Bigr]dt,  \nn
\end{split}
\ee
we obtain, with some constant $C$,  
\be
\small
\begin{split}
\ex\Bigl[|\Del x_T^j|^2+\int_0^T |\Del x_t^j|^2dt\Bigr]\leq C\ex\int_0^T \bigl[|\Del \vp_t|^2+|\Del \vp_t||\Del y_t^j|\bigr]dt. \nn
\end{split}
\ee
Following the same procedures used in Proposition~\ref{prop-convergence-2}, it is simple to obtain
\be
\small
\begin{split}
\sup_{t\in[0,T]}\ex\bigl[|\Del x_t^j|^2+|\Del y_t^j|^2\bigr]\leq C\ex\int_0^T |\Del \vp_t|^2 dt\leq C\vep_{N_1}. \nn
\end{split}
\ee
BDG inequality then provides the desired estimate.
\end{proof}
\end{theorem}

\begin{remark}
If the law-$\mu$ dependence in $(f_1, g_1)$ is just given by its first moment $\int_{\mbb{R}^n}x \mu(dx)$,
then we can replace $\vep_{N_1}$ by $1/\sqrt{N_1}$ since the first part of Lemma~\ref{IID-1} becomes unnecessary.
Therefore, in this case,  we have an explicit convergence speed $\propto 1/\sqrt{N_1}$.
\end{remark}

Theorem~\ref{th-convergence} tells that the market clearing price $\vp_t(y_t^{\mgn},\opb^\mgn_t)$
converges to an $\mbb{F}^0$-adapted process $\pi_t(y_t,\opb_t)$.
This means that the securities prices only respond to the market-wide news and shocks,  
which is consistent with our intuition. This also allows each agent, when the population is large enough,  to focus on the market-wide
and her own idiosyncratic informations  without taking care of the idiosyncratic informations of the others.
Recall the discussion given in  Remark~\ref{remark-info-structure}.
Therefore, we can see that an awkward setup with the perfect information with $\mbb{F}$-adapted controls 
is effectively resolved as the population size grows.
\section{Conclusion and discussion}
\label{sec-conclusion}
In this work, we developed an equilibrium model for price formation in a market composed of 
cooperative and non-cooperative populations.
In the large population limit, we have seen that the problem for the central planner 
becomes an extended mean-field control over the FBSDEs of 
McKean-Vlasov type with common noise. 
In addition to the convexity assumptions, if the relative size of the cooperative population is small enough, 
then we were able to show the existence of a unique equilibrium for both the finite-agent and the mean-field models.
The strong convergence to the mean-field model was also proved under the same conditions.

The development of feasible numerical calculation techniques is an important remaining problem.
If we adopt the linear-quadratic setting as a special example, which has a semi-analytic solution, 
then we may study how the equilibrium is destabilized as the relative population size $\deln$ of the cooperative agents grows. 
Since, in our model,  the control domain $A_1$ can be an arbitrary closed convex subset of $\mbb{R}^n$,
we can fix the direction of trade if necessary. This will be useful when we consider a model of cooperative producers (or buyers)
in the background of a large number of competitive agents.
Extensions of the current model for these economic applications deserve further research.

\subsubsection*{Acknowledgments}
The author thanks Masashi Sekine of Univ.~of Tokyo for useful discussions. 
The author also thanks anonymous referees for valuable comments to improve the paper.
\vspace{-3mm}
\begin{spacing}{0.8}
 
\end{spacing}

\end{document}